\documentclass[prd,reprint,nofootinbib,superscriptaddress,preprintnumbers,showpacs,floatfix]{revtex4-1}
\usepackage{amsfonts}
\usepackage{amssymb}
\usepackage{amsmath}
\usepackage{graphicx,color}
\usepackage{dcolumn}
\usepackage{epsfig}
\usepackage{bm}
\usepackage{bbm}
\usepackage{ulem}
\usepackage{hyperref}
\usepackage{lineno}
\usepackage{multirow}

\usepackage{color}

\def\gtap{\ \raise.3ex\hbox{$>$\kern-.75em\lower1ex\hbox{$\sim$}}\ }
\def\ltap{\ \raise.3ex\hbox{$<$\kern-.75em\lower1ex\hbox{$\sim$}}\ }
\begin{document}

\title{
Global coupled-channel analysis of $e^+e^-\to c\bar{c}$ processes in 
$\sqrt{s}=3.75-4.7$~GeV
}
\author{S.X. Nakamura}
\email{satoshi@sdu.edu.cn}
\affiliation{
Institute of Frontier and Interdisciplinary Science, Shandong
University, Qingdao, Shandong 266237, China
}
\affiliation{
University of Science and Technology of China, Hefei 230026, 
China
}
\affiliation{
State Key Laboratory of Particle Detection and Electronics, Beijing
100049, Hefei 230026, China
}
\author{X.-H. Li}
\author{H.-P. Peng}
\affiliation{
University of Science and Technology of China, Hefei 230026, 
China
}
\affiliation{
State Key Laboratory of Particle Detection and Electronics, Beijing
100049, Hefei 230026, China
}
\author{Z.-T. Sun}
\affiliation{
Institute of Frontier and Interdisciplinary Science, Shandong
University, Qingdao, Shandong 266237, China
}
\author{X.-R. Zhou}
\affiliation{
University of Science and Technology of China, Hefei 230026, 
China
}
\affiliation{
State Key Laboratory of Particle Detection and Electronics, Beijing
100049, Hefei 230026, China
}

\begin{abstract}
Recent high-precision $e^+e^-\to c\bar{c}$ data
from the BESIII and Belle 
are highly useful to understand vector charmonium ($\psi$) pole
 structures and puzzling lineshapes due to the exotic hadron candidates~$Y$.
We thus perform a global coupled-channel analysis
of most of the available data 
(10 two-body, 9 three-body, and 1 four-body final states)
in $\sqrt{s}=3.75-4.7$~GeV.
Not only cross sections but also
invariant-mass distributions of subsystems are fitted.
The $e^+e^-\to \mu^+\mu^-$ cross sections are also predicted. 
Our model includes dozens of (quasi) two-body states
that nonperturbatively couple with each other 
through bare $\psi$ excitations,
particle-exchange,
and short-range mechanisms;
approximate three-body unitarity is considered.
The amplitudes obtained from the fit are
 analytically continued to $\psi$ and $Z_c$ poles.
We find $\psi$ states similar to
those in the Particle Data Group listing and $Y(4320)$.
Moreover, several $\psi$ states, including new ones, are found close to open
 charm thresholds.
Trajectories and compositeness of the near-threshold poles suggest 
dominant hadron-molecule contents in
their internal structures.
Two $Z_c$ poles are found as virtual states 
$\sim$40~MeV below the 
$D^*\bar{D}^{(*)}$ thresholds, 
being consistent with lattice QCD results. 
This work presents the first global analysis
to determine $\psi$ and $Z_c$ poles, 
thereby paving the way to extracting detailed properties of the prominent
 exotic hadron candidates from data.
\end{abstract}

\maketitle

\section{Introduction}

The $Y$-sector of $XYZ$ exotic hadrons was opened with the discovery of
$Y(4260)$ by the BABAR Collaboration~\cite{babar-y4260}
in $e^+e^-\to\gamma_{\rm ISR}\pi^+\pi^- J/\psi$
($\gamma_{\rm ISR}$: initial state radiation $\gamma$),
and subsequent
confirmations by the CLEO and Belle
Collaborations~\cite{cleo-y4260,belle-y4260}.\footnote{
While we basically follow the particle name convention of
the Particle Data Group (PDG)~\cite{pdg},
we sometimes use historical names such as $Y$.
Often,
$D_1(2420)$, $D_1(2430)$, $D_2^*(2460)$, $D_0^*(2300)$, 
and $D_{s1}(2536)$ 
are simplified as
$D_1$, $D'_1$, $D_2$, $D_0$, and $D_{s1}$, respectively.
Either or all of $D\bar{D}$, $D^{*}\bar{D}$, and $D^{*}\bar{D}^{*}$
are collectively denoted 
by $D^{(*)}\bar{D}^{(*)}$.
}
The $Y(4260)$ has been considered an exotic state.
One reason is for its peculiar decay patterns.
Usually, charmonium states above open-charm thresholds dominantly decay
into the open-charm channels.
However, $Y(4260)$ signal were not seen in $e^+e^-\to $ 
(open-charm channels) data, but were seen in hidden-charm channels. 
Second, 
$Y(4260)$ does not have a quark-model counterpart~\cite{barnes}.
The discovery of $Y$ continued:
$Y(4360)$ in $e^+e^-\to\gamma_{\rm ISR}$ $\pi^+\pi^- \psi'$
by the BABAR~\cite{babar-y4360} and Belle~\cite{belle-y4360},
and $Y(4660)$ by the Belle~\cite{belle-y4360}.
The exotic hadrons are considered a key to deepening our understanding of
QCD and thus invited lots of studies; see reviews~\cite{review_chen,review_hosaka,review_lebed,review_esposito,review_ali,review_guo,review_olsen,review_Brambilla}.

The BESIII pursued the precision frontier of the $Y$ sector
with direct $Y$ productions without $\gamma_{\rm ISR}$, 
and found that $Y$ widths appear differently in their different decay
modes ($Y$-width problem)~\cite{bes3_jpsi-pippim,bes3_jpsi-pi0pi0,bes3_jpsi-ksks,bes3_jpsi-kpkm,bes3_psip_pippim,bes3_hc_pippim,bes3_chic0omega,bes3_piDDstar,bes3_jpsi-eta};
see Fig.~4 of \cite{bes3_jpsi-eta}.
The BESIII also found that 
$Y(4260)$ in $e^+e^-\to\pi^+\pi^- J/\psi$
consists of 
$Y(4220)$ and $Y(4320)$~\cite{bes3_jpsi-pippim-prev},
and that $Y(4320)$ does not appear in other final states. 
$Y(4360)$ and $Y(4660)$ were confirmed with higher precision~\cite{bes3_psipipi_Zc,bes3_psip_pippim}.

Actually, the process-dependent $Y$ lineshapes
can be caused by process-dependent
interferences between various charmonia, and by
kinematical effects such as threshold opening/cusp and triangle
singularity.
Thus, the $Y$-width problem indicates the limitation of determining 
the resonance parameters 
by collecting results of 
single-channel analyses;
a single-channel analysis determines
resonance parameters from fitting only one process.
We should understand 
the process-dependent $Y$ lineshapes 
by analyzing the different final states simultaneously
with a unified coupled-channel model;
no need to resort to the 
process-dependent $Y$ widths.

The $Z^+_c(3900)$ is an outstanding exotic $c\bar{c}u\bar{d}$
candidate, and was discovered in the $J/\psi\pi^+$ invariant-mass distribution
of $e^+e^-\to\pi^+\pi^- J/\psi$~\cite{z3900-1-bes3,z3900-1-belle}.
Then, $Z^+_c(4020)$ was discovered in a study of 
$e^+e^-\to\pi^+\pi^- h_c$~\cite{z4020-1-bes3}.
Furthermore, 
$Z^+_c$ signals were also observed in 
the invariant-mass distributions of two-body subsystems in
$e^+e^-\to D^*\bar{D}\pi$~\cite{bes3_DDstarc_zc3900},
$D^*\bar{D}^*\pi$~\cite{bes3_piDstarDstar},
$\psi'\pi\pi$~\cite{bes3_psipipi_Zc}, and
$\eta_c\rho\pi$~\cite{bes3_etac_pippimpi0}.
The properties of $Z_c$ and $Y$ should be correlated
since $Z_c$ appear as $Y\to Z_c\pi$.
Inevitably, the above coupled-channel analysis 
considers the $Z_c$ signals in the data, and addresses their nature.

Regarding previous coupled-channel studies, 
Cleven et al.~\cite{cleven14} fitted
 $e^+e^-\to J/\psi\pi^+\pi^-$, $h_c\pi^+\pi^-$, and
$D\bar{D}^*\pi$ cross sections and invariant-mass lineshapes to study $Y(4260)$.
This pioneering work showed that the asymmetric
$Y(4260)$-lineshapes can be naturally explained with a 
$D_1\bar{D}$-molecule scenario of $Y(4260)$.
However, the data available at that time were rather scarce, compared to
what we have today. 
Recently, Detten et al.~\cite{detten} performed
a similar study including updated data and more final states in the 
$\psi(4230)$ region. To explain the process-dependent lineshapes, they
considered interferences between $\psi(4160)$ and $\psi(4230)$.
They concluded that the data are consistent with the 
$D_1\bar{D}$-molecule interpretation of $\psi(4230)$.
However, as they noted, their study was still exploratory.
Their model did not account for the unitarity that should be
important to describe overlapping resonances. 
Also, they did not include data such as 
$e^+e^-\to D^*\bar{D}^*\pi$~\cite{bes3_piDstarDstar_xs}
and $\psi'\pi\pi$~\cite{bes3_psip_pippim} in their fit.
It remains to be seen whether 
the $D_1\bar{D}$-molecule scenario for $\psi(4230)$
can also explain 
the $e^+e^-\to D^*\bar{D}^*\pi$ and $\psi'\pi\pi$ data showing
$\psi(4230)$ signals.

Chen et al.~\cite{dychen} performed a Breit-Wigner fit to 
$e^+e^-\to  D^*\bar{D}\pi$, $J/\psi\pi\pi$, and $h_c\pi\pi$
cross-section data, and concluded that 
$Y(4320)$ and $Y(4390)$ signals in the data can be explained with 
interferences between 
$\psi(4160)$, $\psi(4230)$, and $\psi(4415)$.
However, their analysis did not include 
$e^+e^-\to\psi'\pi\pi$ data~\cite{bes3_psipipi_Zc}, available at that time,
that clearly shows a $Y(4360/4390)$ signal. 
Zhou et al.~\cite{zyzhou} fitted
$e^+e^-\to D^{(*)}\bar{D}^{(*)}$ and $D\bar{D}\pi$ data with 
a two-body unitary coupled-channel model. 
Their interesting conclusion is that 
$\psi(4160)$ and $\psi(4230)$ are the same state.
However, one of bare charmonium states in their model has an unreasonably
small mass, which 
could be an artifact of not including nonresonant mechanisms. 
Also, the $D\bar{D}\pi$ data were not reasonably fitted.
Overall, the previous analyses used rather limited datasets,
considering the recent data discussed below.
Accordingly, several major coupled-channels and unitarity were not considered, 
which would question
the reliability of their conclusions.

In the last several years, the BESIII has accumulated 
high quality data for various $e^+e^-\to c\bar{c}$ 
cross sections and their invariant-mass distributions
over a
wide energy region~\cite{bes3_DD,bes3_DstDst,bes3_DsDs,bes3_DsstDsst,bes3_piDDstar,bes3_piDstarDstar,bes3_jpsi-eta,bes3_jpsi-eta2,bes3_jpsi-etaprime,bes3_chic0omega_2015,bes3_chicJomega,bes3_chic0omega,bes3_jpsi-pippim,bes3_jpsi-kpkm,bes3_jpsi-ksks,bes3_psip_pippim,bes3_hc_pippim,bes3_etac_pippimpi0,bes3_DDstarc_zc3900,bes3_piDstarDstar_xs,bes3_Zc3900c,bes3_psipipi_Zc,z4020-1-bes3,bes3_jpsi-pippim-prev,bes3_jpsi-pi0pi0,bes3_LL1,bes3_LL2};
see Figs.~\ref{fig:xs-opencc}--\ref{fig:hcpipi-etacrhopi}.
It is timely to analyze these data simultaneously with 
a coupled-channel framework,
and extract vector charmonium properties such as their poles (masses, widths) and
residues (coupling strengths to decay channels).
These resonance properties are a primary basis to study 
the nature of $Y$ and a prerequisite to understand
the $Y$ lineshapes.
They are also
new information for 
well-established charmonia [$\psi(4040)$, $\psi(4160)$, $\psi(4415)$]
since their properties have been mainly from analyzing
the inclusive ($e^+e^-\to$ hadrons) data~\cite{bes2-R}.
In this work, we perform such a global coupled-channel analysis 
over $\sqrt{s}=3.75-4.7$~GeV 
for the first time,
and present the fit results and 
the vector charmonium and $Z_c$ pole positions.

Possible interpretations of the internal
structures of the $Y$ states have been proposed.
$\psi(4230)$ as a $D_1\bar{D}$ hadron molecule has been proposed in
Refs.~\cite{gjding,xkdong1,xkdong2,Ji2022,FZPeng2023}.
Similarly, Refs.~\cite{xkdong2,Ji2022} interpreted 
$\psi(4360)$ as a $D_1\bar{D}^*$ hadron molecule.
Possible other interpretations of $\psi(4230)$ are 
a $c\bar{c}$-gluon hybrid state~\cite{hybrid1,hybrid2,hybrid3} 
and a hadrocharmonium~\cite{hadrocharmonium}.\footnote{
Recent BESIII data indicate that the $Y$ states substantially decay into
open-charm channels, 
disfavoring the hadrocharmonium scenario.
See related discussions in Ref.~\cite{review_olsen}.}
Our analysis model is flexible enough to capture the open-charm hadron
molecule structures 
of the vector charmonium states. 
The internal structures of the extracted states will be explored
 by examining the pole trajectories and the 
compositeness~\cite{weinberg1965,sekihara2015,baru2004}.

We also cross-check our coupled-channel model 
by comparing the model's prediction of 
$e^+e^- \to \mu^+\mu^-$ cross sections with data~\cite{ee-mumu-data}.
Since our model includes all major channels involving $c\bar{c}$
quarks, it provides 
the $c\bar{c}$ contribution to the vacuum polarization (VP) that occurs
in $e^+e^- \to \mu^+\mu^-$.
The light-hadron contribution to the VP is obtainable
through a dispersion relation applied to 
the inclusive ($e^+e^-\to$ hadrons) cross-section data~\cite{bes2-R}
subtracted by
our model's ($e^+e^-\to$ $c\bar{c}$-hadrons) cross sections. 

The organization of this paper is as follows.
In Sec.~\ref{sec:model}, we describe our coupled-channel amplitudes for 
$e^+e^- \to c\bar{c}$ processes, and cross-section formulas.
In Sec.~\ref{sec:fit}, our fit results are presented, and reaction
mechanisms are discussed. 
We compare the inclusive $R$ value from our model with data in 
Sec.~\ref{sec:R}, and predict 
$e^+e^- \to \mu^+\mu^-$ cross sections in Sec.~\ref{sec:ee-mumu}.
In Sec.~\ref{sec:pole},
we present vector-charmonium poles extracted from 
the coupled-channel amplitudes, and examine their pole trajectories and
compositeness. 
The $Z_c$ poles are also presented.
A summary is given in Sec.~\ref{sec:summary}.
Appendices discuss
two-body scattering models that are 
main building blocks of our coupled-channel model,
our pole-uncertainty estimation methods,
and model parameter values.

\section{Model}
\label{sec:model}

\subsection{$e^+e^-\to c\bar{c}$ reaction amplitudes from coupled-channel model}
\label{sec:app1}

\begin{figure}
\begin{center}
\includegraphics[width=.5\textwidth]{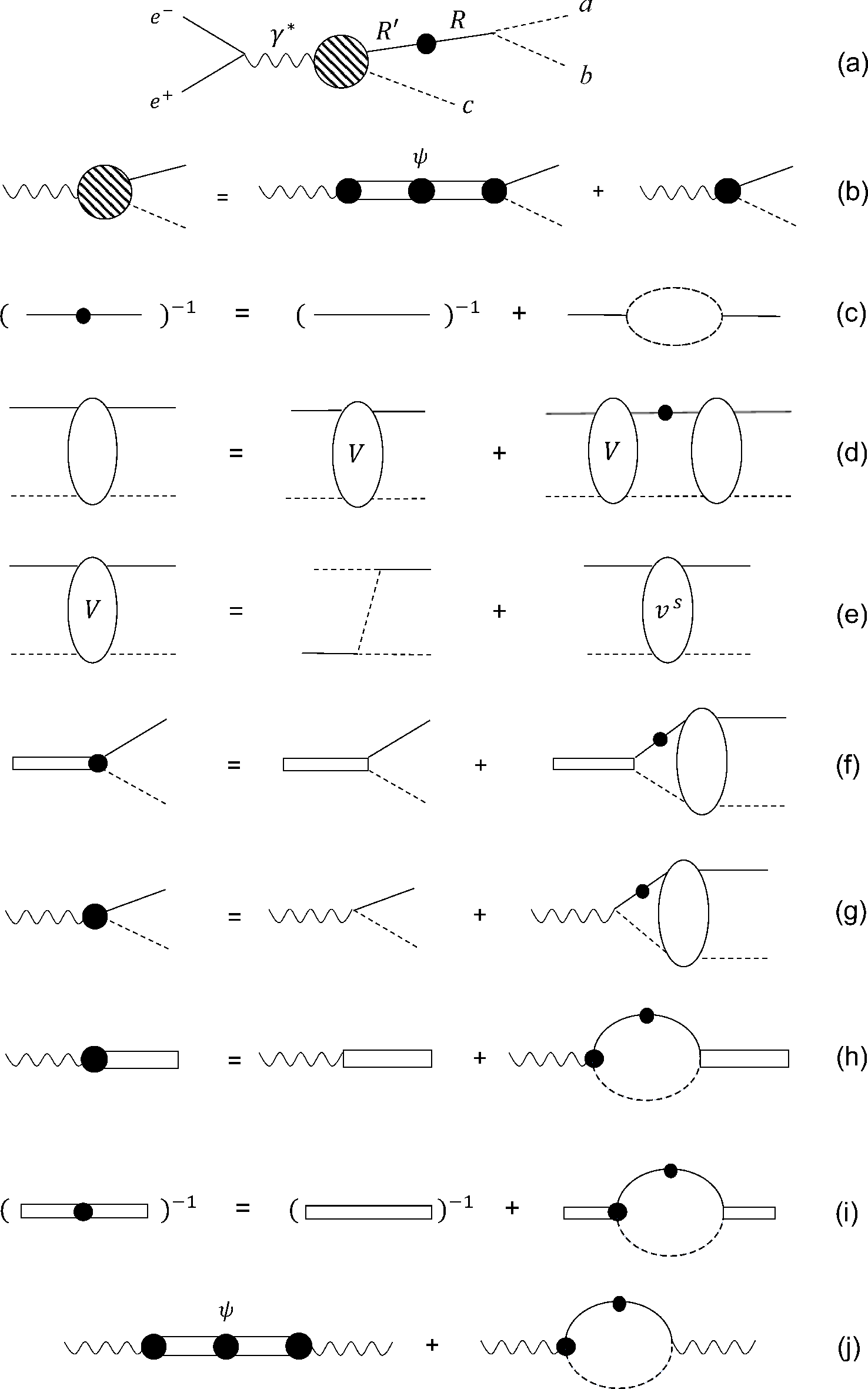}
\end{center}
 \caption{(a) $e^+e^-\to abc$ mechanism in our model.
The dashed lines represent stable particles and 
$abc$ are final three-body states shown in 
Figs.~\ref{fig:xs-opencc}-\ref{fig:xs-hiddencc}.
The solid lines are
bare [Breit-Wigner] resonance states $R$ listed in 
Table~\ref{tab:Rc}(C) [Table~\ref{tab:Rc}(A) and \ref{tab:Rc}(B)].
(b) Resonant and nonresonant mechanisms.
 The double line represents bare charmonium ($\psi$) states.
(c) Dressed $R$ propagator: the first [second] diagram
is a bare $R$ propagator [self energy].
(d) Lippmann-Schwinger-like equation for $Rc$ scattering driven by $V$.
The white oval is a $Rc\to R'c'$ scattering amplitude.
(e) $Rc$ interactions $V$ from particle-exchange and short-range ($v^{\rm s}$) mechanisms.
(f) Dressed $\psi$ decay vertex.
(g) Dressed nonresonant $Rc$ photo-production vertex.
(h) Dressed $\psi$ photo-production vertex.
In (f)-(h), the first [second] diagram
is a bare vertex [rescattering term].
(i) Dressed $\psi$ propagator: the first [second] diagram
is a bare $\psi$ propagator [self energy].
(j) Charm vacuum polarization
 (for $e^+e^-\to \mu^+\mu^-$, not for $e^+e^-\to abc$).
 }
\label{fig:diag}
\end{figure}

Our coupled-channel model for the 
$e^+e^-$ annihilation processes 
is primarily
based on 
the manifestly
three-body unitary formulation presented in Refs.~\cite{3pi,3pi-2,d-decay,etastar}.
For three-body final states, $e^+e^-\to abc$, 
the full amplitude 
[Fig.~\ref{fig:diag}(a,b)]
is given by\footnote{
We denote a particle $x$'s mass, momentum, energy, width, and spin state
in the $abc$ center-of-mass (CM) frame
by $m_x$, $\bm{p}_x$, $E_x$, $\Gamma_x$, and $s_x^z$,
respectively;
$E_x=\sqrt{m_x^2+p_x^2}$ with
$p_x^2=|\bm{p}_x|^2$.
The mass and width values are from the PDG~\cite{pdg}.
Our model is isospin symmetric, and the averaged mass is used for
isospin partners.
}
\begin{eqnarray}
 A_{abc,e^+e^-} &=& 
\sum^{\rm cyclic}_{abc}
\sum_{RR's_R^z}
\Gamma_{ab,R}(\bm{p}_a^*)\,
\tau_{R,R'}(p_c,E-E_{c})\,
\nonumber\\
&&\times \Big[
\sum_{ij}
\bar{\Gamma}^\mu_{R'c,\psi_i}(\bm{p}_c, E)\,
\bar{G}_{ij}(E)\,
\bar\Gamma_{\psi_j,\gamma^*}(E) 
\nonumber \\
&& 
+ \bar\Gamma^\mu_{R'c,\gamma^*}(\bm{p}_c, E) 
\Big]
 {1\over s} l_\mu   ,
\label{eq:amp_full}
\end{eqnarray}
where the first and second terms in the square bracket are
resonant ($\psi$) and nonresonant (NR) parts, respectively.
The symbol $R$ is a two-meson resonance such as $D_1(2420)$;
cyclic permutations $(abc), (cab), (bca)$ are indicated by 
$\sum^{\text{cyclic}}_{abc}$;
$\psi_i$ indicates $i$-th bare $\psi$ state;
$E(=\sqrt{s})$ denotes
the $abc$ invariant mass.
The amplitude includes
$R\to ab$ vertex $\Gamma_{ab,R}$,
dressed $R$ propagator $\tau_{R,R'}$ [Fig.~\ref{fig:diag}(c)], 
dressed $\psi\to Rc$ vertex $\bar{\Gamma}^\mu_{Rc,\psi}$ [Fig.~\ref{fig:diag}(f)],
dressed NR $Rc$ production mechanism 
$\bar\Gamma^\mu_{Rc,\gamma^*}$ [Fig.~\ref{fig:diag}(g)], 
dressed $\psi$ production mechanism 
$\bar\Gamma_{\psi,\gamma^*}$ [Fig.~\ref{fig:diag}(h)], 
and 
dressed $\psi$ propagator $\bar{G}_{ij}$ [Fig.~\ref{fig:diag}(i)].
The virtual photon propagator is $1/s$ and
the lepton current matrix element is 
$l_\mu (= e\bar{v}_{e^+}\gamma_\mu u_{e^-})$.
Amplitudes for two-body final states 
($A_{ab,e^+e^-}$) are obtained
from Eq.~(\ref{eq:amp_full})
by removing 
$\Gamma_{ab,R}\tau_{R,R'}$ and identifying 
$R'c$ with $ab$.

We consider $Rc$ channels summarized in Table~\ref{tab:Rc}.
These channels are understood to be negative $C$-parity ($C=-1$) states.
Taking a convention of
$D_J \xrightarrow{C} \bar{D}_J$ for all charmed mesons $D_J$,
we use a $C=-1$ base for an open-charm channel as
\begin{eqnarray}
\label{eq:c-base}
{1\over\sqrt{2}}
(D_J\bar{D}_{J'}-\bar{D}_JD_{J'}) ,
\end{eqnarray}
where $D_J\ne D_{J'}$ and $m_{D_J} > m_{D_{J'}}$.
We group the $Rc$ channels into (A)--(C) in Table~\ref{tab:Rc}.
A (bare) $R$ state is excited 
in a partial-wave two-body scattering
as $ab\to R\to a'b'$.
The partial wave is specified by 
$\{L,I\}$ or $IJ^{P(C)}$, where
$L$, $I$, $J$, and $P$ are 
the orbital angular momentum, total isospin, total angular momentum, and parity
of the $ab$ (or $a'b'$) system, respectively.
Models for these $ab\to a'b'$ 
two-body scatterings are basic building blocks of the
three-body $e^+e^-\to c\bar{c}$ reaction model described in this subsection,
and are discussed in detail in Appendix~\ref{sec:two-meson}.
In particular, numerical values for 
$R\to ab$ couplings $g^{LS}_{ab,R}$, cutoffs $c_{ab,R}$,
and (bare) $R$ masses $m_R$ are determined from analyzing two-body data
and given in the Appendix~\ref{sec:two-meson};
$g^{LS}_{ab,R}$ and $c_{ab,R}$ will appear in
Eq.~(\ref{eq:pipi-vertex}), 
and $m_R$ in 
Eqs.~(\ref{eq:green-Rc}) and (\ref{eq:bw-Rc}).
For groups (A) and (B), the $R$-propagations are described in a Breit-Wigner
(BW) form.
For group (B), we do not consider $R\to ab$ couplings.
For group (C), 
$R=D_0^*(2300)$, $f_{0(2)}$, and $Z_{c(s)}$ indicate 
bare states that are 
dressed to form poles in
unitary coupled-channel scattering amplitudes 
for 
$\{L,I\}=\{0,1/2\}$ $D\pi$,
$\{0(2),0\}$ $\pi\pi-K\bar{K}$, and
$IJ^{PC}=11^{+-}$ 
$D^*\bar{D}-D^*\bar{D}^*-J/\psi\pi-\psi'\pi-h_c\pi-\eta_c\rho$
($IJ^{P}={1\over 2}1^{+}$ $D_s^*\bar{D}-D_s\bar{D}^*-J/\psi K$), respectively.
We refer to the amplitudes as the $D_0^*(2300)$, $f_{0(2)}$, and
$Z_{c(s)}$ amplitudes, respectively.

\begin{table}[t]
\renewcommand{\arraystretch}{1.3}
\tabcolsep=1.0mm
\caption{\label{tab:Rc}
Quasi two-body ($Rc$) coupled-channels with $J^{PC}=1^{--}$.
See text for grouping (A)-(C).
}
\begin{ruledtabular}
\begin{tabular}{ll}
(A)& $D_1(2420)\bar{D}^{(*)}$, $D_1(2430)\bar{D}^{(*)}$,
     $D_2^*(2460)\bar{D}^{(*)}$, $D^{(*)}\bar{D}^{(*)}$ ,\\
&$D_{s1}(2536)\bar{D}_s$ \\
(B) &
$D_s^{(*)}\bar{D}_s^{(*)}$, 
$J/\psi\eta$, $J/\psi\eta'$, $\omega\chi_{c0}$, $\Lambda_c\bar{\Lambda}_c$
\\
(C) &$D_0^*(2300)\bar{D}^*$, $f_0J/\psi$, $f_2J/\psi$, $f_0\psi'$,
     $f_0 h_c$, $Z_c\pi$, $Z_{cs}\bar{K}$
\end{tabular}
\end{ruledtabular}
\end{table}

The $R\to ab$ vertex is given by
\begin{eqnarray}
\Gamma_{ab,R}(\bm{p}_a^*)&=&
(t_a t^z_a t_b t^z_b |t_R t^z_R)
\sum_{LL^zSS^z}  
(s_a s^z_a s_b s^z_b | S S^z) 
\nonumber\\
&&\times
(L L^z S S^z |s_R s^z_R )
Y_{LL^z}(\hat{p}^*_a) 
\nonumber\\
&&\times
\sqrt{E_a(p_a^*) E_b(p_a^*)
\over E_a(p_a) E_b(p_b)}
f^{LS}_{ab,R}(p_a^*) ,
\label{eq:pipi-vertex0}
\end{eqnarray}
where $\bm{p}_a^*$ is the momentum of a particle $a$
in the $ab$ CM frame and $p_a^*=|\bm{p}_a^*|$.
The spherical harmonics is denoted by $Y_{LL^z}(\hat{p}_a^*)$
with $\hat{p}_a^*=\bm{p}_a^*/|\bm{p}_a^*|$.
The parentheses are Clebsch-Gordan (CG) coefficients where
$t_x$ and $t_x^z$
are the isospin of a particle $x$ and its $z$-component, respectively.
The total spin of $ab$ is denoted by $S$.
We also used the vertex function
\begin{eqnarray}
f^{LS}_{ab,R}(q)&=&
 { g^{LS}_{ab,R} \over
\sqrt{4 E_a(q) E_b(q)}} 
  {q^L / m_\pi^{L-1} 
\over  (1+q^2/c_{ab,R}^{2})^{2+{L\over 2}} } ,
\label{eq:pipi-vertex}
\end{eqnarray}
where $g^{LS}_{ab,R}$ and $c_{ab,R}$ are 
coupling constant and cutoff, respectively, 
as already discussed;
the factor $m_\pi^{1-L}$ is introduced just for making the coupling constant dimensionless.
The phase of $g^{LS}_{\bar{a}\bar{b},\bar{R}}$
(charge-conjugate partner of $g^{LS}_{ab,R}$)
 is fixed to give
$\Gamma_{ab,R}=\Gamma_{\bar{a}\bar{b},\bar{R}}$
in Eq.~(\ref{eq:pipi-vertex0});
for example,
$g^{11}_{D\pi,D^*}=-g^{11}_{\bar{D}\pi,\bar{D}^*}$.
Also in Eq.~(\ref{eq:pipi-vertex0}),
$\Gamma_{ab,R}=-\Gamma_{ba,R}$ can happen due to the CG coefficients. 
To avoid this, our rule of ordering $ab$ in $\Gamma_{ab,R}$ is to satisfy
$m_a > m_b$.
For a $m_a = m_b$ case ($ab=K\bar{K}$), the ordering is particle-antiparticle.

The dressed $R$ propagator [Fig.~\ref{fig:diag}(c)],
used as the dressed $Rc$ Green function in Eq.~(\ref{eq:amp_full}),
is given 
for $R$ listed in Table~\ref{tab:Rc}(C) by
\begin{eqnarray}
[ \tau^{-1}(p,E) ]_{R,R'}
&=& 
( W^2_R - m_R^2) \delta_{R,R'} 
- \Sigma_{R,R'}\left(p, E\right) ,
\label{eq:green-Rc}
\end{eqnarray}
with 
$W_R^2=E^2-p^2$ and 
$m_R$ being a bare mass of $R$.
We also introduced
the $R$ self-energy given as (see Appendix of Ref.~\cite{pipin} for derivation)
\begin{eqnarray}
\Sigma_{R,R'}(p,E) &=& 
\sum_{ab,LS} {\cal B}_{ab}\,
(t_a t^z_a t_b t^z_b |t_R, t^z_a + t^z_b)^2
\int q^2 dq
\nonumber\\
&&\times
{M_{ab}(q)\over \sqrt{M^2_{ab}(q) + p^2}}
  f^{LS}_{R, ab}(q) f^{LS}_{ab,R'}(q) 
\nonumber \\
&&\times
\left\{
{1\over E - \sqrt{M^2_{ab}(q) + p^2} + i{\Gamma_a\over 2}+ i{\Gamma_b\over 2} } 
\right.
\nonumber \\
&&- \left. {1\over E + \sqrt{M^2_{ab}(q) + p^2}} \right\}
,
\label{eq:RR-self}
\end{eqnarray}
with $M_{ab}(q)=E_{a}(q)+E_{b}(q)$;
$s_R=s_{R'}$ is implied in Eq.~(\ref{eq:RR-self}).
Due to the Bose symmetry, we have a factor ${\cal B}_{ab}$:
${\cal B}_{ab}=1/2$ for identical particles $a$ and $b$;
${\cal B}_{ab}=1$ otherwise.
The width of a particle $a(b)$ is denoted by $\Gamma_{a(b)}$.
In our model, 
$\Gamma_{a}=\Gamma_{b}=0$ applies to
most channels
(only the $\eta_c\rho$ channel in 
the $Z_c$ amplitude has $\Gamma_\rho=150$~MeV)
and, in such cases, the above formula reduces to
\begin{eqnarray}
\Sigma_{R,R'}(W^2_R) &=& 
\sum_{ab,LS} {\cal B}_{ab}\,
(t_a t^z_a t_b t^z_b |t_R, t^z_a + t^z_b)^2
\int q^2 dq
\nonumber\\
&&\times
{2 M_{ab}(q) f^{LS}_{R, ab}(q) f^{LS}_{ab,R'}(q) 
\over W^2_R - M^2_{ab}(q) + i \epsilon } .
\label{eq:RR-self2}
\end{eqnarray}

Meanwhile, for $R$ states listed in Table~\ref{tab:Rc}(A) and \ref{tab:Rc}(B),
we use BW propagators
to make the calculations more tractable:
\begin{eqnarray}
 \tau_{R,R}(p,E)
&=& 
{1\over 
2 E_{R}(p)}
{1\over  E - E_{R}(p) + i {\Gamma_R\over 2}
}, 
\label{eq:bw-Rc}
\end{eqnarray}
where $m_{R}$ (hidden in $E_{R}$) and $\Gamma_{R}$ are the BW mass and
width for $R$, respectively.
The use of the BW forms 
causes a partial violation of three-body unitarity.

The dressed $\psi_i\to Rc$ vertex [Fig.~\ref{fig:diag}(f)] is given as
\begin{eqnarray}
\label{eq:dressed-g}
\bar \Gamma^\mu_{Rc ,\psi_i}(\bm{p}_c ,E)&=& 
x_{Rc}
(t_R t^z_R t_c t^z_c |I t^z_R+t^z_c)
\sum_{sll^z}
\nonumber \\
&&\times 
(s_R s_R^z s_c s_c^z|s s_R^z+s_c^z)
(l l^z s s_R^z+s_c^z | J \mu)
\nonumber\\
&&\times 
Y_{l,l^z}(-\hat p_c)
 \bar F_{(Rc)_{ls},\psi_i}(p_c, E), 
\end{eqnarray}
where $l$ ($s$) is the relative orbital angular momentum (total spin) of $Rc$.
For the present case, 
the total angular momentum is $J=1$ and
the total isospin $I=0$, and 
thus $IJ$ indices are suppressed in the notation $(Rc)_{ls}$.
The $Rc$ state in Eq.~(\ref{eq:dressed-g}) decays into a final $abc$
state, as indicated in Eq.~(\ref{eq:amp_full}).
A factor $x_{Rc}$ is introduced in 
Eq.~(\ref{eq:dressed-g}).
If either $D_J\bar{D}_{J'}$ or
$\bar{D}_JD_{J'}$ (but not both) from Eq.~(\ref{eq:c-base}) 
decays into $abc$,
$x_{Rc}=+1/\sqrt{2}$ for $D_J\bar{D}_{J'}$ 
and $-1/\sqrt{2}$ for $\bar{D}_JD_{J'}$;
$x_{Rc}=1$ otherwise.
Similarly,
in Eq.~(\ref{eq:pipi-vertex0}) with $R=Z_c$,
we should have included 
a factor of $+1/\sqrt{2}$ and $-1/\sqrt{2}$
for $ab=D^*\bar{D}$ and $\bar{D}^*D$, respectively,
since the $Z_c$ amplitude includes the
$(D^*\bar{D}-\bar{D}^*D)/\sqrt{2}$ channel.

The dressed $\psi_i\to (Rc)_{ls}$ form factor 
in Eq.~(\ref{eq:dressed-g}) is 
\begin{eqnarray}
\label{eq:dressed-ff}
\bar F_{(Rc)_{ls} ,\psi_i}(p_c ,E)&=&
F_{(Rc)_{ls} , \psi_i}(p_c)
+ \sum_{c'R'R''l's'} \int q^2 dq 
\nonumber \\
&& \times \,
 X_{(Rc)_{ls},(R''c')_{l's'}}(p_c,q;E) 
\nonumber \\
&&\times \,
\tau_{R'',R'}(q,E-E_{c'}) F_{(R'c')_{l's'},\psi_i}(q) ,
\nonumber \\
\end{eqnarray}
where the first and second terms are direct decay and rescattering
mechanisms, respectively.
We use bare dipole form factors parametrized as
\begin{eqnarray}
F_{(Rc)_{ls},\psi_i}(q) &=& 
{C^i_{(Rc)_{ls}} \over \sqrt{4 E_c(q) m_{\psi_i}}}
{q^l/ m_\pi^{l-1}
\over  [1+q^2/(\Lambda^{i}_{(Rc)_{ls}})^2]^{2+{l\over 2}}}, \nonumber\\
\label{eq:bare_mstar}
\end{eqnarray}
where $C^i_{(Rc)_{ls}}$, $\Lambda^i_{(Rc)_{ls}}$, and $m_{\psi_i}$
are coupling constant, cutoff, and bare $\psi_i$ mass, respectively.
We have also introduced the $Rc\to R'c'$ partial wave ($IJ^{PC}=0\,1^{--}$) amplitude 
$ X_{(Rc)_{ls},(R'c')_{l's'}}$ that is obtained by solving the scattering
equation [Fig.~\ref{fig:diag}(d)]:
\begin{eqnarray}
&&X_{(R'c')_{l's'},(Rc)_{ls}} (p',p; E)
= V_{(R'c')_{l's'},(Rc)_{ls}} (p',p; E)
\nonumber\\
&&+ \sum_{c''R''R'''l''s''}
\int q^2dq\, V_{(R'c')_{l's'},(R'''c'')_{l''s''}}(p',q;E) 
\nonumber\\
&&\times
\  \tau_{R''',R''}(q,E-E_{c''}) 
X_{(R''c'')_{l''s''},(Rc)_{ls}}(q,p;E)  ,
\label{eq:pw-tcr}
\end{eqnarray}
with driving terms [Fig.~\ref{fig:diag}(e)]
\begin{eqnarray}
V_{(R'c')_{l's'},(Rc)_{ls}} (p',p; E)
&=&
Z^{\bar{c}}_{(R'c')_{l's'},(Rc)_{ls}} (p',p; E)
\nonumber\\
&&+ 
v^{\rm s}_{(R'c')_{l's'},(Rc)_{ls}} (p',p) .
\label{eq:ptl}
\end{eqnarray}
The so-called $Z$-diagram,
$Z^{\bar{c}}_{(R'c')_{l's'},(Rc)_{ls}}$,
is shown by
the first term of the r.h.s. of Fig.~\ref{fig:diag}(e);
$\bar{c}$ indicates a potentially on-shell exchanged particle.
These long-range interactions couple the
$Rc$ channels listed in Table~\ref{tab:Rc}(A) and \ref{tab:Rc}(C),
but not \ref{tab:Rc}(B).
The formulas and a list of the considered $Z$-diagrams are given in
Appendix~\ref{sec:zgraph} and Table~\ref{tab:z-d} therein.
We also consider 
short-range potentials $v^{\rm s}$ 
between open-charm channels; see Appendix~\ref{sec:supp:DD} for details.
In Refs.~\cite{gjding,xkdong1}, 
the $\rho$-, $\omega$-, and $\sigma$-exchange mechanisms are considered 
to generate $\psi(4230)$ as a $D_1\bar{D}$ hadron-molecule state.
The interactions $v^{\rm s}$ can simulate the sum of such meson-exchange mechanisms.
Also, the resummation of $v^{\rm s}$ and the consequent pole formations
significantly enhance threshold cusps, which
will play an important role in fitting the data.

The dressed NR $Rc$ production amplitude [Fig.~\ref{fig:diag}(g)]
is obtained from Eqs.~(\ref{eq:dressed-g}) and (\ref{eq:dressed-ff})
by replacing the labels ``$\psi_i$'' with ``$\gamma^*$'' and 
using the form factor below:
\begin{eqnarray}
F_{(Rc)_{ls},\gamma^*}(q) &=& 
{C^{\gamma^*}_{(Rc)_{ls}} \over \sqrt{2 E_c(q) }}
{ q^l/ m_\pi^{l-1}
\over [1+q^2/(\Lambda^{\gamma^*}_{(Rc)_{ls}})^2]^{2+{l\over 2}}} .
\label{eq:nr_gamma}
\end{eqnarray}
For $Rc=D_0^*(2300)\bar{D}^*$, $Z_c\pi$, and $Z_{cs}\bar{K}$
in Table~\ref{tab:Rc}(C), 
we assume that these channels are
generated through rescatterings via the $Z$-diagrams,
and set $C^{i(\gamma^*)}_{(Rc)_{ls}}=0$.
Also, we do not enforce SU(3) relations, for example, between
$C^{i(\gamma^*)}_{(D^{(*)}\bar{D}^{(*)})_{ls}}$
and $C^{i(\gamma^*)}_{(D_s^{(*)}\bar{D}_s^{(*)})_{ls}}$.

\begin{figure}
\begin{center}
\includegraphics[width=.43\textwidth]{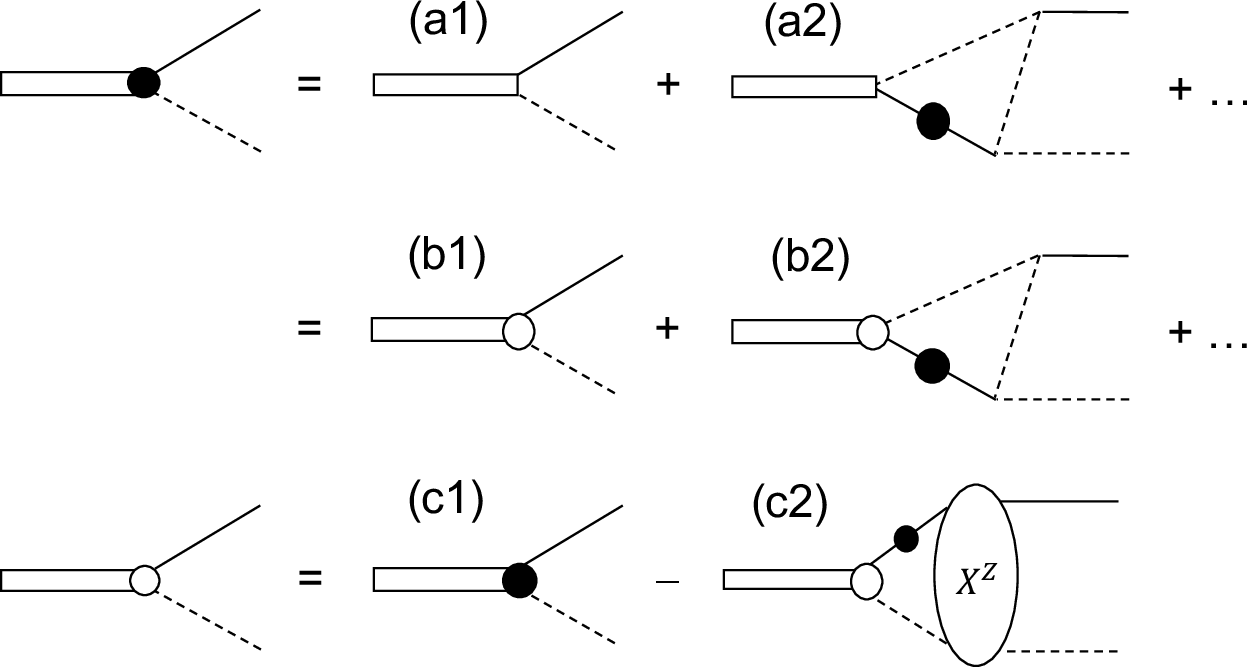}
\end{center}
 \caption{Expansion of dressed $\psi$ ($\gamma^*$) decay vertex [Fig.~\ref{fig:diag}(f,g)].
(a1) direct decay;
(a2) single-triangle rescattering;
(b1) partially dressed (PD) decay;
(b2) PD single-triangle rescattering.
(c1,c2) Definition of the PD decay;
$X^Z$ refers to a scattering amplitude solely driven by 
$Z$-diagrams.
 }
\label{fig:diag3}
\end{figure}

For later purposes, we define several terminologies related to reaction
mechanisms. 
We expand 
the above dressed vertices of 
Eqs.~(\ref{eq:dressed-g}) and (\ref{eq:dressed-ff})
[Fig.~\ref{fig:diag}(f,g)] into 
``direct decay'' [Fig.~\ref{fig:diag3}(a1)] and 
``single-triangle rescattering'' [Fig.~\ref{fig:diag3}(a2)] terms.
The direct decay mechanisms have been defined below Eq.~(\ref{eq:dressed-ff}).
The single-triangle rescattering terms
are obtained by expanding 
the rescattering terms of 
Eq.~(\ref{eq:dressed-ff}) 
using Eqs.~(\ref{eq:pw-tcr}) and (\ref{eq:ptl}) as
\begin{eqnarray}
\label{eq:dressed-ff-expand}
&&\sum_{c'R'R''l's'} \int q^2 dq 
Z^{\bar{c}}_{(Rc)_{ls},(R''c')_{l's'}}(p_c,q;E) 
\nonumber \\
&&\times \,
\tau_{R'',R'}(q,E-E_{c'}) F_{(R'c')_{l's'},\psi_i}(q) .
\end{eqnarray}
The ``partially dressed (PD) decay'' mechanisms [Fig.~\ref{fig:diag3}(b1)] 
are obtained from the dressed vertex of 
Eq.~(\ref{eq:dressed-ff})
by removing all terms in which the last interaction is a $Z$ mechanism;
see Fig.~\ref{fig:diag3}(c1,c2).
This subset of the mechanisms encompasses
the resummed short-range potentials $v^{\rm s}$ in Eq.~(\ref{eq:ptl}).
Finally, 
the ``PD single-triangle rescattering'' mechanisms
[Fig.~\ref{fig:diag3}(b2)] are obtained 
from Eq.~(\ref{eq:dressed-ff-expand})
by replacing the bare form factor $F$ with 
the above-defined PD decay amplitude.

The dressed $\psi$ production amplitude [Fig.~\ref{fig:diag}(h)]
is given as
\begin{eqnarray}
\bar\Gamma_{\psi_i,\gamma^*}(E) &=& 
\Gamma_{\psi_i,\gamma^*}  
+ \sum_{cRR'ls} \int q^2 dq F_{(Rc)_{ls},\psi_i}(q) 
\nonumber \\
&&\times \,
\tau_{R,R'}(q,E-E_{c}) \bar{F}_{(R'c)_{ls},\gamma^*}(q) ,
\label{eq:psi-prod}
\end{eqnarray}
with a bare $\gamma^*\to \psi_i$ amplitude $\Gamma_{\psi_i,\gamma^*}$ 
\begin{eqnarray}
\Gamma_{\psi_i,\gamma^*}  
 &=& 
{1  \over \sqrt{2 m_{\psi_i}}} 
{e\, m^2_{\psi_i}\over g_{\psi_i}} ,
\label{eq:psi-prod2}
\end{eqnarray}
and $g_{\psi_i}$ is a coupling parameter. 
The dressed $\psi$ propagator [Fig.~\ref{fig:diag}(i)]
 is given by 
\begin{eqnarray}
\left[\bar{G}^{-1}(E)\right]_{ij} = (E- m_{\psi_i})\delta_{ij} - \left[\Sigma_{\psi}(E)\right]_{ij}\,,
\label{eq:mstar-g1}
\end{eqnarray}
where the $\psi$ self energy in 
the second term is given by 
\begin{eqnarray}
[\Sigma_{\psi}(E)]_{ij} &=& \sum_{cRR'ls} 
\int q^2 dq\,
F_{(Rc)_{ls} ,\psi_i}(q)
\tau_{R,R'}(q,E-E_c(q)) 
\nonumber\\
&&\times 
\bar F_{(R'c)_{ls} ,\psi_j}(q ,E) \ .
\label{eq:mstar-sigma}
\end{eqnarray}

For fitting the $e^+e^-\to c\bar{c}$ data in $\sqrt{s}\le 4.7$~GeV, 
we need to include resonances heavier than 4.6~GeV such as 
$\psi(4660)$~\cite{bes3_psip_pippim} and
$\psi(4710)$~\cite{bes3_jpsi-ksks}.
See the structure at $\sqrt{s}\sim 4.66$~GeV in
Fig.~\ref{fig:xs-hiddencc}(f), for example. 
However, the currently available data in the $\sqrt{s}>4.6$~GeV region are
insufficient to include these states in the coupled-channel framework. 
More data in $\sqrt{s}>4.6$~GeV are necessary, including
charm-strange final states such as $D^{(*)}\bar{D}_s^{(*)}K$.
Thus, we consider
$\psi(4660)$- and $\psi(4710)$-excitation amplitudes of the BW
form.
Specifically, 
in the square bracket of Eq.~(\ref{eq:amp_full}),
we include additional terms of
\begin{eqnarray}
\sum_{Y=\psi(4660),\psi(4710)}
e^{i\phi_{Y}}\,
{\bar{\Gamma}^\mu_{R'c,Y}(\bm{p}_c)\,
\Gamma_{Y,\gamma^*} 
\over E - m_{Y} + i
{\Gamma_{Y}\over 2}} ,
\label{eq:y4660}
\end{eqnarray}
where $m_Y$ and $\Gamma_Y$ are the BW
mass and width, respectively, 
and $\phi_{Y}$ adjusts the amplitude phase. 
The production and decay vertices
($\Gamma_{Y,\gamma^*}$ and $\bar{\Gamma}^\mu_{R'c,Y}$)
are given in Eqs.~(\ref{eq:psi-prod2}) and (\ref{eq:dressed-g}), respectively.

The formulas in this subsection are based on two-body interactions via bare $R$-excitations,
$ab\to R\to a'b'$.
For a case where two-body interactions also include 
separable contact interactions, 
we can extend the above formulas in a straightforward manner, as
detailed in Ref.~\cite{d-decay}.
In particular, the $Z_{c(s)}$ amplitude is solely from a set of contact
interactions, without $R$ excitations.

The vertex functions in
Eqs.~(\ref{eq:pipi-vertex0}) and (\ref{eq:dressed-g})
can be related
to matrix elements of a Hermitian interaction Hamiltonian
by multiplying the imaginary unit $i$ to Eq.~(\ref{eq:pipi-vertex0})
[Eqs.~(\ref{eq:dressed-g})]
when $s_a+s_b+L+s_R$ [$s_R+s_c+l+J$] is odd.
Or, for our convenience, 
$-i$ is multiplied to Eqs.~(\ref{eq:dressed-g})
for $Rc=D_2^*(2460)\bar{D}$ and $D^*\bar{D}$.
In addition, the factor of $1/\sqrt{2 E_R}$ needs to be moved from 
$\tau_{R,R}$ in Eqs.~(\ref{eq:green-Rc}) or (\ref{eq:bw-Rc}) to the
vertex functions.
When considering the imaginary unit in the vertex functions as above,
we must consistently multiply $i$ [$-i$] to
the r.h.s. of Eq.~(\ref{eq:z-pw})
 when $s_R+s_c+l+J$ [$s_{R'}+s_{c'}+l'+J$] is
odd; $i\leftrightarrow -i$
for $R^{(\prime)}c^{(\prime)}=D_2^*(2460)\bar{D}$ and $D^*\bar{D}$.
Then, no change happens to our results including parameter values.

\subsection{Cross section formulas}

The cross section for 
a three meson ($abc$) production from 
an $e^+e^-$ annihilation ($e^+e^-\to abc$)
is given by
\begin{eqnarray}
d\sigma_{e^+e^-\to abc}
&=&
\sum_{\bar{i}f}
{\cal B}
{(2 \pi)^4 \delta^{(4)}(p_i-p_f)\over 4 v_{\rm rel} E_{e^+}E_{e^-}}  |{\cal M}_{fi}|^2
(2 m_e)^2
\nonumber\\
&&\times 
 {d^3p_a\over (2 \pi)^3 2 E_a}
 {d^3p_b\over (2 \pi)^3 2 E_b}
 {d^3p_c\over (2 \pi)^3 2 E_c} ,
\label{eq:xs-formula1}
\end{eqnarray}
with a Bose factor
${\cal B}=1/3!$ for three identical particles $abc$,
${\cal B}=1/2!$ for identical two particles among $abc$, and
${\cal B}=1$ otherwise;
$\sum_{\bar{i}f}$ indicates the average (sum) of initial (final) spin states.
The invariant amplitude 
${\cal M}_{fi}$ is
 related to Eq.~(\ref{eq:amp_full}) by
\begin{eqnarray}
{\cal M}_{fi} &=& 
 - (2\pi)^3 \sqrt{8 E_a E_b E_c} \;
 A_{abc,e^+e^-}
\nonumber \\
&=&   \tilde{\cal M}_{abc}^{\mu} {1\over s} l_\mu  .
\label{eq:inv-amp}
\end{eqnarray}
Then, the cross section in the total CM frame can be written as:
\begin{eqnarray}
d\sigma_{e^+e^-\to abc}
&=&
\sum_{f}
{\cal B}
{\alpha \over 512 \pi^3 s^3}
(|\tilde{\cal M}_{abc}^x|^2+|\tilde{\cal M}_{abc}^y|^2)
\nonumber \\
&&\times
d m^2_{ab}\, d m^2_{ac} 
d\cos\theta_c
d\bar{\phi}_a ,
\label{eq:xs-formula2}
\end{eqnarray}
where $\alpha$ is the fine structure constant and
the $z$-axis is taken along the $e^+e^-$ beam direction;
$m_{ab}$ and $m_{ac}$ are the invariant masses of the
 $ab$ and $ac$ subsystems, respectively; 
$\theta_c$
is the polar angle of $c$ in the total CM frame;
$\bar{\phi}_a$
is the azimuthal angle of $a$ 
in the $ab$ CM frame, 
relative to the azimuthal angle of $c$. 

Similarly, 
for a two meson ($ab$) production, the invariant amplitude is
\begin{eqnarray}
{\cal M}_{fi} &=& 
 -  \sqrt{ (2\pi)^3 4 E_a E_b} \;
 A_{ab,e^+e^-}
\nonumber \\
&=&   \tilde{\cal M}_{ab}^{\mu} {1\over s} l_\mu  ,
\label{eq:inv-amp2}
\end{eqnarray}
and its cross section in the total CM frame is
\begin{eqnarray}
d\sigma_{e^+e^-\to ab}
&=&
\sum_{\bar{i}f}
{\cal B}
{(2 \pi)^4 \delta^{(4)}(p_i-p_f)\over 4 v_{\rm rel} E_{e^+}E_{e^-}}  |{\cal M}_{fi}|^2
(2 m_e)^2
\nonumber\\
&&\times 
 {d^3p_a\over (2 \pi)^3 2 E_a}
 {d^3p_b\over (2 \pi)^3 2 E_b}
\nonumber \\
&=& \sum_{f}
{{\cal B}\,
\alpha\, p_a \over 8 (\sqrt{s})^5}
(|\tilde{\cal M}_{ab}^x|^2+|\tilde{\cal M}_{ab}^y|^2)
d\cos\theta_a .
\nonumber\\
\label{eq:xs-formula3}
\end{eqnarray}

For $e^+e^-\to \Lambda_c\bar{\Lambda}_c$,
the $\Lambda_c\bar{\Lambda}_c$ scattering due to the Coulomb force
may significantly enhance the cross section near the threshold. 
We thus multiply Eq.~(\ref{eq:xs-formula3})
by the Sommerfeld factor~\cite{sommerfeld,sakharov}:
\begin{eqnarray}
{\pi\alpha \over \beta}
{ \sqrt{1-\beta^2}
\over 
1-\exp(-\pi\alpha\sqrt{1-\beta^2}/\beta)
} ,
\label{eq:sommerfeld}
\end{eqnarray}
with $\beta=p_{\Lambda_c}/E_{\Lambda_c}$
in the CM frame. 
Because of this factor, 
the $e^+e^-\to \Lambda_c\bar{\Lambda}_c$ cross section
is nonzero at the threshold. 

For $e^+e^-\to \eta_c \rho \pi$ followed by $\rho\to\pi\pi$, 
the $e^+e^-\to \eta_c \rho \pi$ cross section of Eq.~(\ref{eq:xs-formula2})
is multiplied by the $\rho$ propagator and $\rho$-decay vertex. 
See Eq.~(10) of Ref.~\cite{sxn_x3872} for a similar formula.

Near thresholds,
the available phase-spaces in our isospin-symmetric model can be
nonnegligibly different from
those in data. 
To correct this, for $e^+e^-\to D^{*+}D^{*-}$ as an example, 
the phase-space factor in Eq.~(\ref{eq:xs-formula3})
is calculated after modifying $\sqrt{s}$ as
\begin{eqnarray}
\sqrt{s} \to \sqrt{s} + 2(m_{D^{*}_{\rm iso}}-m_{D^{*\pm}}) ,
\label{eq:E-modified}
\end{eqnarray}
with 
$m_{D^{*}_{\rm iso}} = (m_{D^{*\pm}}+m_{D^{*0}})/2$.
The amplitude ${\cal M}_{fi}$ is calculated with the unmodified $\sqrt{s}$.

\begin{figure*}
\includegraphics[width=1.0\textwidth]{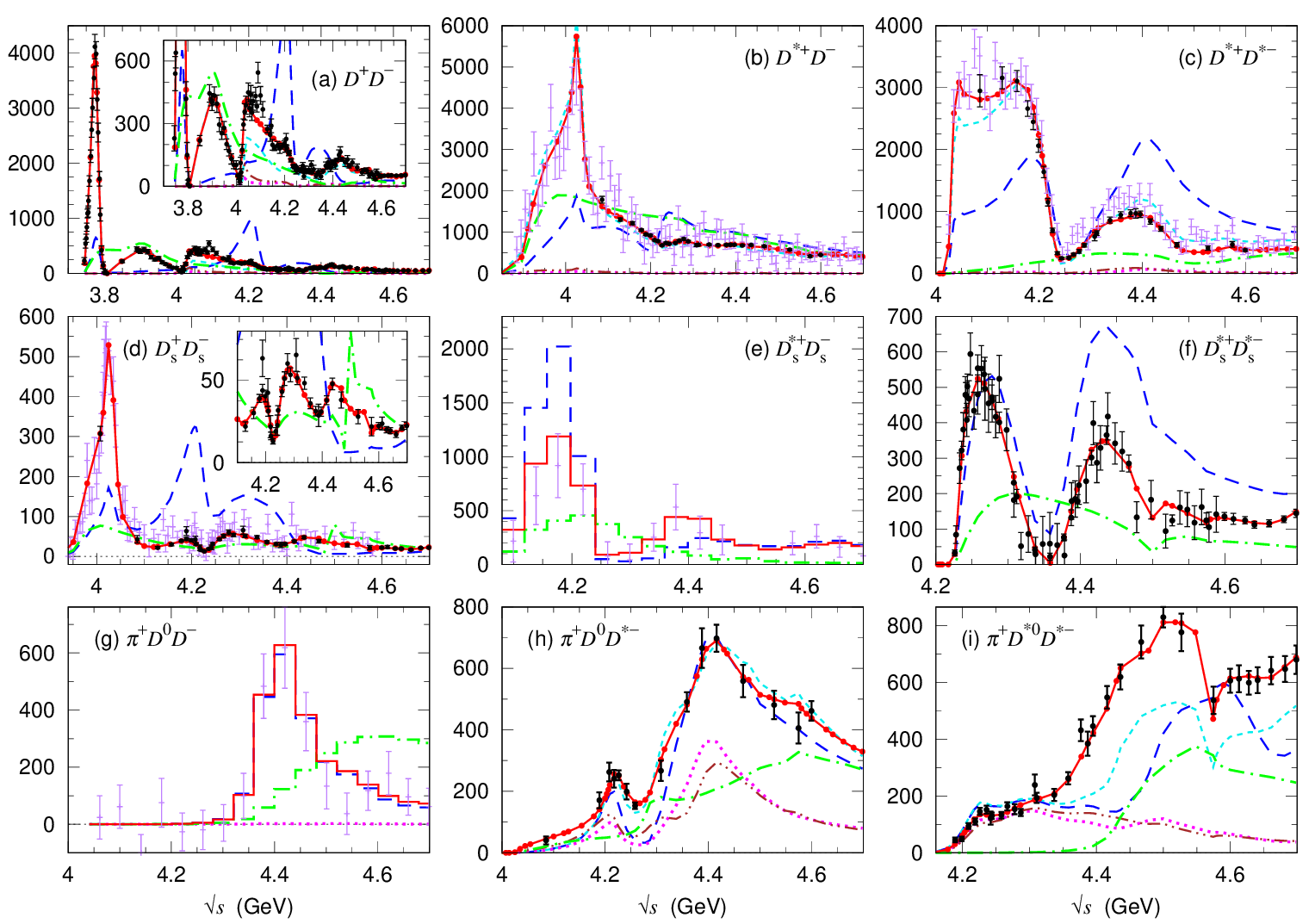}
 \caption{\label{fig:xs-opencc}
Cross sections (unit:pb) for 
$e^+e^-$ annihilations into open-charm final states 
(indicated in each panel; charge-conjugates included)
as functions of the total energy $\sqrt{s}$.
The red points are from our coupled-channel calculation;
the lines are just for guiding eyes.
Various contributions are shown such as 
direct decays of Fig.~\ref{fig:diag3}(a1) (blue dashed),
single-triangle rescattering of Fig.~\ref{fig:diag3}(a2) (magenta dotted),
partially dressed decays of Fig.~\ref{fig:diag3}(b1) (cyan short-dashed),
partially dressed single-triangle rescattering of Fig.~\ref{fig:diag3}(b2) (brown dash-two-dotted),
and nonresonant mechanisms (green dash-dotted).
The BESIII (Belle ISR) data are shown by the black circles (purple
 bars) with error bars, and they
are from Refs.~\cite{bes3_DD,bes3_DD2} in the panel (a);
\cite{bes3_DstDst} (black) and \cite{belle_DstDst} (purple) in (b) and (c);
\cite{bes3_DsDs} (black and purple) in (d);
\cite{belle_DsDs} in (e);
\cite{bes3_DsstDsst} in (f);
\cite{belle_DDpi} in (g);
\cite{bes3_piDDstar} in (h);
\cite{bes3_piDstarDstar_xs} in (i).
The experimental uncertainties include statistical and systematic ones.
In the panels (e) and (g), the calculated
cross sections have been averaged within each bin
to compare with the ISR data.
 }
\end{figure*}

\begin{figure*}
\includegraphics[width=1.0\textwidth]{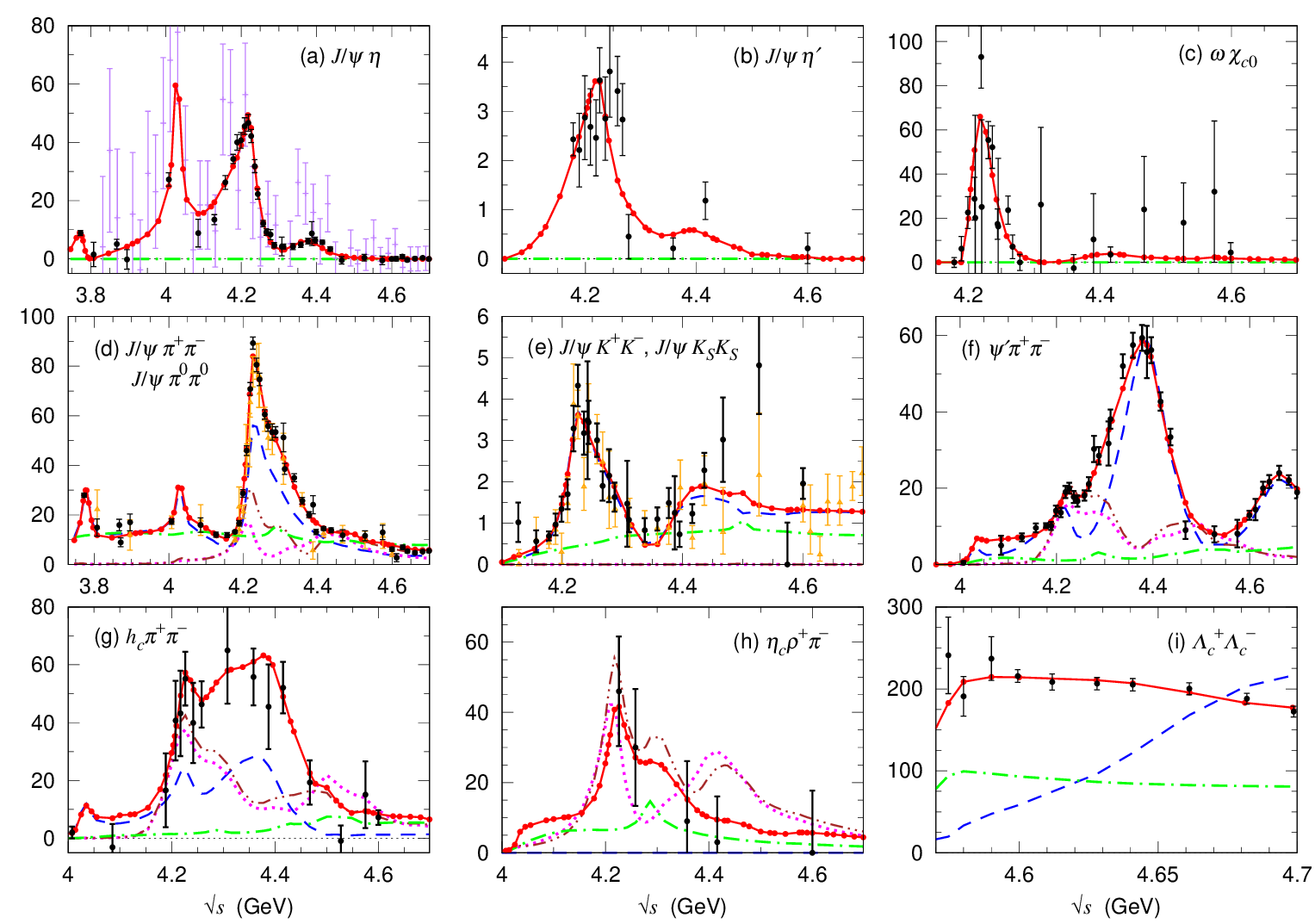}
 \caption{\label{fig:xs-hiddencc}
Cross sections (unit:pb) for 
$e^+e^-$ annihilations into hidden-charm and baryonic final states.
$\rho\to\pi\pi$ is considered in (h).
The data are from 
Refs.~\cite{bes3_jpsi-eta,bes3_jpsi-eta2} (black)
and \cite{belle_jpsi-eta} (purple) in (a);
\cite{bes3_jpsi-etaprime} in (b);
\cite{bes3_chic0omega_2015,bes3_chicJomega,bes3_chic0omega} in (c);
\cite{bes3_jpsi-pippim} for $J/\psi \pi^+\pi^-$ (black) and 
\cite{bes3_jpsi-pi0pi0} for $J/\psi \pi^0\pi^0$ (orange triangles, doubled) in (d);
\cite{bes3_jpsi-kpkm} for $J/\psi K^+K^-$ (black) and 
\cite{bes3_jpsi-ksks} for $J/\psi K^0_SK^0_S$ (orange, doubled) in (e);
\cite{bes3_psip_pippim} in (f);
\cite{bes3_hc_pippim} in (g);
\cite{bes3_etac_pippimpi0} in (h);
\cite{bes3_LL1,bes3_LL2} in (i).
Other features are the same as those in Fig.~\ref{fig:xs-opencc}.
 }
\end{figure*}

\section{Fit results}
\label{sec:fit}

\subsection{Fitting parameters, fitting method, $\chi^2$}
\label{sec:fit-a}

Our couple-channel model is fitted 
to $e^+e^-\to c\bar{c}$ cross-section data (20 final states) as shown in 
Figs.~\ref{fig:xs-opencc} and \ref{fig:xs-hiddencc},
and also to currently available
invariant-mass and angular distribution data
shown in Figs.~\ref{fig:xdxs-jpsi}--\ref{fig:hcpipi-etacrhopi}.
We include five bare $\psi$ states that are minimally needed to obtain a
reasonable fit. 
In addition, 
we also consider $\psi(4660)$ and $\psi(4710)$ as BW amplitudes.

We have 200 fitting parameters in total from:
$m_{\psi_i}$;
real coupling constants in 
$\Gamma^\mu_{Rc ,\psi_i}$,
$\Gamma^\mu_{Rc,\gamma^*}$,
$\Gamma_{\psi_i,\gamma^*}$, $Z_{c(s)}$ amplitude,
and $v^{\rm s}_{R'c',Rc}$;
BW masses, widths, and complex $\psi\to Rc$ couplings for
$\psi(4660)$ and $\psi(4710)$.
We adjust cutoffs
in $\Gamma^\mu_{Rc,\gamma^*}$ 
for
$Rc=D_{(s)}^{(*)}\bar{D}_{(s)}^{(*)}$ and $\Lambda_c\bar{\Lambda}_c$
to control the energy dependences of
the NR contributions (green dash-dotted curves) in
Figs.~\ref{fig:xs-opencc}(a-f) and \ref{fig:xs-hiddencc}(i).
Most (some) of the other cutoffs in $\Gamma^\mu_{Rc ,\psi_i(\gamma^*)}$
are fixed to 1 (0.7)~GeV.

The parameters are adjusted to
minimize the $\chi^2$:
\begin{eqnarray}
\chi^2 = 
\sum_{i=1}^{N_{\rm data}}\chi^2_i
=
\sum_{i=1}^{N_{\rm data}}
{ [O_i({\rm model}) - O_i({\rm data})]^2
\over
[\delta O_i({\rm data})]^2
} ,
\label{eq:chi2}
\end{eqnarray}
where 
$O_i({\rm data})$ 
and $\delta O_i({\rm data})$ 
are the $i$th experimental data and its error (statistical and systematic
errors are quadratically summed), respectively, and 
$O_i({\rm model})$ is the corresponding model calculation.
The number of the data points is 
$N_{\rm data}=1635$.
To obtain a physically reasonable solution,
it is necessary to apply 
weighting factors $w_i$ to some of the data
by $\chi^2_i\to w_i\times\chi^2_i$ 
in the above $\chi^2$ function.
Otherwise, the fit will be poor for data with
relatively large errors
such as $e^+e^-\to \pi D\bar{D}$ [Fig.~\ref{fig:xs-opencc}(g)].
Also, sharp structures, such as the peaks at 
$M_{J/\psi\pi^+}\sim 3.9$~GeV in 
Figs.\ref{fig:xdxs-jpsi}(b,e),
require weights, as the data points are much more scarce compared to
the other smooth regions.
No established weighting scheme exists for this type of problem.
Thus, we apply the weighting in an ad hoc manner to obtain an overall
reasonable fit. 
Furthermore, we set allowed ranges of the fitting parameters to avoid
unreasonably large cancellations among different mechanisms. 
We obtained an overall reasonable fit with unweighted
$\chi^2/{\rm ndf}=2320/(1635-200)\simeq 1.6$ [default fit].
The parameter values for the default fit are given in 
Appendix~\ref{app:para}, 
Tables~\ref{tab:Zc},
\ref{tab:h}, and \ref{tab:psi_param}--\ref{tab:Y4660_param}.
Also, a discussion is given in Appendix~\ref{app:para2} on whether 
the parameter values could be consistent with
the heavy quark spin symmetry (HQSS).

\subsection{Comparison with data}

The full calculations of the $e^+e^-\to c\bar{c}$ cross sections
from the default-fit model are shown 
by the red circles connected by lines
in Figs.~\ref{fig:xs-opencc}-\ref{fig:xs-hiddencc}
along with the data.
The NR contributions [last term of Fig.~\ref{fig:diag}(b)] are shown by the green
dash-dotted curves.
We also show contributions from various
subsets of the $\psi$ and $\gamma^*$ decay mechanisms,
defined around Eq.~(\ref{eq:dressed-ff-expand}),
such as 
the direct decay [Fig.~\ref{fig:diag3}(a1); blue dashed], 
single-triangle rescattering [Fig.~\ref{fig:diag3}(a2); magenta dotted], 
PD decay [Fig.~\ref{fig:diag3}(b1); cyan short-dashed], and 
PD single-triangle rescattering [Fig.~\ref{fig:diag3}(b2); brown
dash-two-dotted].

Let us explain why some theoretical curves appear to be missing in 
Figs.~\ref{fig:xs-opencc} and \ref{fig:xs-hiddencc}.
In Figs.~\ref{fig:xs-opencc}(d-f) and \ref{fig:xs-hiddencc}(a-c,i),
(PD) single-triangle rescattering contributions do not
exist, and 
the PD decay contributions are the same as the full calculations.
This is because the final two-body channels for these processes
belong to
Table~\ref{tab:Rc}(B) that do not {\it directly} couple with the $Z$
diagrams. 
Furthermore, $J/\psi\eta^{(\prime)}$ and $\omega\chi_{c0}$ channels do
not directly couple with the short-range potentials $v^{\rm s}$ either and,
in this case,
the direct-decay contributions are the same as the full calculations. 
In Figs.~\ref{fig:xs-opencc}(g) and \ref{fig:xs-hiddencc}(d-g),
the direct-decay and the PD decay 
contributions are the same.
This is because the final quasi two-body channels\footnote{
{\it Final} quasi-two-body channels
directly decay to a final three-body state. 
For example, $f_0J/\psi$, $f_2J/\psi$, and $Z_c\pi$ are,
among those listed in Table~\ref{tab:Rc},
possible final quasi two-body channels for $e^+e^-\to J/\psi \pi\pi$.}
do not directly couple with $v^{\rm s}$.
In Fig.~\ref{fig:xs-hiddencc}(h),
direct-decay mechanisms are absent.

\begin{figure*}
\includegraphics[width=1\textwidth]{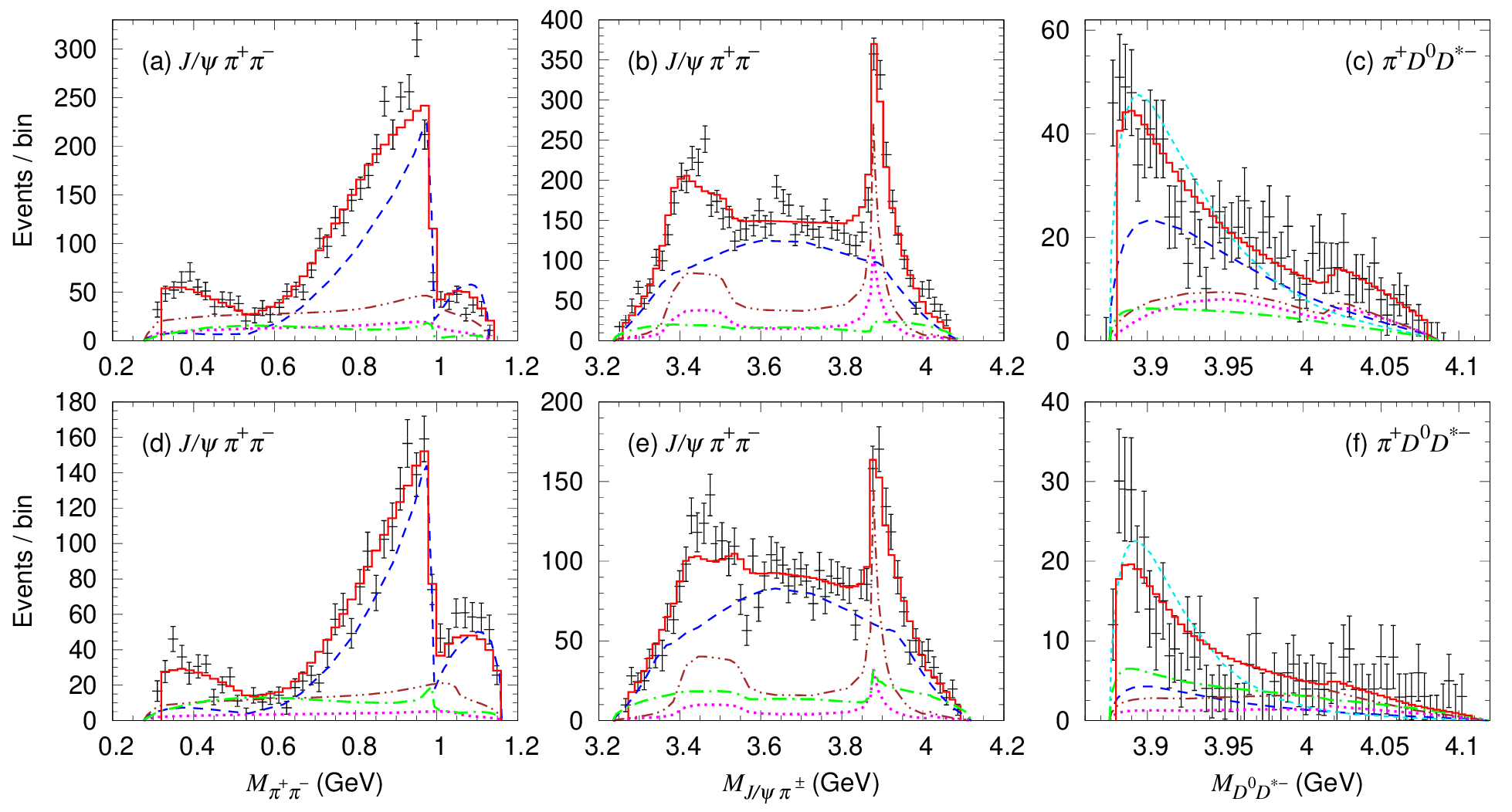}
 \caption{\label{fig:xdxs-jpsi}
Invariant-mass distributions at
$\sqrt{s}=4.23$~GeV for (a)-(c) and
$\sqrt{s}=4.26$~GeV for (d)-(f).
Data are from Ref.~\cite{bes3_Zc3900c} for (a,b,d,e)
and Ref.~\cite{bes3_DDstarc_zc3900} for (c,f).
For the other features, see the caption of Fig.~\ref{fig:xs-opencc}.
 }
\end{figure*}

We also show our fit results for
invariant-mass and angular distributions
in Figs.~\ref{fig:xdxs-jpsi}-\ref{fig:xdxs-psip-jpsipi}.
The various contributions discussed above are also shown.
The red histograms are obtained by averaging the differential cross sections in each of
the bins.
In each panel of the figures, 
the histogram is normalized so that
the total event number from the histogram 
equals that of the data (central value).
The same normalization factor is also multiplied to 
the other curves in the same panel.

\subsection{Remarks on reaction mechanisms}
\label{sec:remark}

We begin with general remarks that are common to several processes.

\begin{itemize}
 \item The coherent sum of the PD decay and
PD single-triangle rescattering contributions mostly saturates the full
calculation. An exception is 
$e^+e^-\to \rho\eta_c\pi$ [Fig.~\ref{fig:xs-hiddencc}(h)]
for which direct decays are absent and 
 double-triangle mechanisms 
[one-order higher than Fig.~\ref{fig:diag3}(b2)] give a sizable
contribution. 
\item For two-body open-charm final states, 
$e^+e^-\to D_{(s)}^{(*)}\bar{D}_{(s)}^{(*)}, \Lambda_c\bar{\Lambda}_c$,
the PD decay mechanisms dominate, and
the PD single-triangle rescatterings 
contribute only slightly or not at all.
[Figs.~\ref{fig:xs-opencc}(a-f) and \ref{fig:xs-hiddencc}(i)].
The resummed $v^{\rm s}$ significantly contributes to the processes, 
as indicated by the differences between 
the PD decay (cyan short-dashed)
  and the direct decay (blue dashed) in 
Figs.~\ref{fig:xs-opencc}(a-c),
and those between the full (red solid) and the direct decay (blue dashed) in 
Figs.~\ref{fig:xs-opencc}(d-f) and \ref{fig:xs-hiddencc}(i).
Actually, the resummed $v^{\rm s}$ generates several hadron molecules
near open-charm thresholds, as will be
discussed later in Sec.~\ref{sec:pole-traj}.
Thus, the resummed $v^{\rm s}$ contributions can be viewed as effects
from the hadron molecules, and the effects are larger near the thresholds.
\end{itemize}
\begin{figure*}
\includegraphics[width=1.0\textwidth]{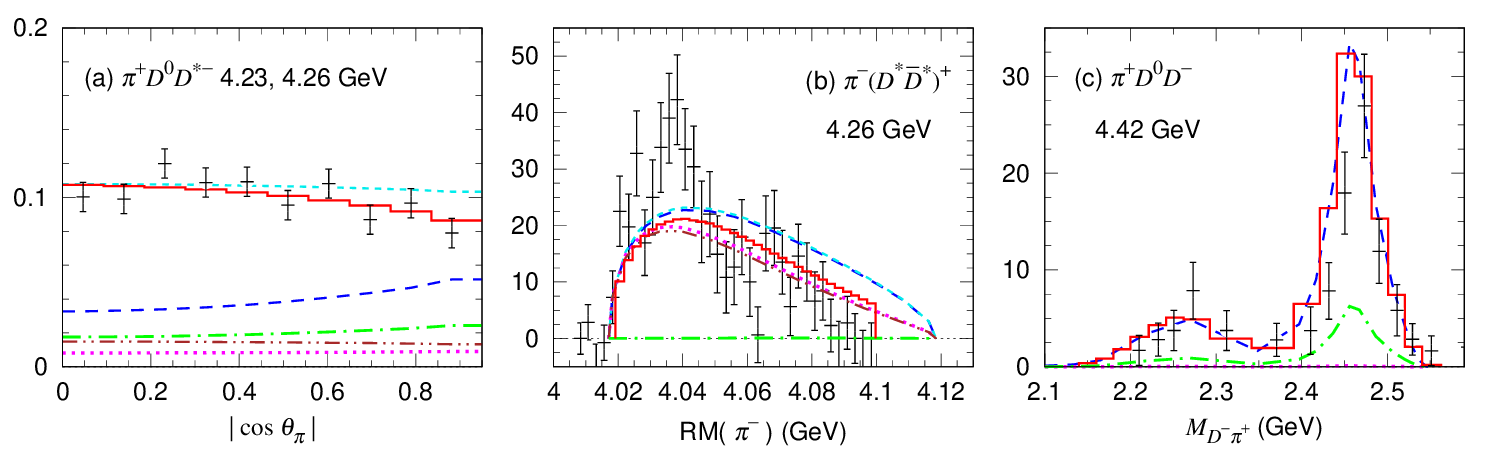}
 \caption{\label{fig:m_dstardstar}
(a) $\cos\theta_\pi$ distribution (fractional yield) where
$\theta_\pi$ is the $\pi$ angle from the beam direction in the CM frame;
(b) $\pi^-$ recoil mass spectrum (events per bin);
(c) $D\pi$ invariant-mass distribution (events per bin). 
The final state and $\sqrt{s}$ are indicated in each panel.
The data,
from which backgrounds have been subtracted,
are from 
\cite{bes3_DDstarc_zc3900} in (a);
\cite{bes3_piDstarDstar} in (b);
\cite{belle_DDpi} in (c).
 }
\end{figure*}
\begin{itemize}
\item The resummed $v^{\rm s}$ significantly contributes to
      three-body open-charm final states $\pi D^*\bar{D}^{(*)}$ 
[Figs.~\ref{fig:xs-opencc}(h,i)]. 
The effect can also be seen in the invariant-mass distributions
[Figs.~\ref{fig:xdxs-jpsi}(c,f)] by the difference between 
the PD decay (cyan short-dashed)
 and the direct decay (blue dashed) contributions.
\item For three-body hidden-charm final states in 
Figs.~\ref{fig:xs-hiddencc}(d-h), 
the resummed $v^{\rm s}$ effects appear 
as the difference between 
the PD single-triangle rescattering (brown dash-two-dotted) and 
the single-triangle rescattering (magenta dotted) contributions.
In particular, the effect is large for the $J/\psi\pi\pi$ final state,
and significantly enhancing the $Z_c(3900)$ peak in 
Figs.~\ref{fig:xdxs-jpsi}(b,e).
\item The data show several cusp structures at
$\sqrt{s}=4431$~MeV ($D_1\bar{D}^*$ threshold) in Fig.~\ref{fig:xs-opencc}(a);
4289~MeV ($D_1\bar{D}$) in Fig.~\ref{fig:xs-opencc}(b);
4503~MeV ($D_{s1}\bar{D}_s$) in Fig.~\ref{fig:xs-opencc}(f);
4573~MeV ($\Lambda_c\bar{\Lambda}_c$) in Fig.~\ref{fig:xs-opencc}(i).
Our calculation fits these structures with
the threshold cusps caused and enhanced by
the resummed $v^{\rm s}$.
Conversely speaking, 
the strengths of $v^{\rm s}$ and the associated molecule poles
are constrained by fitting the cusps.
\item As a consequence of the coupled-channel fit,
our model creates 
common structures in different processes,
even when not necessarily required by the data.
For example, $\psi(4040)$ peaks appear in 
$D^*\bar{D}$ [Fig.~\ref{fig:xs-opencc}(b)] and 
$D_s\bar{D}_s$ [Fig.~\ref{fig:xs-opencc}(d)] to fit the data,
and they also appear in other processes [Fig.~\ref{fig:xs-hiddencc}(a,d,f,g)]
for which data are lacking at the peak.
\item  For two-body open-charm final states, 
the fit quality is sensitive to 
the $\sqrt{s}$-dependence of the NR contributions, as seen in  
Figs.~\ref{fig:xs-opencc}(a-f) and \ref{fig:xs-hiddencc}(i).
Therefore, we adjusted the cutoffs in $\Gamma^\mu_{Rc,\gamma^*}$ 
with $Rc=D_{(s)}^{(*)}\bar{D}_{(s)}^{(*)}, \Lambda_c\bar{\Lambda}_c$.
Although not used, $s$-dependent form factors may be an alternative option
for these NR photon couplings to the open-charm hadron pairs.
\end{itemize}
\begin{figure*}
\includegraphics[width=1.0\textwidth]{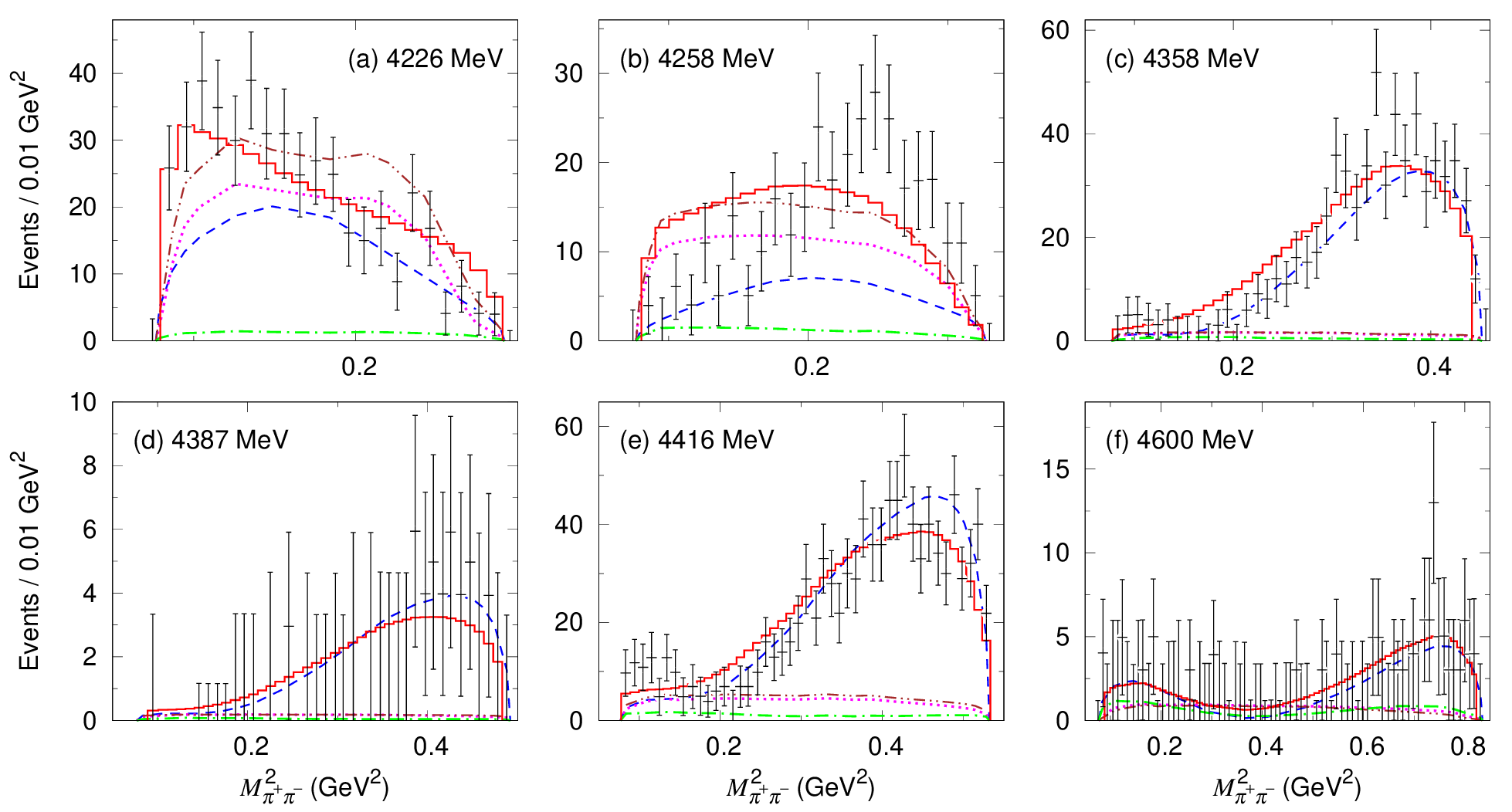}
 \caption{\label{fig:xdxs-psip-pipi}
$\pi^+\pi^-$ invariant-mass distributions in
$e^+e^-\to \psi'\pi^+\pi^-$;
$\sqrt{s}$ is indicated in each panel.
Only for panel (e), the vertical axis is `Events/0.02~GeV$^2$'.
Data are from Ref.~\cite{bes3_psipipi_Zc}.
See the caption of Fig.~\ref{fig:xs-opencc}
for other features.
 }
\includegraphics[width=1.0\textwidth]{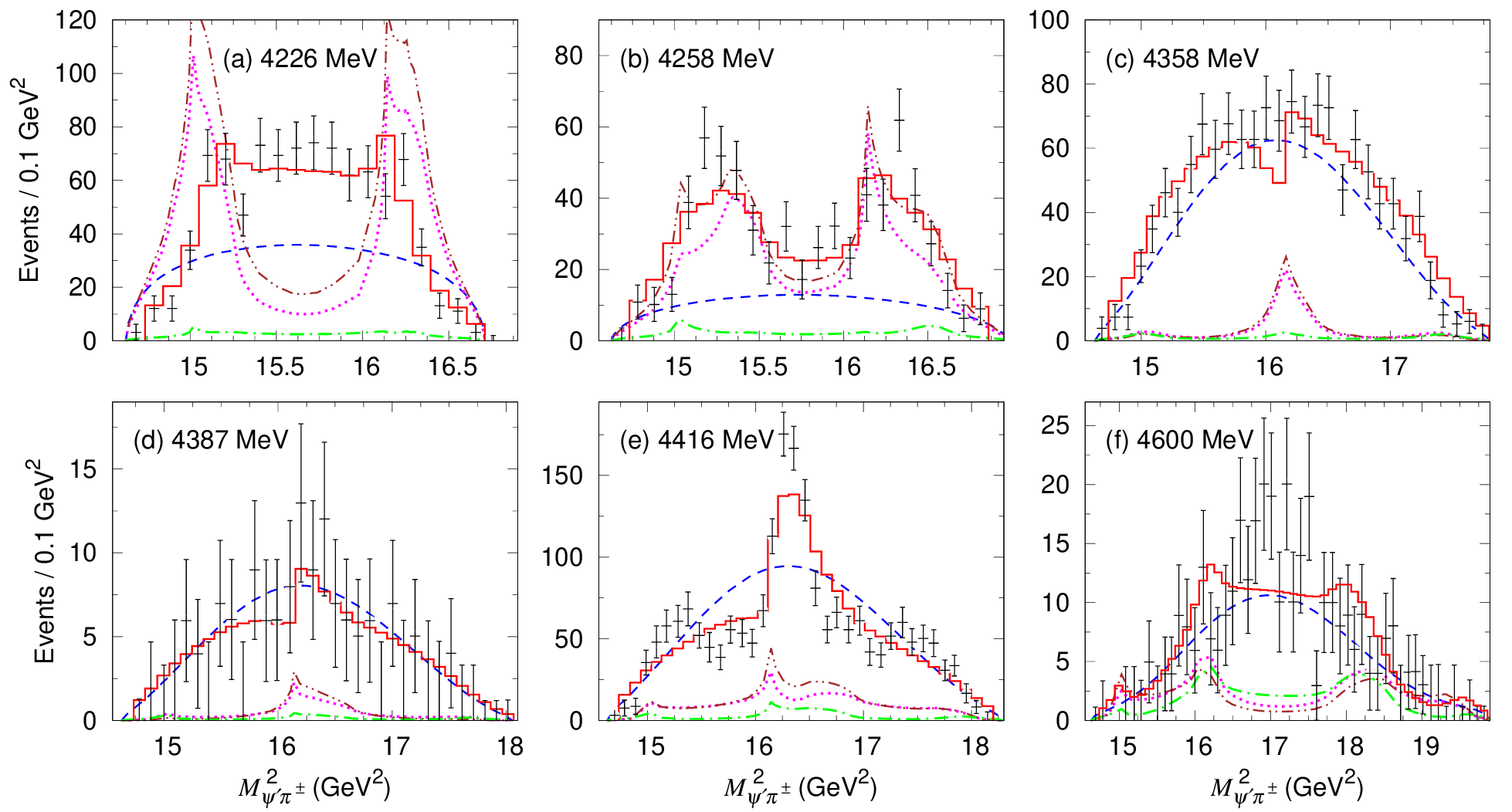}
 \caption{\label{fig:xdxs-psip-jpsipi}
$\psi'\pi^\pm$ invariant-mass distributions in
$e^+e^-\to \psi'\pi^+\pi^-$.
Other features are the same as Fig.~\ref{fig:xdxs-psip-pipi}.
 }
\end{figure*}
\begin{itemize}
\item  For understanding lineshapes of cross-section data
and then correctly extracting vector charmonium properties,
it is essential to consider the openings of the final quasi-two-body ($Rc$) channels at their 
       quasi-thresholds.\footnote{For a given quasi-two-body channel,
       its quasi-threshold is the sum of the channel-particle nominal
       masses. Because of finite widths, their exact thresholds do not exist.}
This is because the threshold effects can significantly alter lineshapes caused by vector charmonium resonances.
For $e^+e^-\to \pi D\bar{D}$ as an example, 
the final $d$-wave $D_2\bar{D}$ contribution equals the full
calculation.
This $\psi$ ($\gamma^*$) decay sequence is supported by 
the $D\pi$ invariant-mass distribution data 
[Fig.~\ref{fig:m_dstardstar}(c)].
The cross section is suppressed
below the $D_2\bar{D}$ quasi-threshold ($\sqrt{s}=4.33$~GeV),  
as seen in Fig.~\ref{fig:xs-opencc}(g).
Furthermore, the suppression near the quasi-threshold occurs due to the centrifugal
       barrier.
We find that the lineshape peak position (4.42~GeV) is shifted from the
resonance mass position (4.39~GeV from Table~\ref{tab:pole})
by these threshold effects.
On the other hand, in Ref.~\cite{belle_DDpi}, the data of Fig.~\ref{fig:xs-opencc}(g) was
       fitted without considering the centrifugal barrier effect,
       resulting in $4.411\pm 0.007$~GeV for the $\psi(4415)$ mass. 
The threshold effects are also important to
understand the $e^+e^-\to \pi D^*\bar{D}^{(*)}$ lineshapes in
       Figs.~\ref{fig:xs-opencc}(h,i).
This can be seen in Fig.~\ref{fig:xs-opencc2} where 
the final $D_1\bar{D}^{(*)}$ and $D_2\bar{D}^{(*)}$ contributions 
rapidly grow around their quasi-thresholds but are strongly suppressed below them.
In Ref.~\cite{dychen}, 
the cross-section data are fitted with the squared charmonium (BW) propagators
multiplied by the three-body phase space, missing the threshold effects.
This prescription cannot be justified since
the lineshapes of Figs.~\ref{fig:xdxs-jpsi}(c,f) are very
       different from the phase-space shape, indicating the dominant
       open-charm $Rc$ contributions.
\item Various final $Rc$ contributions 
to $\pi D^*\bar{D}^{(*)}$ are shown in Fig.~\ref{fig:xs-opencc2}.
Their coherent sums are well constrained by the
$e^+e^-\to \pi D^*\bar{D}^{(*)}$ cross-section data.
To reliably control the individual final $Rc$ contributions,
we need detailed experimental information such as invariant-mass
      distributions or Dalitz plots at various $\sqrt{s}$.
Currently, they are available only near
the $\psi(4230)$ region, as shown in Figs.~\ref{fig:xdxs-jpsi}(c,f) and
      \ref{fig:m_dstardstar}(a,b).
As remarked above, 
understanding the individual final $Rc$ contributions
is important for correctly determining 
vector charmonium properties.
\item  To examine resonance contributions and understand the
       process-dependent $Y$-lineshapes,
       we need to construct resonance amplitudes with 
       poles and residues extracted from the coupled-channel amplitudes.
       This study will be done in the future.
\item  The NR contributions (green dash-dotted curves) to some processes
       exhibit resonant structures,
e.g., near $D_1\bar{D}$ threshold (4289~MeV) in
       Figs.~\ref{fig:xs-opencc}(h) and \ref{fig:xs-hiddencc}(d,f-h).
The structures are mainly caused by hadron-molecule poles generated by
       the interactions of
Eq.~(\ref{eq:ptl}) without coupling to bare $\psi$ states; see
Fig.~\ref{fig:cc-traj}.
In a unitary model like what we use, 
these molecule poles in the NR amplitude [second term in the square bracket of
Eq.~(\ref{eq:amp_full})] are canceled in the full amplitude of Eq.~(\ref{eq:amp_full}).
This point is well discussed in Ref.~\cite{NR-pole}.
\end{itemize}

In the following, remarks are made about process-specific reaction mechanisms.

\begin{figure}
\includegraphics[width=.5\textwidth]{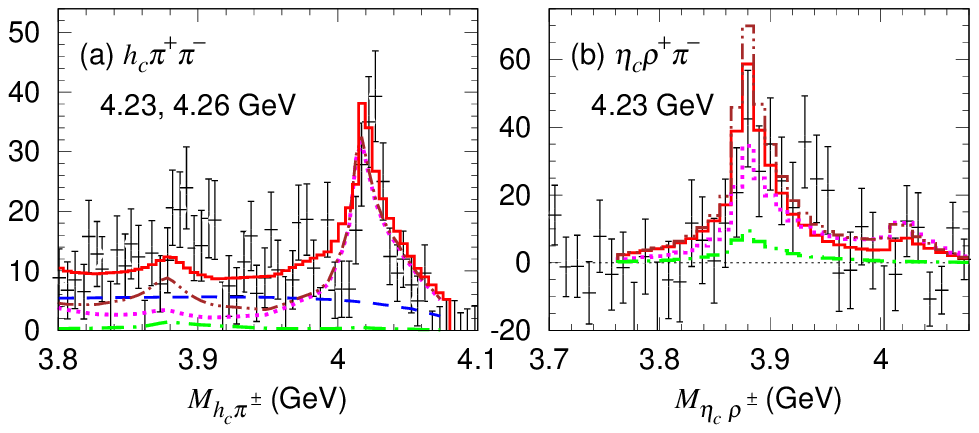}
 \caption{\label{fig:hcpipi-etacrhopi}
 Invariant-mass distributions (unit: events per bin).
The final state and $\sqrt{s}$ are indicated in each panel.
The data,
from which backgrounds have been subtracted,
are from 
Ref.~\cite{z4020-1-bes3} in (a) and 
\cite{bes3_etac_pippimpi0} in (b).
See the caption of Fig.~\ref{fig:xs-opencc}
for other features.
 }
\end{figure}

\subsubsection{$e^+e^-\to \pi D^* \bar{D}$}

\begin{figure}
\includegraphics[width=.5\textwidth]{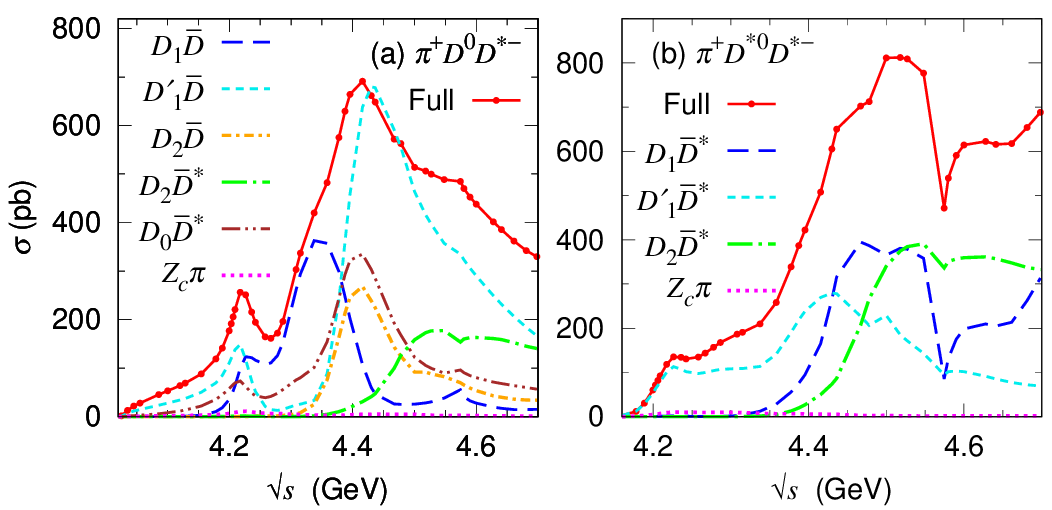}
 \caption{\label{fig:xs-opencc2}
Contributions from various final quasi two-body channels. 
The quasi thresholds are at 
4289~MeV for $D_1\bar{D}$, 
4328~MeV for $D_2\bar{D}$, 
4431~MeV for $D_1\bar{D}^{*}$, and
4470~MeV for $D_2\bar{D}^{*}$,
with the widths of 
$\Gamma_{D_1}=31$~MeV and $\Gamma_{D_2}=47$~MeV.
 }
\end{figure}

Various final $Rc$ states contribute to this process as seen in 
Fig.~\ref{fig:xs-opencc2}(a).
The final $D_1\bar{D}$ contribution is enhanced at $\sim$4230~MeV even
below its quasi-threshold, and rapidly grow above its quasi-threshold.
This indicates that $D_1\bar{D}$ is an important decay channel of
$\psi(4230)$.
In the $\psi(4230)$ region, 
the contributions from the broad $D_1'\bar{D}$ and $D_0\bar{D}^*$
channels are comparable to the $D_1\bar{D}$ contribution,
while the $Z_c\pi$ contribution is rather small.
We note that the $D_0\bar{D}^*$ and $Z_c\pi$ contributions arise
from triangle mechanisms of Fig.~\ref{fig:diag3}(b2),
which is a consequence of the coupled-channel unitarity,
and the model has no bare $\psi_i\to D_0\bar{D}^*, Z_c\pi$ couplings.
 
Examining the reaction mechanisms in Fig.~15 of Ref.~\cite{detten},
their ``Tree-level'' and ``Triangle'' contributions are similar to 
our counterparts ($D_1\bar{D}$ and $Z_c\pi$) in magnitude.
Their model has large contributions from 
contact $\psi(4160,4230)\to Z_c\pi$ mechanisms
that they argue are from
$\psi(4160,4230)\to D_1'\bar{D} - {\rm loop}\to Z_c\pi$.
However, in our analysis, such mechanisms are 
included in the final $Z_c\pi$ contribution and thus small.
On the other hand, they do not consider tree-level
$\psi(4160,4230)\to D_1'\bar{D}\to \pi D^* \bar{D}$
contributions, which is inconsistent with the above argument for the
contact $Z_c$ mechanisms. 
They also do not consider the $D_0\bar{D}^*$ contribution required by
the unitarity. 
In addition, in Ref.~\cite{detten},
the interference between overlapping $\psi(4160)$ and $\psi(4230)$ is
not constrained by the unitarity.
Thus, the above comparison between the model of Ref.~\cite{detten} and ours
clarifies the crucial importance of unitarity in describing this
coupled-channel system.

The enhanced lineshapes near the $D^0D^{*-}$ threshold
in Figs.~\ref{fig:xdxs-jpsi}(c,f) are largely caused by
the final $D_1\bar{D}$ contribution.
However, in Ref.~\cite{bes3_DDstarc_zc3900}, 
the BESIII analysis found that
the $D_1\bar{D}$ contribution is very small. 
More experimental information (Dalitz plots, amplitude analysis results) is
necessary to further test our model and examine the BESIII's finding.

\subsubsection{$e^+e^-\to \pi D^* \bar{D}^*$}

In the $\psi(4230)$ region, the final $D_1'\bar{D}^*$ contribution
dominates, as seen in Fig.~\ref{fig:xs-opencc2}(b).
Since the $D_1'$ width is broad (314~MeV), no significant kinematical
suppression happens even if 
the $D_1'\bar{D}^*$ quasi-threshold is $\sim$210~MeV above.
On the other hand,
the other open-charm channels $D_1\bar{D}^*$ and $D_2\bar{D}^*$ are
kinematically much suppressed due to their narrow widths. 
The $Z_c\pi$ contribution is also small, but its interference with other
mechanisms is not so small (at most $\sim$20\% of the cross section at $\sqrt{s}\sim 4.3$~GeV).
As seen in Fig.~\ref{fig:xs-opencc}(i),
the final $D_1'\bar{D}^*$ contribution is comparably from 
the direct-decay [Fig.~\ref{fig:diag3}(a1)] and 
single-triangle-rescattering [Fig.~\ref{fig:diag3}(a2)] mechanisms.
This is possibly because 
the $D_1'\to D^*\pi$ coupling is large.
The resummed $v^{\rm s}$ enhancement is small because the 
$D_1'\bar{D}^*$ channel does not directly couple to $v^{\rm s}$.

The recoil pion spectrum peak in Fig.~\ref{fig:m_dstardstar}(b)
does not seem to be well fitted.
This problem might be due to a significant detection efficiency effect
not considered in our fit. 
Indeed, while the kinematical endpoint is $\sim$4.12~GeV,
the data extends only up to $\sim$4.09~GeV, 
suggesting considerable loss of low-momentum pion events.
Efficiency-corrected data is highly desirable for analyzing this and
also other data overall.

\subsubsection{$e^+e^-\to J/\psi \pi^+\pi^-, J/\psi \pi^0\pi^0$}

An issue is to understand the asymmetric lineshape of the $\psi(4230)$
in the cross-section data. 
The BESIII introduced $\psi(4320)$ in addition to $\psi(4230)$~\cite{bes3_jpsi-pippim}.
However, $\psi(4320)$ is not necessary to fit other processes, which
casts doubt about its existence. 
In Ref.~\cite{detten}, a $D_1\bar{D}D^*$ triangle diagram
and $D_1\bar{D}D^{*}\bar{D}^{(*)}$ box diagrams get enhanced at the 
$D_1\bar{D}$ threshold (4289~MeV), causing the asymmetric lineshape; 
$Y(4320)$ is unnecessary.
Our coupled-channel model has 
a similar triangle mechanism of Fig.~\ref{fig:diag3}(a2).
However, the box diagrams are absorbed by the bare $\psi_i\to f_{0(2)} J/\psi$
vertices, losing the ability to cause the threshold enhancement. 
Possibly due to this partial lack of the threshold enhancement,
we find a $Y(4320)$ pole in our coupled-channel amplitude; see Sec.~\ref{sec:vec-pole}.
In the future, we will introduce direct
couplings between the hidden-charm and open-charm channels via $v^{\rm s}$
to account for the threshold effects,
as done in Ref.~\cite{detten} via the box diagrams, and examine if 
$Y(4320)$ still exists.
Also, this development could make one bare $\psi$ state redundant.
Yet,
it is unclear whether the conclusion of Ref.~\cite{detten}
remains valid after they include a dataset as comprehensive as ours.

The $J/\psi\pi$ invariant-mass distributions are
well fitted in Figs.~\ref{fig:xdxs-jpsi}(b,e).
The figures show that the $Z_c(3900)$ peak is caused by the (PD)
single-triangle rescattering [Fig.~\ref{fig:diag3}(b2)] that mainly includes
$D_1^{(\prime)}\bar{D}D^*\ {\rm triangle\ loop}\to Z_c\pi$. 
The triangle loops causes the $D^*\bar{D}$ threshold cusp 
that is further enhanced by a pole in the $Z_c$ amplitude.
While the $D_1\bar{D}D^*$ triangle singularity also occurs near the 
$D_1\bar{D}$ quasi-threshold, 
$\sqrt{s}$ for Figs.~\ref{fig:xdxs-jpsi}(b,e) are somewhat lower, and the
enhancement due to the triangle singularity would not be drastic.

\subsubsection{$e^+e^-\to \psi' \pi^+\pi^-$}

The $\psi'\pi$ invariant-mass lineshape in 
Fig.~\ref{fig:xdxs-psip-jpsipi} sensitively depends on 
$\sqrt{s}$.
Let us see how this happens.
The $Z_c\pi$ contribution forms
$D^*\bar{D}^{(*)}$ threshold cusps while 
the $\psi'f_0$ contribution resembles a phase-space shape. 
This can be seen in Fig.~\ref{fig:xdxs-psip-jpsipi} because 
the PD single-triangle contribution (brown dash-two-dotted curves) 
equals the $Z_c\pi$ contribution, 
and the direct-decay contribution (blue dashed curves)
equals the $\psi'f_0$ contribution.
The relative magnitude of the
$Z_c\pi$ and $\psi'f_0$ contributions
varies significantly with $\sqrt{s}$.
As a result, their interference pattern (lineshape) 
sensitively depends on $\sqrt{s}$.

Let us see how the $Z_c\pi$ and $\psi'f_0$ contributions interfere with each other.
In Fig.~\ref{fig:xdxs-psip-jpsipi}(a), for example, 
the $Z_c\pi$ contribution shows 
the $D^*\bar{D}$ 
and $D^*\bar{D}^{*}$ threshold cusps
at $M^2_{\psi'\pi}\sim 15$~GeV$^2$
and 16.1~GeV$^2$, respectively, 
and their reflections (the cusps from another $\psi'\pi$ pair)
appear in 
$M^2_{\psi'\pi}=16.2-16.3$~GeV$^2$
and 15.1--15.2~GeV$^2$, respectively, 
In $M^2_{\psi'\pi}=15.4-16$~GeV$^2$, 
the $Z_c\pi$ and $\psi'f_0$ contributions interfere constructively.
In the cusp region, the phase of the $Z_c\pi$ amplitude changes rapidly,
and the interference becomes destructive. 
As a result of these interferences, the cusp structures are mostly wiped
out.

In Fig.~\ref{fig:xdxs-psip-jpsipi}(b) where
$\sqrt{s}$ is $\sim$30~MeV larger
than that in Fig.~\ref{fig:xdxs-psip-jpsipi}(a),
the $D^*\bar{D}$ and $D^*\bar{D}^{*}$ threshold cusps stay at the same 
$M^2_{\psi'\pi}$, but their reflections appear in 
$M^2_{\psi'\pi}=16.4-16.6$~GeV$^2$
and 15.3--15.5~GeV$^2$, respectively.
Compared with Fig.~\ref{fig:xdxs-psip-jpsipi}(a),
the $\psi'f_0$ contribution relative to 
the $Z_c\pi$ contribution is significantly smaller. 
This can be seen in Fig.~\ref{fig:xs-hiddencc}(f)
where, from $\sqrt{s}=4.22$~GeV to $4.26$~GeV, 
the $\psi'f_0$ contribution (blue dashed curve)
sharply drops 
while the $Z_c\pi$ contribution (brown dash-two-dotted curve) decreases
only slightly. 
As a result, both constructive and destructive interferences 
between the $Z_c\pi$ and $\psi'f_0$ contributions are
significantly smaller compared to Fig.~\ref{fig:xdxs-psip-jpsipi}(a),
and the cusp structures remain.
This explains the rather rapid change of the lineshape from 
Figs.~\ref{fig:xdxs-psip-jpsipi}(a) to \ref{fig:xdxs-psip-jpsipi}(b).
The fairly constant $Z_c\pi$ contribution in this region is related to
the flat $D'_1\bar{D}^*$ contribution to $e^+e^-\to D^*\bar{D}^*\pi$
[Fig.~\ref{fig:xs-opencc2}(b)]
because the $D'_1\bar{D}^*D^*$ triangle rescattering is a main 
$Z_c\pi$ production mechanism. Different processes are related in this
way by the coupled-channel dynamics.

In Fig.~\ref{fig:xdxs-psip-jpsipi}(e), a prominent peak is formed.
This is partly because the $D^*\bar{D}^*$ threshold cusp is enhanced by a
$D_1\bar{D}^*D^*$ triangle singularity, as can be
inferred from the $D_1\bar{D}^*$ contribution in Fig.~\ref{fig:xs-opencc2}(b).
Still, the peak height seems lower than the data. 
This might indicate that 
the $D^*\bar{D}^*$ interaction in the $Z_c$ amplitude needs to be more
attractive. 
In our current $Z_c$ amplitude, 
the $D^*\bar{D}$ and $D^*\bar{D}^*$ interaction strengths are the same 
following the HQSS. 
In the future, we relax this constraint to see if the fit improves. 
This might also improve the fit in Fig.~\ref{fig:m_dstardstar}(b).

The invariant-mass distribution data in 
Figs.~\ref{fig:xdxs-psip-pipi} and \ref{fig:xdxs-psip-jpsipi}
are fitted using $\sqrt{s}$-dependent coupling parameters in
Ref.~\cite{psip-mainz}.
The cross-section data of Fig.~\ref{fig:xs-hiddencc}(f) were not fitted.
Therefore, the reason for the $\sqrt{s}$-dependence of the 
$M^2_{\psi'\pi}$-lineshape was not clarified.
The $Z_c$-propagations were treated with BW amplitudes.

In Ref.~\cite{psip-lanzhou}, both cross-section and invariant-mass
distribution data were fitted; single-channel analysis.
The $D^*\bar{D}^{(*)}$ threshold cusps (or $Z_c$ structures)
 are generated with the initial single pion emission (ISPE) mechanisms.
The ISPE mechanisms correspond to our single-triangle rescattering
mechanisms shown in Fig.~\ref{fig:diag3}(a2), 
 but unstable charm mesons (solid line with a solid circle in the
 figure) shrunk to a point.
Thus, kinematical effects such as 
$D_1\bar{D}^{(*)}$ threshold cusps and 
$D_1\bar{D}^{(*)}D^*$ triangle singularities are lost in the ISPE mechanisms.
Also, the ISPE mechanisms in Ref.~\cite{psip-lanzhou}
do not cause $Z_c$ poles.

\subsubsection{$e^+e^-\to h_c \pi^+\pi^-$}

For the $M_{h_c\pi}$-lineshape in Fig.~\ref{fig:hcpipi-etacrhopi}(a), 
the final $Z_c\pi$ contribution (brown dash-two-dotted curve)
causes the $D^*\bar{D}$ and $D^*\bar{D}^{*}$ threshold cusps.
In particular,  
the $D^*\bar{D}^{*}$ cusp is prominent, and has been interpreted as
$Z_c(4020)$.
The final $h_c f_0$ contribution (blue dashed curve) is relatively
small, compared to 
the final $J/\psi f_0$ and $\psi' f_0$ contributions 
in $e^+e^-\to J/\psi\pi\pi$ and $\psi'\pi\pi$, respectively.

As discussed in Ref.~\cite{detten}, this process is a HQSS-violating
process;
however, its cross section is comparable to that of $e^+e^-\to J/\psi \pi\pi$, which
is HQSS-conserving.
For a similar case in the $b$-quark sector, 
it is argued in Ref.~\cite{hqss-h} that 
comparable cross sections of HQSS-conserving
$\Upsilon(10860)\to\pi Z_b^{(\prime)}\to \Upsilon\pi\pi$
and HQSS-violating $\Upsilon(10860)\to \pi Z_b^{(\prime)}\to h_b\pi\pi$
can be understood with two parameters $m_{Z_b}-m_{Z'_b}$ and 
$\Gamma_{Z_b^{(\prime)}}$, and the HQSS restores at 
$m_{Z_b}=m_{Z'_b}$.
An assumption in Ref.~\cite{hqss-h} is that 
$Z_b$ and $Z'_b$ are almost equally excited in both processes;
data support this.
For the $c$-quark sector, this assumption is not valid. 
As seen in Figs.~\ref{fig:xdxs-jpsi}(b,e) and
\ref{fig:hcpipi-etacrhopi}(a), 
$Z_c(3900)$ and $Z_c(4020)$ are strongly excited in 
$e^+e^-\to J/\psi \pi\pi$ and $h_c \pi\pi$, respectively, but not vise
versa. 
Thus, the argument of Ref.~\cite{hqss-h} cannot be directly applied to
the $c$-quark sector.

\begin{figure}[b]
\includegraphics[width=.5\textwidth]{xs_KK_jpsi}
 \caption{\label{fig:m_KKjpsi}
Predictions for
$K^+K^-$ (left) and
$J/\psi K^+$ (right)
invariant-mass distributions of
$e^+e^-\to J/\psi K^+K^-$ from
$\sqrt{s}=4.1$ to 4.6~GeV. 
Data are from Ref.~\cite{bes3_jpsi-kpkm}.
 }
\end{figure}

\subsubsection{$e^+e^-\to J/\psi K^+K^-, J/\psi K_SK_S$}

The $J/\psi K^+K^-$ data [Fig.~\ref{fig:xs-hiddencc}(e)] shows
an enhancement suggesting $Y(4500)$~\cite{bes3_jpsi-kpkm}.
However, our model does not fit it since
the data is rather fluctuating in this region, and 
the $J/\psi K_SK_S$ data does not indicate the same enhancement. 

For $\sqrt{s}$ slightly above the $D_{s1}\bar{D}_s$ threshold (4503~MeV), 
the $D_{s1}\bar{D}_sD^*$ triangle diagram
[Fig.~\ref{fig:diag3}(b2)] causes a triangle singularity, and 
attractive $D_{s1}\bar{D}_s$ and 
$\bar{D}_sD^*$ ($Z_{cs}$ amplitude) interactions further
enhance the triangle amplitude.
However, this contribution alone is rather small, as indicated by
the brown dash-two-dotted curve in Fig.~\ref{fig:xs-hiddencc}(e),
and causes a modest change in the lineshape.
This triangle contribution sensitively depends on bare
$\psi_i\to D_{s1}\bar{D}_s$ coupling strengths.
We constrained the couplings
using $e^+e^-\to D_{s1}\bar{D}_s$ cross-section
data near the threshold~\cite{bes3_ds1}.
In Ref.~\cite{FZPeng2023}, 
a $D_{s1}\bar{D}_s$ bound state is predicted and assigned to $Y(4500)$.
However, considering the smallness of the above triangle contribution,
it seems difficult to fit 
the $Y(4500)$ fluctuation in the $J/\psi K^+K^-$ data
with this $D_{s1}\bar{D}_s$ molecule's contribution
under the constraint from the $e^+e^-\to D_{s1}\bar{D}_s$ data~\cite{bes3_ds1}.

We also present in Fig.~\ref{fig:m_KKjpsi}
our model's prediction for 
$J/\psi K^+$ and $K^+K^-$ invariant-mass distributions of
$e^+e^-\to J/\psi K^+K^-$ over $\sqrt{s}=4.1-4.6$~GeV. 
The agreement with data is fair.

\section{Inclusive cross sections}
\label{sec:R}

\begin{figure}
\begin{center}
\includegraphics[width=.5\textwidth]{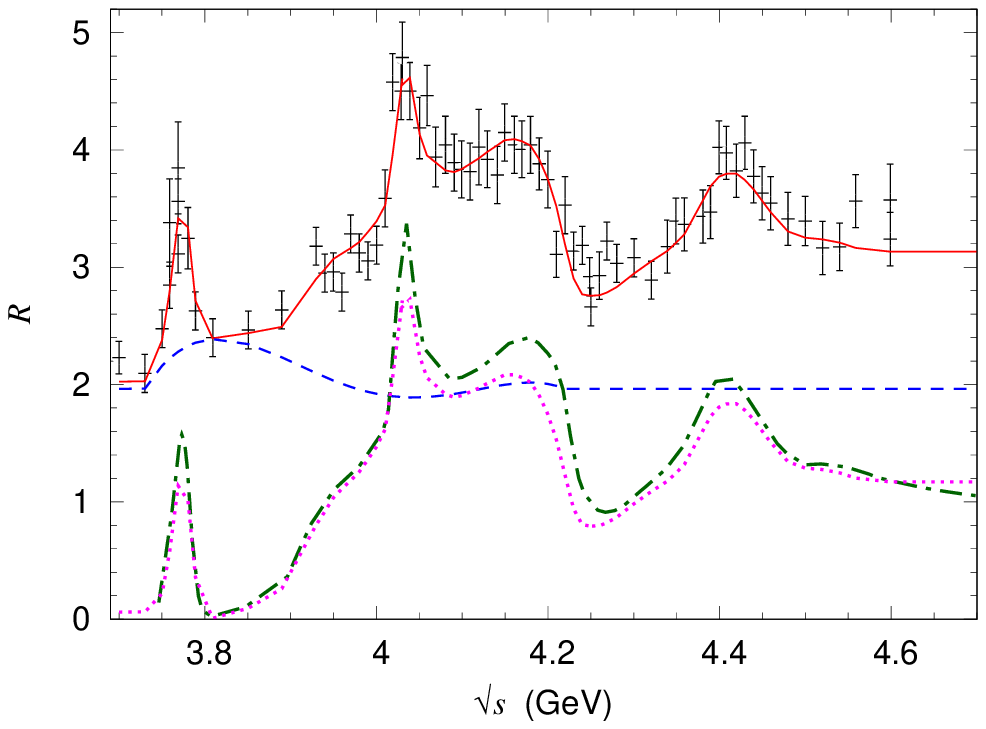}
\end{center}
 \caption{The $R$ value.
The sum of $\sigma(e^+e^-\to c\bar{c}\ {\rm hadrons})$
from our coupled-channel model gives the magenta dotted curve.
The light-hadron contribution is shown by the blue dashed curve.
Their sum is shown by the red solid curve. 
The green dash-dotted curve is 
calculated with an optical theorem of Eq.~(\ref{eq:opt}).
Data is from Ref.~\cite{bes2-R}.
 }
\label{fig:Rvalue}
\end{figure}

Let us see if our coupled-channel model reasonably gives 
the inclusive $e^+e^-\to c\bar{c}$ cross sections or the conventional
$R$ value defined by
\begin{eqnarray}
R(s) &=&
{\sigma(e^+e^-\to {\rm hadrons}) \over
\sigma_{e^+e^-\to\mu^+\mu^-}^{\rm tree}(s)}
\nonumber\\ 
&=&
R_{c}(s) + 
R_{uds}(s),
\label{eq:R-def}
\end{eqnarray}
where we have separated the $R$ value into 
the contribution from the open- and hidden-charm channels ($R_c$) and 
that from light-hadron channels ($R_{uds}$), assuming that 
the couplings between the separated channels are small.
We have also introduced 
the tree-level $e^+e^-\to\mu^+\mu^-$ cross sections,\footnote{
We neglect the lepton mass, which is a good approximation for $\sqrt{s}>3.7$~GeV.}
\begin{eqnarray}
\sigma_{e^+e^-\to\mu^+\mu^-}^{\rm tree}(s)
&=&
{4\pi\alpha^2\over 3s}.
\label{eq:xs-eemumu-tree}
\end{eqnarray}
With our coupled-channel model, 
$R_{c}\times \sigma_{e^+e^-\to\mu^+\mu^-}^{\rm tree}$
is obtained by summing all the calculated cross sections
in Figs.~\ref{fig:xs-opencc} and \ref{fig:xs-hiddencc}
and their isospin partners. 
The obtained $R_{c}$ is shown by the magenta dotted curve in 
Fig.~\ref{fig:Rvalue} along with the experimental $R$ values~\cite{bes2-R}. 
We assume that the difference between them is from light-hadron
contributions, $R_{uds}$.
We express $R_{uds}$ by
\begin{eqnarray}
R_{uds}(s) &=&
(\sqrt{s}-E_1)
(\sqrt{s}-E_2)
(\sqrt{s}-E_3) \nonumber\\
&&\times(c_1{s}+c_2\sqrt{s}+c_3)+R^0_{uds} ,
\label{eq:R-uds1}
\end{eqnarray}
for $E_1< \sqrt{s}< E_3$, and 
\begin{eqnarray}
R_{uds}(s) =   R^0_{uds},
\label{eq:R-uds2}
\end{eqnarray}
otherwise, and then adjust the parameters 
as in Table~\ref{tab:Ruds}
to reproduce the experimental $R$.
\begin{table}[b]
\renewcommand{\arraystretch}{1.4}
\caption{\label{tab:Ruds}
Numerical parameter values in Eq.~(\ref{eq:R-uds1}).
 }    
\begin{ruledtabular}
\begin{tabular}{lclclc}
$E_1$ (GeV) & 3.73& $c_1$ (GeV$^{-5}$) &$-$30.1 &$R^0_{uds}$ & 1.96 \\
$E_2$ (GeV) & 3.98& $c_2$ (GeV$^{-4}$) &$-$7.93 &&\\
$E_3$ (GeV) & 4.22& $c_3$ (GeV$^{-3}$) & 545 &&\\
\end{tabular}
\end{ruledtabular}
\end{table}
The $R_{uds}$ (blue dashed) as well as $R_c+R_{uds}$ (red
solid) obtained from the fit are shown in Fig.~\ref{fig:Rvalue}.
The resonant structures in the data are well reproduced by our
coupled-channel model.

\begin{figure*}
\begin{center}
\includegraphics[width=.495\textwidth]{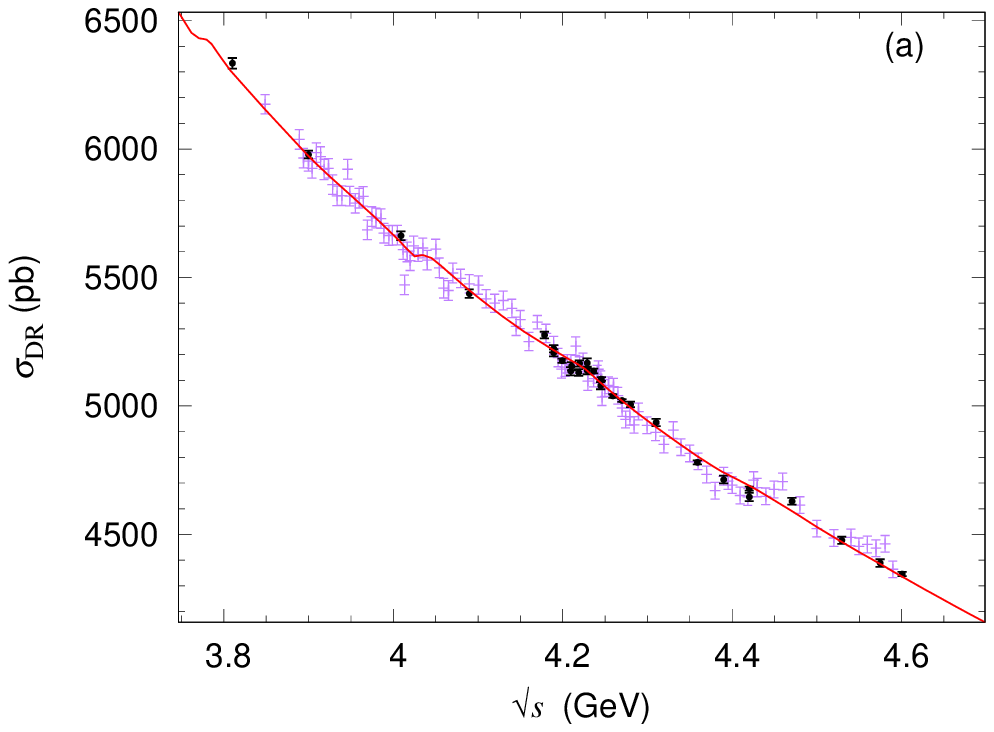}
\includegraphics[width=.495\textwidth]{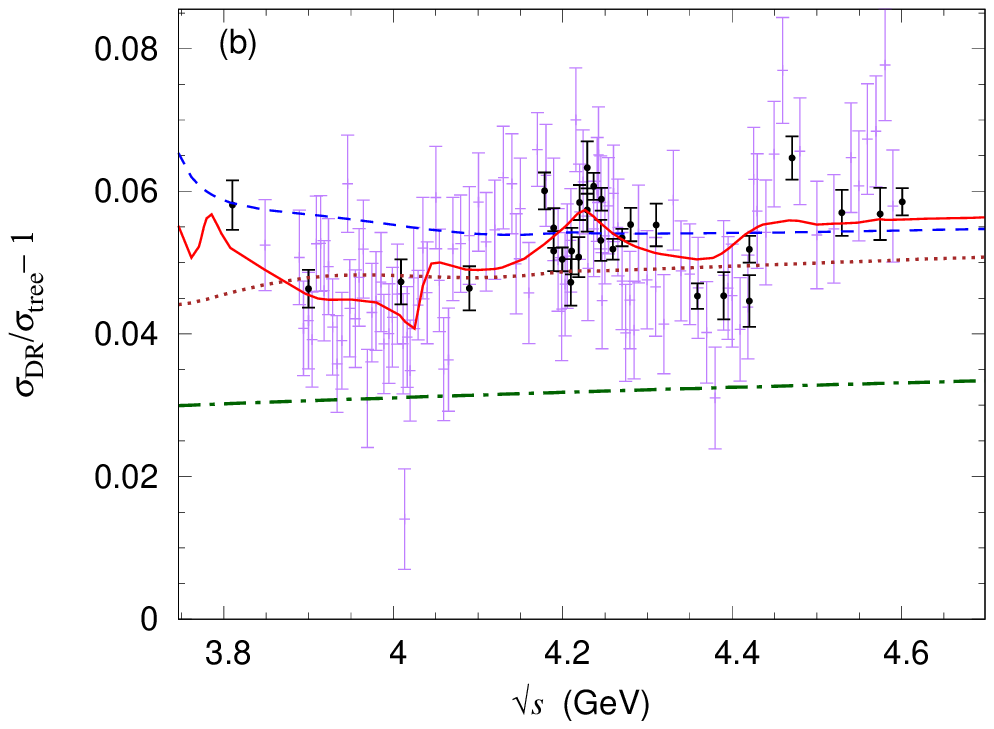}
\end{center}
 \caption{\label{fig:xs-ee-mumu}
(a) Dressed $e^+e^-\to \mu^+\mu^-$ cross sections.
The red solid curve is our prediction from Eq.~(\ref{eq:xs-eemumu})
 including the pure leptonic tree and VP contributions.
The black circles (purple bars) are higher (lower) luminosity data 
 from Ref.~\cite{ee-mumu-data}.
The data have been slightly shifted as
Eq.~(\ref{eq:ee-mumu-shift}).
(b) Ratios of the DR $e^+e^-\to \mu^+\mu^-$ cross sections
 shown in the left panel to the tree-level
$e^+e^-\to \mu^+\mu^-$ cross sections (red solid curve and data).
The contributions from $\Pi_{\rm lepton}$ (green dash-dotted), 
$\Pi_{\rm lepton}$ + $\Pi_{uds}$ (brown dotted), 
and $\Pi_{\rm lepton}$ + $\Pi_{uds}$ + $\Pi_{J/\psi}$ + $\Pi_{\psi'}$ (blue dashed)
are also shown.
See Eq.~(\ref{eq:vp}) for the notations.
 }
\end{figure*}

If our coupled-channel model were three-body unitary, 
$R_{c}$ can also be calculated using an optical theorem
\begin{eqnarray}
R_c(s)= -{3\over\alpha} {\rm Im}[\Pi_{c}(s)] ,
\label{eq:opt}
\end{eqnarray}
where we have introduced a charm vacuum polarization (VP) by
\begin{eqnarray}
\Pi_{c}(s) &=&
{1\over s}
(\Sigma_{\gamma^*}
+\sum_{ij}
\bar\Gamma_{\gamma^*,\psi_i}
\bar{G}_{ij} 
\bar\Gamma_{\psi_j,\gamma^*} ),\
\label{eq:Pi_c}
\end{eqnarray}
where the NR contribution $\Sigma_{\gamma^*}$ is obtained from
Eq.~(\ref{eq:mstar-sigma}) by replacing labels ``$\psi_{i(j)}$'' with ``$\gamma^*$''.
The resonant part consists of 
$\bar\Gamma_{\psi_i,\gamma^*} (=\bar\Gamma_{\gamma^*,\psi_i})$ and
$\bar{G}_{ij}$ that have been
defined in Eqs.~(\ref{eq:psi-prod}) and (\ref{eq:mstar-g1}),
respectively.
A diagrammatic representation for $\Pi_{c}$ is Fig.~\ref{fig:diag}(j).
The $R_c$ calculated with Eq.~(\ref{eq:opt}) is shown in Fig.~\ref{fig:Rvalue}
(green dash-dotted).
The difference between the magenta dotted and green dash-dotted curves
is a measure of the three-body unitarity violation in our model.
The violation is modest, which should allow our model to predict 
$e^+e^-\to\mu^+\mu^-$ cross sections in the next subsection.

\section{$e^+e^-\to\mu^+\mu^-$ cross sections}
\label{sec:ee-mumu}

The BESIII measured 
the $e^+e^-\to\mu^+\mu^-$ cross sections
for $\sqrt{s}=3.8-4.6$~GeV, and subtracted 
initial state radiation effects to obtain the dressed (DR) cross
sections~\cite{ee-mumu-data}.
Theoretically, 
the DR $e^+e^-\to\mu^+\mu^-$ cross section up to and including ${\cal O}(\alpha^3)$
is given by
\begin{eqnarray}
\sigma^{\rm DR}_{e^+e^-\to \mu^+\mu^-}(s)
&=&
\sigma_{e^+e^-\to\mu^+\mu^-}^{\rm tree}
\left( 1 +2 {\rm Re}[\Pi(s)]
\right) ,
\label{eq:xs-eemumu}
\end{eqnarray}
where $\sigma_{e^+e^-\to\mu^+\mu^-}^{\rm tree}$ has been given in 
Eq.~(\ref{eq:xs-eemumu-tree}), 
and $\Pi$ is VP including various intermediate states:
\begin{eqnarray}
\Pi(s) &=& \Pi_{\rm lepton}
+ \Pi_{uds}
+ \Pi_{J/\psi}
+ \Pi_{\psi'}
+ \Pi_{c},
\label{eq:vp}
\end{eqnarray}
where $\Pi_{\rm lepton}$ includes
the $e^+e^-, \mu^+\mu^-,\tau^+\tau^-$ one-loop contribution;
we use Eq.~(214) of Ref.~\cite{lep-VP} to calculate $\Pi_{\rm lepton}$.
The light hadron contributions $\Pi_{uds}$ are calculated
with the once-subtracted dispersion relation:
\begin{eqnarray}
{\rm Re}[\Pi_{uds}(s)] = 
{s-s_0\over \pi} 
P\!\!\int^\infty_{4m_\pi^2}ds'{(-\alpha/3) R_{uds}(s) \over (s'-s_0)(s'-s)},
\end{eqnarray}
where $R_{uds}$ has been introduced in 
Eqs.~(\ref{eq:R-uds1}) and (\ref{eq:R-uds2});
 $P$ indicates the principal value integral.
We take the subtraction point $s_0=0$ and set the subtraction constant
to zero.\footnote{A possibly nonzero subtraction constant could be absorbed by the
parameter $f_0$ in Eq.~(\ref{eq:ee-mumu-shift}), as far as 
the difference between the calculation and the shifted data is concerned.}
Regarding the $J/\psi$ and $\psi'$ contributions 
$\Pi_{\psi}$ ($\psi=J/\psi$ or $\psi'$), 
we use
\begin{eqnarray}
\Pi_{\psi}(s) = {3s\over \alpha} {\Gamma^\psi_{e^+e^-}\over m_\psi}{1\over s-m_\psi^2+im_\psi\Gamma_\psi},
\label{eq:vp-psi}
\end{eqnarray}
where $\Gamma_\psi$ and $\Gamma^\psi_{e^+e^-}$ are the $\psi$ total
width and partial width to $e^+e^-$, respectively, and their values are
from Ref.~\cite{pdg}.
The other charm contributions $\Pi_{c}$ is from our coupled-channel
model as has been given in Eq.~(\ref{eq:Pi_c}).

Our prediction from Eq.~(\ref{eq:xs-eemumu}) is compared with the
corresponding data~\cite{ee-mumu-data} in Fig.~\ref{fig:xs-ee-mumu}(a).
For this comparison, 
we followed Ref.~\cite{ee-mumu-analysis} and shifted the data as 
\begin{eqnarray}
\sigma^{\rm DR, exp}_i
\to
(f_0 + f_1\sqrt{s_i})\,
\sigma^{\rm DR, exp}_i, 
\label{eq:ee-mumu-shift}
\end{eqnarray}
where $\sigma^{\rm DR, exp}_i$ is the $i$th data point at $s=s_i$, 
and $f_0$ and $f_1$ are fitting parameters.
We find $f_0=0.917$ and $f_1=0.0259$~GeV$^{-1}$ for the best fit. 
The shift amounts to 1.59\% at $\sqrt{s}=3.81$~GeV
and 3.64\% at $\sqrt{s}=4.6$~GeV, which is fairly comparable to
the systematic uncertainty of 2.91\%.

For more detailed comparisons, we divide the cross sections in 
Fig.~\ref{fig:xs-ee-mumu}(a) by 
$\sigma_{e^+e^-\to\mu^+\mu^-}^{\rm tree}$, and show the ratios
subtracted by one in Fig.~\ref{fig:xs-ee-mumu}(b).
Contributions from various VP terms in Eq.~(\ref{eq:vp}) are also shown.
Our full result (red solid curve) does not have a sharp dip at
$\sqrt{s}\sim 4.2$~GeV that the data seems to indicate. 
Also, the structure at $\sqrt{s}\sim 4.45$~GeV from our model is smaller
than the data. 
Similar results were also obtained in previous theoretical
studies~\cite{ee-mumu-analysis}(\cite{Lyon})
where the full $R$ ($R_c$) values of Eq.~(\ref{eq:R-def}) were converted
to the hadronic VP ($\Pi_c$).\footnote{
See Fig.~1(bottom) of Ref.~\cite{Lyon}. 
The notation $h_c$ of Ref.~\cite{Lyon} and our notation $\Pi_c$ are related by 
$h_c=-{\pi\over\alpha}\Pi_c$.}
Farrar et al.~\cite{ee-mumu-analysis} 
suggested the possibility of undetected hadronic final states and
statistical fluctuations in the data.

On the other hand, the model of Detten et al.~\cite{detten} reproduces the
dip at $\sqrt{s}\sim 4.2$~GeV.
It is noted, however, that their model is not unitary and
there is no clear relation with the inclusive $R$ values.
Furthermore, 
a fitting parameter, 
$c_{\rm mix}$ in Eq.~(44) of \cite{detten}, 
is introduced exclusively for 
$e^+e^-\to\mu^+\mu^-$,
and it plays an important role in fitting the data, as seen in Fig.~18 of \cite{detten} (``Mixing'' contribution).
However, the unitarity dictates that such mixing occurs not
only in $e^+e^-\to\mu^+\mu^-$ but also in all the other 
$e^+e^-\to c\bar{c}$ processes.
As discussed in the previous paragraph, our model and the dispersive
approaches~\cite{ee-mumu-analysis,Lyon} do not have any freedom to
fit the dip, once they begin with the $R$ values.
Thus it is fair to say that the structures in 
the $e^+e^-\to\mu^+\mu^-$ cross section data are not well-understood.
On the experimental side,
more precise data would be highly desirable to verify the structures.

\section{Pole extraction and resonance properties}
\label{sec:pole}

\subsection{Vector-charmonium states}
\label{sec:vec-pole}

We analytically continue
the coupled-channel amplitude
fitted to the dataset to the complex energy plane,
using the method for three-body unitary models
discussed in Refs.~\cite{kpp-ikeda,a1-gwu}.
We can then find 
a pole location $E=E_\psi$ where 
${\rm det}[\bar{G}^{-1}(E_\psi)]=0$
for Eq.~(\ref{eq:mstar-g1}).
We search for vector charmonium poles
on the relevant Riemann sheets:\footnote{
See Sec.~50 in Ref.~\cite{pdg} for the definition of the (un)physical
sheet.}
unphysical sheets of the open channels and 
physical sheets of the closed channels, or sheets slightly deviating from this condition;
$3.75 < M < 4.7$~GeV ($M\equiv {\rm Re} [E_{\psi}]$),
and $\Gamma\equiv -2\times{\rm Im} [E_\psi] < 0.2$~GeV.
\begin{table}
\renewcommand{\arraystretch}{1.25}
\tabcolsep=1.05mm
\caption{\label{tab:pole}
Vector charmonium poles and 
$\psi(4660)$ BW parameters.
See the text for the notations.
}
\begin{ruledtabular}
\begin{tabular}{rrccl}
  \multicolumn{2}{c}{This work} &
 \multicolumn{3}{l}{PDG($\psi$)~\cite{pdg}, BESIII~\cite{bes3_r3780,bes3_DD,bes3_jpsi-pippim}}
 \\
 $M$ (MeV) & $\Gamma$ (MeV) & $M$ (MeV) & $\Gamma$ (MeV) & \\\hline
$ 3764.2\pm  2.0$ &$   47.3\pm  2.6$ &$3751.9\pm 3.8$ & $32.8 \pm 5.8$&${}^{\rm r}D\bar{D}$\\
$ 3780.2\pm  1.2$ &$   29.9\pm  2.3$ &$3778.1\pm 0.7$ & $27.5\pm 0.9$ &$\psi(3770)$\\      
$ 3898.4\pm  0.9$ &$  127.5\pm  6.7$ &$3872.5\!\pm\!\! 14.2$& $179.7\!\pm\!\! 14.1$&${}^{\rm r}D^*\bar{D}$\\
$ 3956.1\pm  1.0$ &$   96.8\pm 10.4$ & -- & -- & ${}^{\rm r}D_{s}\bar{D}_s$\\
$ 4029.2\pm  0.4$ &$   26.3\pm  1.0$ &$4039\pm 1$     & $80\pm 10$    &$\psi(4040)$\\      
$ 4052.4\pm  0.4$ &$   49.0\pm  0.3$ & -- & -- & ${}^{\rm v}D_{s}^*\bar{D}_s$\\
$ 4192.2\pm  2.2$ &$  129.3\pm  4.2$ &$4191\pm 5$     & $70\pm 10$    &$\psi(4160)$\\      
$ 4216.2\pm  0.5$ &$   40.3\pm  1.0$ & -- & -- & ${}^{\rm v}D_{s}^*\bar{D}^*_s$\\
$ 4229.9\pm  0.9$ &$   46.4\pm  2.6$ &$4222.5\pm 2.4$ & $48\pm 8$     &$\psi(4230)$\\      
$ 4308.1\pm  2.2$ &$  138.2\pm  4.4$ &$4298\pm 12$    & $127\pm 17$   &$Y(4320)$\\
$ 4346.2\pm  3.8$ &$  122.8\pm  6.7$ &$4374\pm 7$     & $118\pm 12$   &$\psi(4360)$\\      
$ 4390.1\pm  2.0$ &$  106.5\pm  4.1$ &$4421\pm 4$     & $62\pm 20$    &$\psi(4415)$\\
$ 4496.3\pm  3.1$ &$   16.4\pm  2.1$ & -- & -- & ${}^{\rm b}D_{s1}\bar{D}_s$\\
$ 4579.6\pm  1.7$ &$   -5.2\pm  7.6$ & -- & -- & ${}^{\rm b}\Lambda_c\bar{\Lambda}_c$\\
\hline
$ 4655.9\pm 3.0$ &$  134.9\pm 5.9$ &$4630\pm 6$  & $72^{+14}_{-12}$ &$\psi(4660)$
\end{tabular}
\end{ruledtabular}
\end{table}

Pole uncertainty estimates
are generally difficult
in global coupled-channel analyses, 
and simplified methods 
have been used~\cite{ystar,jb22}.
For statistical uncertainty estimate,
we introduce complex parameters $\delta m_{\psi_i}$ 
as $m_{\psi_i} \to m_{\psi_i} + \delta m_{\psi_i}$ in Eq.~(\ref{eq:mstar-g1}).
We also select parameters to which pole
locations are sensitive
such as: bare couplings for ``$\psi, \gamma^*\to$ open-charm channels'' that
mainly dress bare $\psi$ states thereby shifting their masses and generating widths;
diagonal and/or large couplings of $v^{\rm s}$ in Eq.~(\ref{eq:ptl})
that generate hadron-molecule states.
We then vary these parameters, 85 in total, around the default fit 
for the uncertainty estimate.
This time, we neither weight the data nor limit the parameter ranges as we did to
obtain the default fit. 
See Appendix~\ref{sec:vec-error} for more details.

Regarding systematic uncertainty (model dependence) of the poles, 
there are certainly many possible sources
such as the choice of form factors (cutoffs), the number of bare states, 
whether parameters are constrained by the HQSS or SU(3), etc. 
For a substantial model variation, which is crucial for the systematic
uncertainty estimate, finding a solution comparable to the default fit
requires considerable effort that warrants an independent paper.
We thus do not go into this task here.
This important issue can be addressed when updating the model 
in the future by including more data and theoretical inputs.

\begin{figure*}
\begin{center}
\includegraphics[width=1\textwidth]{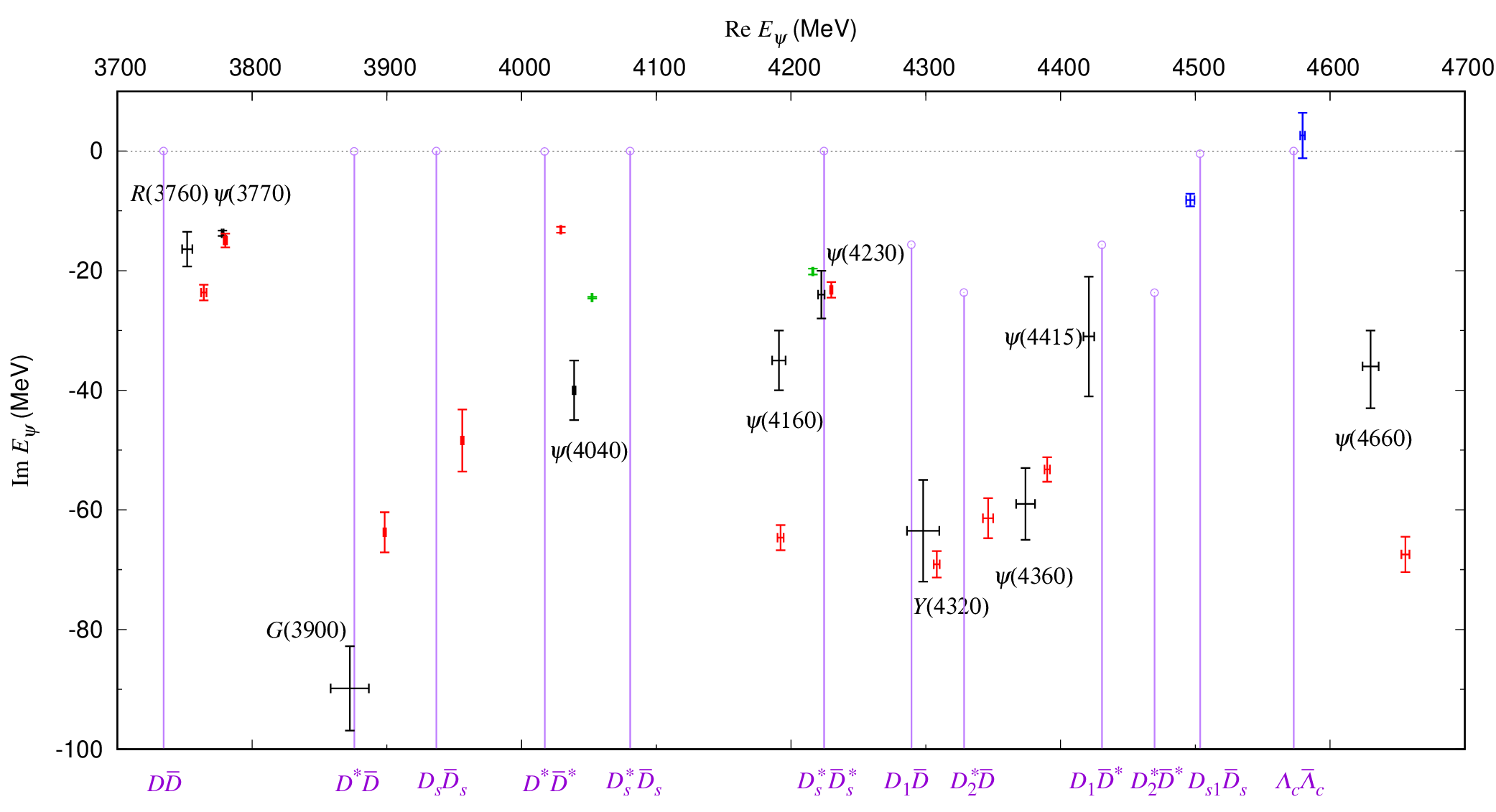}
\end{center}
 \caption{Vector charmonium poles ($E_\psi$) with uncertainties. 
Red points indicate resonance poles [located on unphysical (physical) sheets of
 open (closed) channels], while
blue and green points indicate bound and virtual
 poles, respectively, of the nearest-threshold channels.
Black points are $\psi$ states listed in PDG~\cite{pdg}, 
${\cal R}(3760)$~\cite{bes3_r3780}, 
$G(3900)$~\cite{bes3_DD}, and 
$Y(4320)$~\cite{bes3_jpsi-pippim}.
Open circles and accompanying vertical lines are 
branch points [Eq.~(\ref{eq:bp})]
and cuts, respectively, for open-charm channels indicated 
at the bottom.
 }
\label{fig:cc}
\end{figure*}

We find 14 states, as listed in Table~\ref{tab:pole} where 
experimental analysis results are also shown for comparison.
A graphical presentation of this table is given in Fig.~\ref{fig:cc}.
Overall, the pole uncertainties from our coupled-channel analysis are
smaller than those from the experimental single-channel analyses. This
can be expected since, in the former, the data of the various processes
constrain the pole locations, and some data are very precise.
Our analysis finds states that
can be identified with all of the vector charmonia
($M>3.75$~GeV) listed in the PDG~\cite{pdg}.
However, there are sizable differences with the PDG average
such as the $\psi(4040)$ width and the $\psi(4415)$ mass and width. 
One possible cause of the differences is threshold effects,
as discussed in Sec.~\ref{sec:remark},
that can shift a lineshape peak position from a resonance mass. 
The $\psi(4415)$ as well as $\psi(4040)$ and $\psi(4160)$
resonance parameters in the PDG are basically from the BW fit to
the $R$ values~\cite{bes2-R}
without considering any thresholds and coupled channels.
The previous simple analyses might have introduced
artifacts in the extracted resonance parameters.

\begin{figure*}
\begin{center}
\includegraphics[width=1\textwidth]{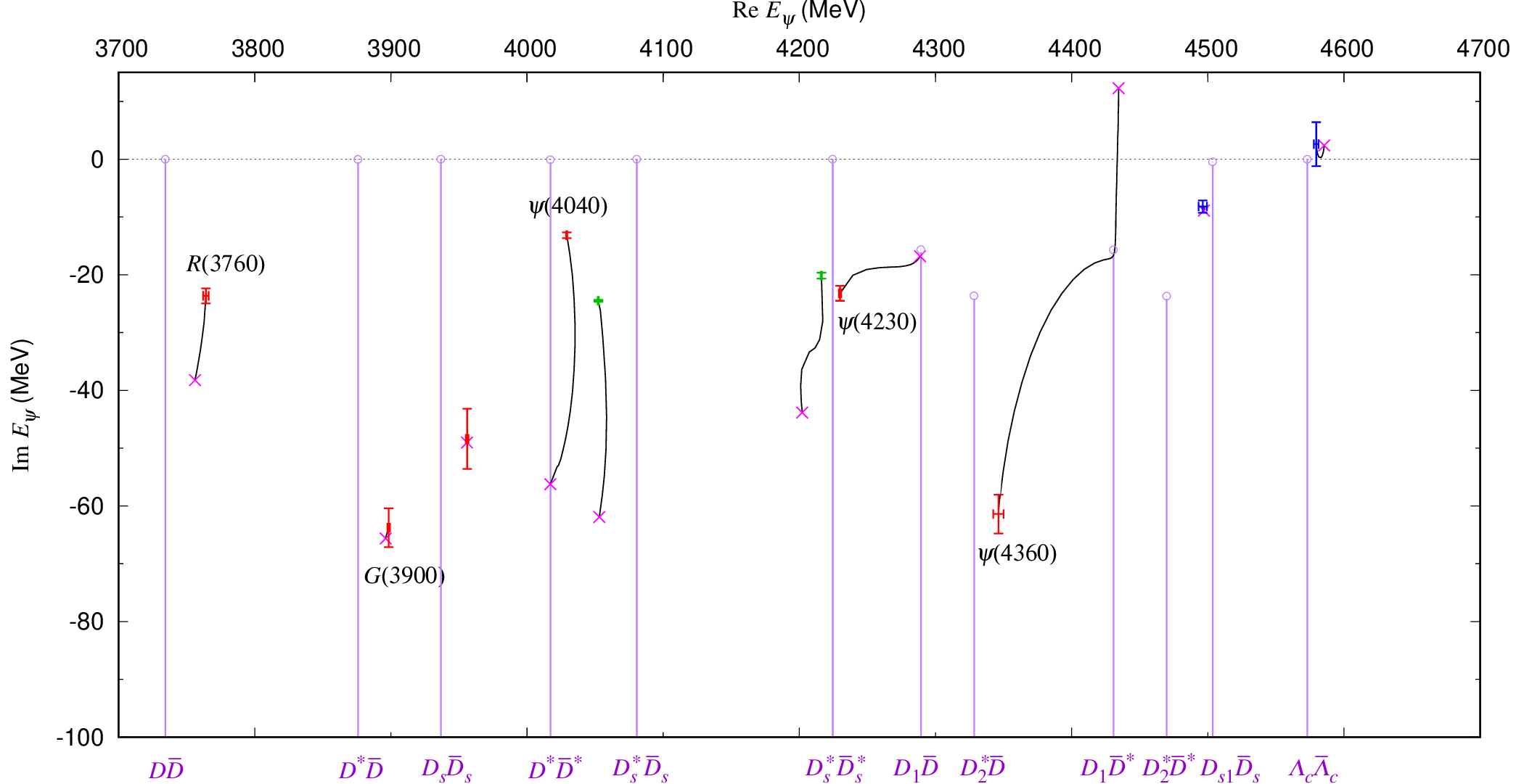}
\end{center}
 \caption{Pole trajectories of hadron-molecule dominated states. 
The magenta crosses are the locations of poles generated 
by the hadron-hadron interactions of Eq.~(\ref{eq:ptl}), without
couplings to bare $\psi$ states. 
The poles move along the black curves (trajectories) as 
bare $\psi_i\to Rc$ couplings are gradually turned on. 
The other ends of the trajectories are the pole locations,
listed in Table~\ref{tab:pole},
from the default model.
 }
\label{fig:cc-traj}
\end{figure*}

Moreover, several states are found close to 
quasi-thresholds of open-charm ($H\bar{H}'$) channels.
We denote these states as ${}^{\rm x}H \bar{H}'$
with ${\rm x=b,r,v}$ (bound, resonant, virtual),
based on the pole locations, regardless of their internal structures.
Given the branch point of
\begin{eqnarray}
 E^{\rm BP}_{H\bar{H}'} = (m_H+m_{\bar{H}'}) -{i\over 2}(\Gamma_H+\Gamma_{\bar{H}'}),
\label{eq:bp}
\end{eqnarray}
${}^{\rm b}H \bar{H}'$ 
with ${\rm Im} [E_\psi] \le {\rm Im} [E^{\rm BP}_{H\bar{H}'}]$
(${\rm Im} [E_\psi] > {\rm Im} [E^{\rm BP}_{H\bar{H}'}]$
and ${\rm Re} [E_\psi] > {\rm Re} [E^{\rm BP}_{H\bar{H}'}]$)
is located on the physical (unphysical) sheet of the 
$H \bar{H}'$ channel. 
${}^{\rm r(v)}H \bar{H}'$ 
is located on the unphysical sheet of the 
$H \bar{H}'$ channel, 
and 
${\rm Re} [E_\psi] \ge {\rm Re} [E^{\rm BP}_{H\bar{H}'}]$
(${\rm Re} [E_\psi] < {\rm Re} [E^{\rm BP}_{H\bar{H}'}]$).

The ${}^{\rm r}D\bar{D}$ state from our analysis is
similar to ${\cal R}(3760)$ claimed by the BESIII's analyses 
of $e^+e^-\to$ hadrons~\cite{bes3_r3780}
and $e^+e^-\to$ non-open-charm hadrons~\cite{R3810}.
The BESIII also found ${\cal R}(3810)$
with $M\sim 3805$~MeV and $\Gamma\sim 10$~MeV
in Refs.~\cite{bes3_r3780,R3810}.
Our analysis does not find ${\cal R}(3810)$ because our dataset does not
show any structure associated with it. 
We find a ${}^{\rm r}D^*\bar{D}$ state
similar to $G(3900)$ from the BESIII analysis on $e^+e^-\to D\bar{D}$~\cite{bes3_DD}.
Coupled-channel $K$-matrix analyses were done for 
the $e^+e^-\to D^{(*)} \bar{D}^{(*)}$ cross section data [Fig.~\ref{fig:xs-opencc}(a-c)]
and inclusive data for $\sqrt{s}<4.2$~GeV~\cite{bes3_DD2,peking_G3900}.
A $G(3900)$ pole was found 
at $(3869.2\pm 6.7)-(29.0\pm 5.2)i$~MeV in Ref.~\cite{peking_G3900} 
but not in Ref.~\cite{bes3_DD2}.
A similar analysis of older data was done in Ref.~\cite{Du-hqss} using a
HQSS-based coupled-channel model, and 
$G(3900)$ was claimed at $3879-35i$~MeV.
The $G(3900)$ widths from Refs.~\cite{peking_G3900,Du-hqss} are
significantly narrower than our result. 
These theoretical analyses~\cite{bes3_DD2,Du-hqss,peking_G3900} did not
find ${\cal R}(3760)$ and ${\cal R}(3810)$.
A $\Lambda_c\bar{\Lambda}_c$ bound state was claimed at $\sim$38~MeV
below the threshold from a single-channel analysis of the
$e^+e^-\to\Lambda_c\bar{\Lambda}_c$ data~\cite{LLbar}.
Our analysis found a similar pole but located above the threshold due to
a coupled-channel effect.
A $D_{s1}\bar{D}_s$ bound state predicted 
with a HQSS and SU(3) symmetric model in Ref.~\cite{FZPeng2023} 
is similar to our ${}^{\rm b}D_{s1}\bar{D}_s$.
The authors of Ref.~\cite{FZPeng2023} interpreted this state as
$Y(4500)$ with $\Gamma\sim 111$~MeV appearing in $e^+e^-\to J/\psi K^+K^-$.
In our analysis, ${}^{\rm b}D_{s1}\bar{D}_s$
is not $Y(4500)$ but a much narrower state 
that causes a dip in the $e^+e^-\to D^*_s\bar{D}_s^*$ cross section [Fig.~\ref{fig:xs-opencc}(f)].
The other ${}^{\rm x}H \bar{H'}$ states in Table~\ref{tab:pole}
are found for the first time in
the present analysis.

Two reasons why our model can accommodate
more poles than the five bare states are:
(i) The short-range interactions among the open-charm channels 
can generate hadron-molecule states;
(ii) A bare state
may cause more than one resonance by
coupling with hadronic continuum states (Table~\ref{tab:Rc});
see a demonstration in Ref.~\cite{ebac_suzuki}.

There exist 4 poles with $M\sim 4.23$~GeV,
and 2 poles with $M\sim 4.38$~GeV.
It is speculated that
these overlapping resonances interfere with each other differently in various
processes, resulting in the process-dependent $Y$ lineshapes. 
Indeed, the previous analyses
considered the interference of 
$\psi(4160)$ and $\psi(4230)$ to explain
the seemingly process-dependent $\psi(4230)$ width
for $e^+e^-\to J/\psi\eta$~\cite{tcpeng2024} 
and for 
$e^+e^-\to J/\psi\pi\pi, \pi D^*\bar{D}, J/\psi\eta$ and more in Ref.~\cite{detten}. 
We will address this issue in the future, taking into account the unitarity that
was not considered in the previous analyses.

\subsection{Pole trajectories}
\label{sec:pole-traj}

In our coupled-channel model, 
vector charmonium states can be formed from 
bare $\psi$ states dressed by quasi-two-body continuum
states of Table~\ref{tab:Rc}, as can be seen in Eq.~(\ref{eq:mstar-g1}).
Another pole formation is from hadron-hadron interactions of
Eq.~(\ref{eq:ptl}).
In this case, 
the $Rc\to R'c'$ partial wave amplitude
[$X_{(Rc)_{ls},(R'c')_{l's'}}$ in Eq.~(\ref{eq:pw-tcr})]
include hadron-molecule poles and, as a consequence, 
so does the NR amplitude in Eq.~(\ref{eq:amp_full}). 
We find such hadron-molecule poles 
near most of the open-charm thresholds,
as shown by the magenta crosses
in Fig.~\ref{fig:cc-traj};
$v^{\rm s}$ terms in Eq.~(\ref{eq:ptl}) play a dominant role for these
hadron-molecule formations.

The hadron-molecule poles further couple with bare $\psi$ states via continuum
states, yielding some of the poles in Table~\ref{tab:pole}.
This development can be visualized in Fig.~\ref{fig:cc-traj}
as pole trajectories (black curves).
To draw the trajectories, we multiply a common parameter $\lambda$ to
all of bare $\psi_i\to Rc$ couplings 
[$C^i_{(Rc)_{ls}}$ in Eq.~(\ref{eq:bare_mstar})] of the default model.
As we vary $\lambda$ from 0 to 1, the pole locations move from the magenta
crosses to the other ends (pole locations from the default model)
 following the black curves.
An exception is 
the trajectory connecting to $\psi(4360)$
because we do not find
a practical integral path for the analytic continuation 
to the pole location for $\lambda=0$.
Thus we introduced one more
parameter $\lambda'$ by which the $Z$ terms in
Eq.~(\ref{eq:ptl}) are multiplied.
The trajectory is drawn by
varying the parameters as 
$\lambda=0\to 1$ with $\lambda'=0$, and then 
$\lambda'=0\to 1$ with $\lambda=1$.
We note that the trajectories shown in Fig.~\ref{fig:cc-traj} are not
unique but dependent on how the bare $\psi_i\to Rc$ couplings
are turned on;
even the endpoints of the trajectories may change.

Previous theoretical studies~\cite{gjding,xkdong1,xkdong2,Ji2022,FZPeng2023}
speculated that $\psi(4230)$ is 
a $D_1\bar{D}$ bound state formed by short-range interactions without
coupling to charmonium states.
References~\cite{xkdong2,Ji2022} also assigned 
a $D_1\bar{D}^*$ bound state to $\psi(4360)$.
A HQSS model of Ref.~\cite{FZPeng2023} predicted a $D_1\bar{D}^*$
virtual state, which is more consistent with our result. 
Our analysis found not only such molecular poles but also their
significant shifts due to couplings with (bare) charmonium states.
The $G(3900)$ pole was described as a $p$-wave $D^*\bar{D}$ molecule 
from an one-boson-exchange model in Ref.~\cite{peking_G3900}.
Our analysis found a similar $D^*\bar{D}$ molecular state that is
slightly shifted by couplings with (bare) charmonium states.
We also found a $p$-wave $D^*\bar{D}^*$ molecular state that
is shifted to become $\psi(4040)$, one of the well-established vector
charmonium states.

\subsection{Compositeness}

The pole trajectories shown in Fig.~\ref{fig:cc-traj} imply that some
of the vector charmonium states from the default model may have substantial hadron-molecule
contents.
To explore the internal structure, 
the compositeness~\cite{weinberg1965,sekihara2015,baru2004}
might serve as a qualitative measure. 
When a resonance state (Gamow state) $|\psi)$ is normalized as 
$(\psi|\psi)=1$, 
a compositeness $X_{Rc}$ is the contribution from a continuum (quasi-two-body)
$Rc$ channel to this normalization. 
With an elementariness $Z_a$ from a bare state $a$, 
the sum rule $\sum_{Rc} X_{Rc} + \sum_a Z_a=1$ is satisfied.

Here, we calculate the compositeness in a manner similar to
Ref.~\cite{sekihara2015}.
The formulas in Ref.~\cite{sekihara2015} are for models including bare
states and two-body continuum states.
Since our coupled-channel model includes three-body channels that enter the
calculations in the forms of the $Z$-diagrams of Eq.~(\ref{eq:ptl}) and 
the $R$ self energies and widths of Eqs.~(\ref{eq:green-Rc})-(\ref{eq:RR-self2}), 
we extend the formulas in Ref.~\cite{sekihara2015}.

The compositeness of a state $\psi$ at $E=E_\psi$
can be expressed with its residue. 
The residue for an $Rc\to R'c'$ transition is given by
\begin{eqnarray}
&& r^\psi_{(R'c')_{l's'},(Rc)_{ls}}(p_{c'},p_{c})
\nonumber\\
&=& 
\lim_{E\to E_\psi} (E-E_\psi)\sum_{ij}
 \bar F_{(R'c')_{l's'},\psi_i}(p_{c'}, E) 
\nonumber\\
&&\times \bar{G}_{ij}(E)
 \bar F_{(Rc)_{ls},\psi_j}(p_c, E) 
\nonumber\\
&=&
\sum_{ij}
 \bar F_{(R'c')_{l's'},\psi_i}(p_{c'}, E_\psi) [\widetilde{\bar{G}^{-1}}(E_\psi)]_{ij} [\Delta'(E_\psi)]^{-1}
\nonumber\\
&&\times 
 \bar F_{(Rc)_{ls},\psi_j}(p_c, E_\psi),
\label{eq:residue}
\end{eqnarray}
where 
$\bar F_{(Rc)_{ls},\psi_j}$ and $\bar{G}_{ij}$ have been defined in
Eqs.~(\ref{eq:dressed-ff}) and (\ref{eq:mstar-g1}), respectively;
$[\widetilde{\bar{G}^{-1}}]_{ij}$ is the adjugate matrix of $[\bar{G}^{-1}]_{ij}$,
$\Delta(E)$ is the determinant of 
the matrix $\left[\bar{G}^{-1}(E)\right]_{ij}$, 
 and $\Delta'(E_\psi)=d\Delta(E)/dE|_{E=E_\psi}$.
The compositeness is given 
with the residue as
\begin{eqnarray}
X_{Rc} &=& 
\int {q^2 dq\over 4 E^2_R(q)}
 {  \sum_{ls}r^\psi_{(Rc)_{ls},(Rc)_{ls}}(q,q)
\over [E_\psi - E_R(q) - E_c(q) +i{\Gamma_R\over 2} ]^2} .
\label{eq:comp}
\end{eqnarray}
The above compositeness formula can be reduced to Eq.~(93) of  
Ref.~\cite{sekihara2015} 
by turning off $Z$-diagrams and $R$ self energies and widths,
by non-relativistic reductions, and by using the same normalizations of
the form factors.

\begin{table*}
\renewcommand{\arraystretch}{1.3}
\caption{\label{tab:comp1} 
Compositeness $X_{Rc}$
of vector charmonium states $\psi$
listed in Table~\ref{tab:pole}.
Contributions from open-charm channels are shown. 
Hyphens indicate $|X_{Rc}|<0.01$.
}    
\begin{ruledtabular}
\begin{tabular}{lccccccc}
$Rc\ \backslash\ \psi$
& 
${}^{\rm r}D\bar{D}$                 &
$\psi(3770)$                         &
${}^{\rm r}D^*\bar{D}$               & 
${}^{\rm r}D_{s}\bar{D}_s$	     &
$\psi(4040)$                         & 
${}^{\rm v}D_{s}^*\bar{D}_s$	     &
$\psi(4160)$      		     
 \\\hline
$D\bar{D}$ &                $ 1.66+ 0.42i$&$-0.66-0.41i$&$-0.03-0.01i$&--&--&--&--\\
$D^*\bar{D}$ &	       $ 0.01-0.01i$&--&$ 1.10+ 0.04i$&--&$-0.01+ 0.01i$&--&--\\
$D^*\bar{D}^*$ &	       --&--&--&--&$ 0.86+ 0.22i$&$ 0.06+ 0.12i$&$ 0.26-0.03i$\\
${D}_1\bar{D}$ & 	       --&--&--&--&--&--&$ 0.01-0.02i$\\
${D}_1\bar{D}^*$ & 	       $ 0.01-0.01i$&--&--&--&--&--&$-0.00+ 0.01i$\\
${D}_2^*\bar{D}^*$&	       --&--&--&--&--&--&$-0.01-0.02i$\\
$D_s\bar{D}_s$ &	       --&--&--&$ 1.01+ 0.01i$&--&--&--\\
$D_s^*\bar{D}_s$ &	       --&--&--&--&$ 0.01-0.03i$&$ 0.91-0.09i$&$-0.03-0.25i$\\
$D_s^*\bar{D}_s^*$ &	       --&--&--&--&--&--&$-0.10-0.04i$\\
$D_{s1}\bar{D}_s$ & 	       --&--&--&--&--&--&--\\
$\Lambda_c\bar{\Lambda}_c$& --&--&--&--&--&--&--\\
Sum&			       $ 1.68+ 0.38i$&$-0.68-0.40i$&$ 1.05+ 0.02i$&$ 1.00+ 0.00i$&$ 0.86+ 0.19i$&$ 0.97+ 0.03i$&$ 0.14-0.36i$\\
  \end{tabular}
\end{ruledtabular}
\end{table*}

\begin{table*}
\renewcommand{\arraystretch}{1.2}
\caption{\label{tab:comp2} 
Continued from Table~\ref{tab:comp1}.
}    
\begin{ruledtabular}
\begin{tabular}{lccccccc}
$Rc\ \backslash\ \psi$
& 
${}^{\rm v}D_{s}^*\bar{D}^*_s$	     &
$\psi(4230)$                         &
$Y(4320)$			     &
$\psi(4360)$      		     &
$\psi(4415)$			     &
${}^{\rm b}D_{s1}\bar{D}_s$	     &
${}^{\rm b}\Lambda_c\bar{\Lambda}_c$ \\\hline
$D\bar{D}$ &                  --&--&--&--&--&--&--\\
$D^*\bar{D}$ &	         --&--&--&--&--&--&--\\
$D^*\bar{D}^*$ &	         $ 0.01+ 0.02i$&--&$ 0.02-0.15i$&--&--&--&--\\
${D}_1\bar{D}$ & 	         $ 0.23-0.12i$&$ 0.18+ 0.13i$&--&$ 0.31-0.06i$&$-0.03+ 0.04i$&--&$ 0.03-0.00i$\\
${D}_1\bar{D}^*$ & 	         $ 0.04-0.06i$&$ 0.09+ 0.05i$&$ 0.01-0.01i$&$ 0.29-0.16i$&$ 0.08+ 0.13i$&--&$ 0.01-0.01i$\\
${D}_2^*\bar{D}^*$&	         $ 0.02-0.01i$&--&$-0.01-0.00i$&--&$ 0.06-0.03i$&--&--\\
$D_s\bar{D}_s$ &	         --&--&--&--&--&$ 0.02-0.03i$&--\\
$D_s^*\bar{D}_s$ &	         $ 0.04+ 0.05i$&$ 0.02-0.05i$&$ 0.05+ 0.06i$&--&$ 0.01-0.02i$&--&--\\
$D_s^*\bar{D}_s^*$ &	         $ 0.50+ 0.25i$&$ 0.35-0.27i$&$-0.11-0.03i$&--&$-0.03-0.00i$&--&--\\
$D_{s1}\bar{D}_s$ & 	         --&--&--&--&--&$ 0.99+ 0.00i$&--\\
$\Lambda_c\bar{\Lambda}_c$&   --&--&--&$ 0.02-0.04i$&$ 0.01+ 0.01i$&--&$ 0.97+ 0.01i$\\
Sum&			         $ 0.83+ 0.12i$&$ 0.65-0.13i$&$-0.06-0.14i$&$ 0.61-0.27i$&$ 0.10+ 0.13i$&$ 1.01-0.01i$&$ 1.00-0.02i$\\
\end{tabular}
\end{ruledtabular}
\end{table*}

Caveats are in order regarding the 
compositeness calculated with Eq.~(\ref{eq:comp}).
The compositeness is generally complex, and it is difficult to interpret 
its imaginary part.
Furthermore, interpreting $X_{Rc}$ is difficult for cases with
${\rm Re}[X_{Rc}] < 0$ or ${\rm Re}[X_{Rc}] > 1$.
Therefore, only when the imaginary part is significantly smaller than the real part
and $0 \le {\rm Re}[X_{Rc}]\le 1$,
we may interpret $X_{Rc}$ as an approximate probability of finding $Rc$ continuum
states (or a $Rc$ molecule) in a resonance state. 
Also, Eqs.~(\ref{eq:residue}) and (\ref{eq:comp}) indicate that 
the compositeness depends on the form factors, and this dependence would be
more pronounced for $l>0$. 
Thus, 
the compositeness of $p$-wave states from $D^{(*)}_{(s)}\bar{D}^{(*)}_{(s)}$
scattering should be viewed with more caution.
The compositeness is given model-independently 
for $s$-wave states
in the weak binding limit only~\cite{weinberg1965,sekihara2015,baru2004}.

Tables~\ref{tab:comp1} and \ref{tab:comp2} present
the compositeness for the vector 
charmonium states listed in Table~\ref{tab:pole}.
The result confirms what the trajectory analysis suggested:
the states shown in Fig.~\ref{fig:cc-traj} have large compositeness. 
A noteworthy case is 
$\psi(4040)$ with $X_{D^*\bar{D}^*}\sim 0.86$.
This well-established state has been assumed to be the $\psi(3S)$ state in
quark models, and its experimentally determined mass has been used as an
input in determining the quark-model parameters~\cite{barnes}.
However, our comprehensive analysis might suggest reconsidering this conventional assumption.

As mentioned earlier, 
$\psi(4230)$ and $\psi(4360)$ 
have often been speculated to be 
$D_1\bar{D}$ and $D_1\bar{D}^*$ molecules, respectively~\cite{gjding,xkdong1,xkdong2,Ji2022,FZPeng2023}.
However, our compositeness analysis suggests more complex structures than these expectations.
Hadron dynamics cause
large mixings among nearby molecular states 
[$D_1\bar{D}$, $D_1\bar{D}^*$, and $D_s^*\bar{D}^*_s$] 
and also bare $\psi$ states to form 
$\psi(4230)$, $\psi(4360)$, and 
${}^{\rm v}D_{s}^*\bar{D}^*_s$ states.
Because of these mixings, it is not straightforward to relate 
quark-model states to the vector charmonium states in this region. 
A possible idea is to introduce quark-model states as bare $\psi$ states
in our coupled-channel model; a related work can be found in Ref.~\cite{heft}.

\subsection{$Z_c$ poles}

\begin{figure}
\includegraphics[width=.49\textwidth]{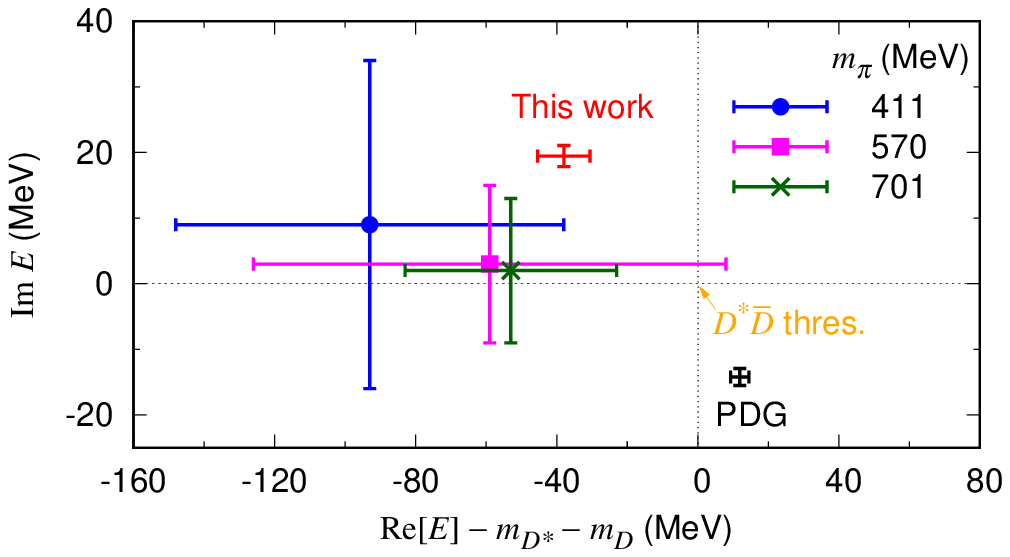}
 \caption{\label{fig:zc}
$Z_c(3900)$ poles on the $D^*\bar{D}$ unphysical sheet from
this work, PDG~\cite{pdg}, and 
lattice QCD~\cite{ikeda_zc_pole}.
}
 \end{figure}

\begin{figure}[b]
\includegraphics[width=.5\textwidth]{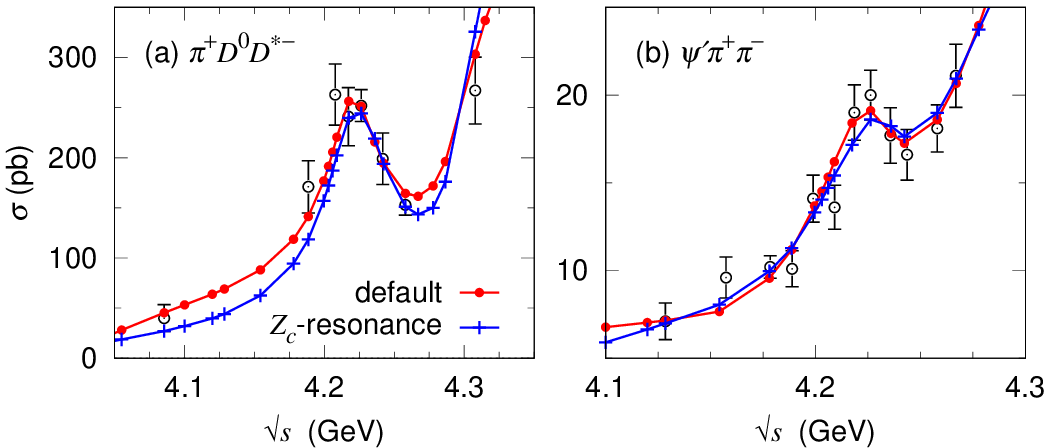}[b]
 \caption{\label{fig:zc-res-fit}
Comparison of the default model ($Z_c$-virtual-state solution)
 and $Z_c$-resonance solution on fitting $e^+e^-$ annihilation
 cross-section data. Data are from Refs.~\cite{bes3_piDDstar,bes3_psip_pippim}.
 }
\end{figure}

Finally, Table~\ref{tab:pole2} presents the $Z_c$ poles
in $D^*\bar{D}-D^*\bar{D}^*-J/\psi\pi-\psi'\pi-h_c\pi-\eta_c\rho$
coupled-channel scattering amplitude ($IJ^{PC}=11^{+-}$)
implemented in our three-body scattering model.\footnote{
Since we did not analyze data showing
a $Z_{cs}(3985)$ structure~\cite{Zcs}, 
we do not discuss a pole in our 
$D_s^*\bar{D}-\bar{D}^*D_s-J/\psi K$
coupled-channel amplitude.}
See Appendix~\ref{sec:zc-error} for the uncertainty estimation method.
One pole (the other) is a $D^*\bar{D}$ ($D^*\bar{D}^*$) virtual state,
located at $\sim$40~MeV below the threshold.
The previous analyses fitted the $M_{\pi J/\psi}$ and $M_{D^*\bar{D}}$
lineshape data [Figs.~\ref{fig:xdxs-jpsi}(b,c,e,f)] 
where the $Z_c(3900)$ signals are clearest, but not fitting
the cross-section data that can
test $Z_c$ production mechanisms and $Z_c$-pole residues.
While some analyses~\cite{Yu_zc_pole,peking_G3900,Zc3900-mldu,Zc3900-Albaladejo,Zc3900-jhe,gong_zc3900,ortega_zc3900}
obtained virtual poles,
the others~\cite{Zc3900-yhchen,Zc3900-jpac,Zc3900-mldu,Zc3900-Albaladejo}
and the experimental ones~\cite{bes3_Zc3900c,bes3_DDstarc_zc3900,z3900-1-belle}
found resonance poles near the PDG value.
Lattice QCD
results~\cite{Prelovsek13,ychen14,Prelovsek15,ikeda16,cheung17,ikeda_zc_pole}
favor the virtual-state-solution, providing $Z_c(3900)$ virtual poles\cite{ikeda_zc_pole}\footnote{
The poles listed in Ref.~\cite{ikeda_zc_pole} have their conjugates 
that are physically more relevant, as pointed out in Ref.~\cite{yamada_zc_pole}.
Figure~\ref{fig:zc} shows the conjugates.
} as shown in Fig.~\ref{fig:zc} where our result compares fairly well.
We also searched a $Z_c$ resonance solution.
The $Z_c$ amplitude in the default-fit model is replaced
by that including energy-dependent $D^*\bar{D}^{(*)}$
interactions~\cite{Zc3900-mldu}.
Then we refitted the full dataset with 201 parameters, 
under a constraint that 
the $Z_c$ amplitude has a resonance pole above the 
$D^*\bar{D}^{(*)}$ threshold.
We could achieve $\chi^2\sim 2510$ while $\chi^2=2320$ for the default model.
Compared to the default model, the invariant-mass distributions are
fitted equally well, but some of the cross-section data are not fitted
comparably. Some cases are shown in Fig.~\ref{fig:zc-res-fit}. This may suggest the
importance of fitting the cross-section data to discriminate between the
$Z_c$ pole locations.

\begin{table}
\renewcommand{\arraystretch}{1.2}
\tabcolsep=.6mm
\caption{\label{tab:pole2}
$IJ^{PC}=11^{+-}$ $D^*\bar{D}-D^*\bar{D}^*-J/\psi \pi-\psi'\pi-h_c\pi-\eta_c\rho$
coupled-channel scattering amplitude poles (unit:MeV).
$Z_c(3900)$ and $Z_c(4020)$ are 
$D^*\bar{D}$ and $D^*\bar{D}^*$ virtual (resonance) poles in this work (PDG~\cite{pdg}).
}
\begin{ruledtabular}
\begin{tabular}{rccc}
 \multicolumn{1}{c}{$E^{\rm This\, work}_{Z_c}$}  & $M^{\rm PDG}_{Z_c}$
     & $\Gamma^{\rm PDG}_{Z_c}$ &\\\hline
 $(3837.7\! \pm\! 7.4)\!+\!(19.4\! \pm\! 1.6)i$ &\ $3887.1\pm 2.6$ & $28.4\pm 2.6$ &$Z_c(3900)$ \\
 $(3989.9\! \pm\! 5.6)\!+\!(26.1\! \pm\! 4.3)i$ &\ $4024.1\pm 1.9$ & $13\pm 5$     &$Z_c(4020)$
\end{tabular}
\end{ruledtabular}
\end{table}

\section{Summary and outlook}
\label{sec:summary}
We performed a global coupled-channel analysis of most of the available
$e^+e^-\to c\bar{c}$ data (20 channels) in $\sqrt{s}=3.75-4.7$~GeV for the first time, 
considering three-body unitarity approximately and all relevant
coupled-channels.
Channel-couplings are caused by bare
$\psi$-excitations, long-range particle exchanges, and short-range
contact interactions. 
We obtained overall reasonable fits to both 
cross-section and invariant-mass distribution data
with $\chi^2/{\rm ndf}\simeq 1.6$.
We clarified mechanisms that generate various structures in the data,
paying special attentions to open-charm threshold cusps enhanced by
nearby poles. 

We predicted the $e^+e^-\to \mu^+\mu^-$ cross sections
using the vacuum polarization due to light hadrons, leptons, and charm
contributions.
The charm contribution was calculated with the tails of the $J/\psi$ and $\psi'$,
as well as with our coupled-channel model. 
While 
our prediction is consistent with previous calculations based on
dispersion relations, it
does not reproduce the fine structure in the data at
$\sqrt{s}\sim 4.2$~GeV.

We analytically continued the coupled-channel amplitude to extract
vector-charmonium poles. 
We obtained not only familiar vector charmonia, but
also those near the open-charm thresholds;
some of the states were found for the first time.
We examined pole trajectories and compositeness of the poles to explore
the internal structures of the vector-charmonium states. 
This study suggested that open-charm hadron-molecular structures
dominated in many states.
The $\psi(4230)$ and $\psi(4360)$ states are not simple 
$D_1\bar{D}$ and $D_1\bar{D}^*$ molecules, respectively, 
as proposed in the literature,
but rather mixtures of them plus 30-40\% elementariness; 
$\psi(4230)$ further includes substantial 
$D_s^*\bar{D}_s^*$ compositeness.
Also, our analysis suggested a large $D^*\bar{D}^*$ compositeness in
$\psi(4040)$, 
although $\psi(4040)$ has been assumed to be the $\psi(3S)$ quark-model state.
While we estimated statistical uncertainties of the pole locations, 
we did not address systematic uncertainties (see Sec.~\ref{sec:vec-pole}),
leaving this task to the future work.

We found $Z_c$ poles as $D^*\bar{D}^{(*)}$ virtual states;
similar conclusions are also from lattice QCD analyses.
We suggested the importance of 
analyzing cross-section data in addition to the invariant-mass
distribution data to discriminate whether 
$Z_c$ is a resonance or a virtual state.

In the future, we will examine how the vector-charmonium states contribute to
each of the $e^+e^-\to c\bar{c}$ processes through interferences with
one another,
causing the process-dependent $Y$-lineshapes.
The BESIII will provide more data covering higher energy and more
channels. We will update the current analysis by including such data,
and address the properties of vector-charmonium states heavier than 4.6
GeV and those of $Z_{cs}$. 
More detailed data such as Dalitz plots are also expected to be
available from the BESIII.
It is important to verify the existence
and properties of the presented
vector-charmonium states by analyzing such detailed experimental
information, while also incorporating and/or varying theoretical inputs.

\begin{acknowledgments}
We acknowledge T. Doi, F.-K. Guo, C. Hanhart, T. Sato, Q. Wang,
 J.-J. Wu, W. Yamada, M.-J. Yan, and Z.-Y. Zhou for useful discussions.
This work is in part supported by 
National Natural Science Foundation of China (NSFC) under contracts 
U2032103 and 11625523, 
and also by
National Key Research and Development Program of China under Contracts 2020YFA0406400.
\end{acknowledgments}

\appendix

\section{Two-meson scattering models}
\label{sec:two-meson}

Basic components [Eqs.~(\ref{eq:pipi-vertex0})-(\ref{eq:bw-Rc}) and (\ref{eq:ptl})]
in our three-body coupled-channel model
are two-meson scattering models for
$ab\to a'b'$ via (bare) $R$-excitations [see Fig.~\ref{fig:diag}(a) for the notation],
and also via contact interactions.
The (bare) $R$ states
are categorized into groups (A)--(C) of Table~\ref{tab:Rc};
no $R\to ab$ couplings are considered for group (B) in our model.
We also consider short-range $Rc\to R'c'$ interactions 
[$v^{\rm s}$ in Eq.~(\ref{eq:ptl})]
between open-charm channels.
These two-meson scattering models and parameters therein are described below.

\subsection{Groups (A) and (B) of Table~\ref{tab:Rc}}

The $R$-propagations are described by the BW form
of Eq.~(\ref{eq:bw-Rc}), 
and the BW mass and width values 
are taken from the PDG~\cite{pdg} except that
$\Gamma_{J/\psi}=\Gamma_{D_s^{*}}=0$.
The $R\to ab$ decay vertices ($\Gamma_{ab,R}$) are 
Eqs.~(\ref{eq:pipi-vertex0}) and (\ref{eq:pipi-vertex}). 
The coupling constants ($g^{LS}_{ab,R}$) in $\Gamma_{ab,R}$, 
whose numerical values are listed in Table~\ref{bw-param},
are determined, assuming that 
$D_1(2420)\to D^*\pi$ (mainly $d$-wave),
$D_1(2430)\to D^*\pi$ ($s$-wave),
$D_2^*(2460)\to D^*\pi + D\pi$
[$\Gamma(D\pi)/\Gamma(D^*\pi)\sim 1.5$~\cite{pdg}], 
$D^*\to D\pi$,
and
$D_{s1}(2536)\to D^*K$ ($d$-wave)
saturate their widths. 
The $D^*\to D\pi$ and $D^*\to D^*\pi$ 
coupling constants are related by the HQSS~\cite{hqss_wise};
$g^{11}_{D^*\pi,D^*}/g^{10}_{D\pi,D^*}=\sqrt{2 m_{D^*}/m_D}$
within our definition.
A small $s$-wave decay of $D_1(2420)$
is also included to reproduce the helicity angle
distribution~\cite{d1-helicity}.
The $D_1(2420)$ $d$-wave decay partial width is 98.7\% of the total 
 decay width.
This is rather different from 48\% in Ref.~\cite{detten}
where their $d$-wave $D_1(2420)\to D^{*}\pi$ coupling constant was
determined using a HQSS relation with 
$D_2^*(2460)\to D^{(*)}\pi$ couplings;
the $D_2^*(2460)$ couplings
were fitted to experimental $\Gamma_{D^*_2(2460)}$.

\begin{table}
\renewcommand{\arraystretch}{1.2}
\caption{\label{bw-param}
Parameter values for $R$ listed in Table~\ref{tab:Rc}(A):
the BW mass $m_R$ and width $\Gamma_R$
in Eq.~(\ref{eq:bw-Rc});
couplings $g^{LS}_{ab,R}$ in Eq.~(\ref{eq:pipi-vertex}).
For the cutoff in  Eq.~(\ref{eq:pipi-vertex}), 
$c_{ab,R}=1$~GeV is commonly used for all $R$ in this table.
}
\begin{ruledtabular}
\begin{tabular}{lr}
$m_{D_1(2420)}$ (MeV)          &  2422\\
$\Gamma_{D_1(2420)}$ (MeV)     &  31	  \\
$g^{01}_{D^*\pi,D_1(2420)}$       & $-$0.860  \\
$g^{21}_{D^*\pi,D_1(2420)}$       &   1.25    \\ \hline
$m_{D_1(2430)}$ (MeV)          & 2412\\
$\Gamma_{D_1(2430)}$ (MeV)     & 314 \\
$g^{01}_{D^*\pi,D_1(2430)}$       &   23.5    \\
$g^{21}_{D^*\pi,D_1(2430)}$       & 0 \\\hline
$m_{D^*_2(2460)}$ (MeV)          & 2461 \\
$\Gamma_{D^*_2(2460)}$ (MeV)     & 47   \\
$g^{21}_{D^*\pi,D^*_2(2460)}$       & 0.840    \\
$g^{20}_{D\pi,D^*_2(2460)}$         & 0.569    \\\hline
$m_{D}$ (MeV)          &   1867\\
$\Gamma_{D}$ (MeV)     &      0\\
$g^{11}_{D^*\pi,D}$       &   4.62    \\\hline
$m_{D^*}$ (MeV)          & 2009 \\
$\Gamma_{D^*}$ (MeV)     & 0.069\\
$g^{10}_{D\pi,D^*}$         &  2.67    \\
$g^{11}_{D^*\pi,D^*}$       &3.92  \\ \hline
$m_{D_{s1}(2536)}$ (MeV)          &  2535 \\
$\Gamma_{D_1(2420)}$ (MeV)     &  0.92	  \\
$g^{01}_{D_{s1}(2536)}$       &   0   \\
$g^{21}_{D_{s1}(2536)}$       &   1.14   \\
\end{tabular}
\end{ruledtabular}
\end{table}

\subsection{Group (C) of Table~\ref{tab:Rc}}

\begin{table*}
\renewcommand{\arraystretch}{1.6}
\tabcolsep=2.mm
\caption{\label{tab:R} Description of two-meson scattering models.
Partial waves are specified by 
the orbital angular momentum $L$ and the isospin $I$;
 $J^{P(C)}$ is used for $R=Z_{c}$ and $Z_{cs}$.
}
\begin{ruledtabular}
\begin{tabular}{cccccc}
$R$  & $\{L,I\}$ & coupled-channels ($ab$)&\# of bare $R$ states   & contact interactions & \# of poles\\\hline
$f_0$        & $\{0,0\}$   & $\pi\pi$, $K\bar{K}$ & 2  &  included & 3\\
$f_2$        & $\{2,0\}$   & $\pi\pi$, $K\bar{K}$ & 1  & not included & 1\\
$D_0^*(2300)$& $\{0,1/2\}$ & $D\pi$               & 1  & included & 1\\
\end{tabular}

\vspace{5mm}
\begin{tabular}{cccccc}
$R$  & $I, J^{P(C)}$ & coupled-channels ($ab$) &\# of bare $R$ states   & contact interaction & \# of poles\\\hline
$Z_c$  & $1, 1^{+-}$  & ${D^*\bar{D}-\bar{D}^*D\over \sqrt{2}},D^*\bar{D}^*,J/\psi \pi, \psi'\pi,h_c\pi,\eta_c\rho$
  & 0
& included & 2\\
$Z_{cs}$  & ${1\over 2}, 1^{+}$  & ${D_s^*\bar{D}-\bar{D}^*D_s\over \sqrt{2}},J/\psi K$
  & 0
& included & 1\\
\end{tabular}
\end{ruledtabular}
\end{table*}

\begin{figure*}
\begin{center}
\includegraphics[width=.32\textwidth]{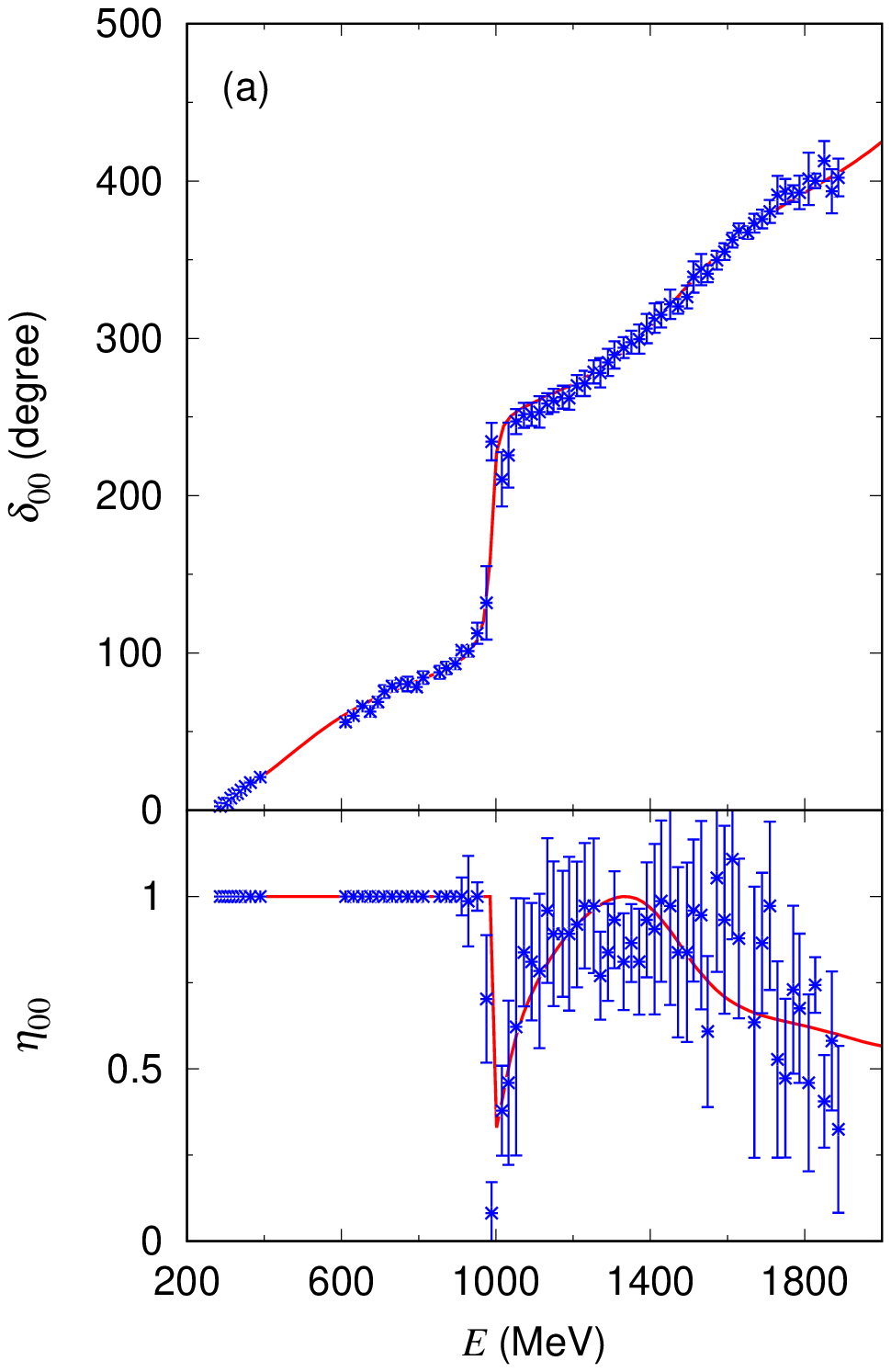}
\includegraphics[width=.32\textwidth]{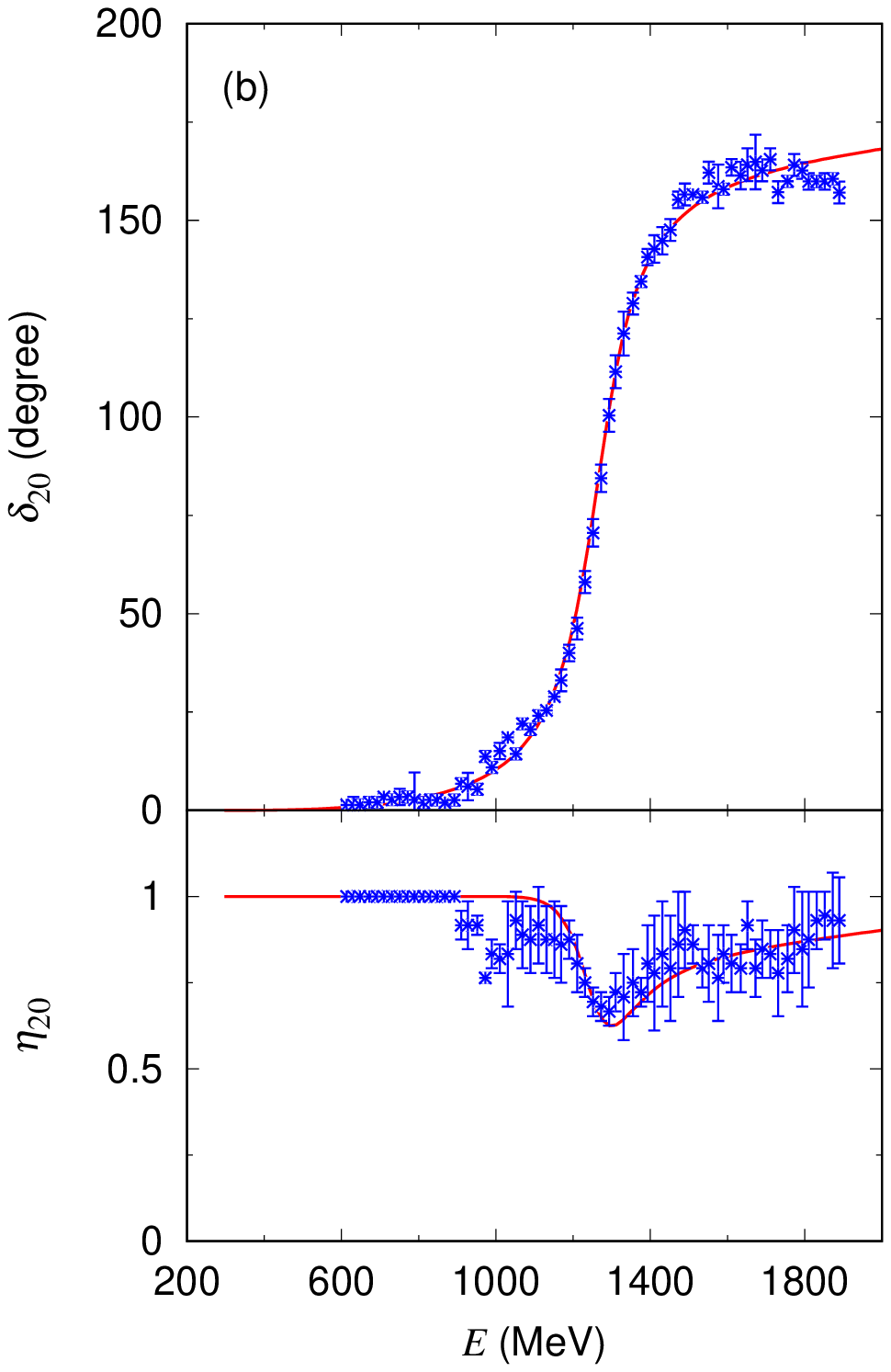}
\includegraphics[width=.32\textwidth]{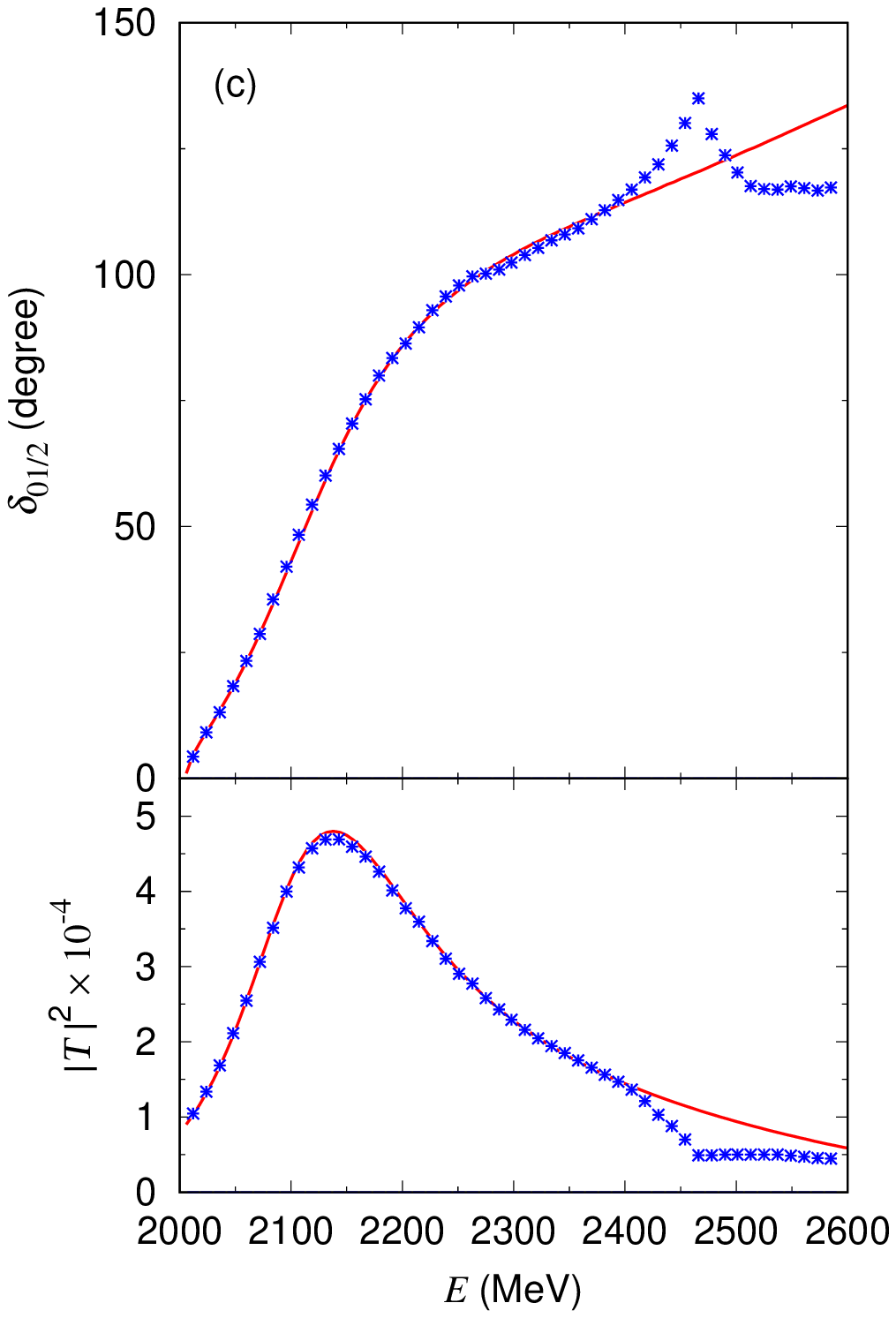}
\end{center}
 \caption{
(a,b) The $\pi\pi$ scattering amplitudes. 
Phase shifts and inelasticities are shown in the upper and lower panels,
 respectively. Data are from Ref.~\cite{pipi-data2}.
(a) $\{L, I\}=\{0,0\}$; (b) $\{L, I\}=\{2,0\}$.
(c) The $D\pi$ scattering amplitude for $\{L, I\}=\{0,1/2\}$.
The phase shifts and modulus of the amplitudes are shown in the upper and lower panels,
 respectively. 
The blue points are from Ref.~\cite{D0-lqcd} (errors are not shown) and
the red curve is our fit.
 }
\label{fig:two-body}
\end{figure*}

Bare $R$ ($D^*_0(2300)$, $f_0$, $f_2$, or $Z_{c(s)}$) states and/or 
contact interactions are implemented in 
a two-body unitary coupled-channel model, and 
are dressed to form pole(s) in the scattering amplitude.
These two-body models are described below.

We consider an $ab\to a'b'$ partial-wave scattering
with the total energy $E$, 
the orbital angular momentum $L$, and the total isospin $I$.
We denote the incoming and outgoing momenta by $q$ and $q'$,
respectively.
First, we introduce a contact interaction for the scattering:
\begin{eqnarray}
v^{LI}_{a'b',ab} (q',q) = 
w^{LI}_{a'b'}(q') h^{LI}_{a'b',ab}\; w^{LI}_{ab}(q) ,
\label{eq:cont-ptl}
\end{eqnarray}
where $h^{LI}_{a'b',ab}$ is a coupling constant.
A vertex function $w^{LI}_{ab}(q)$ is also introduced in the form of:
\begin{eqnarray}
w^{LI}_{ab}(q) = 
{1\over \sqrt{ 4 E_a(q)E_b(q)}} 
{ (q/ m_\pi)^L 
\over
[1 +(q/b^{LI}_{ab})^2]^{2+{L\over 2}} 
} , 
\nonumber\\
\label{eq:vf-cont}
\end{eqnarray}
with a cutoff $b^{LI}_{ab}$.
The corresponding partial wave amplitude is 
\begin{eqnarray}
t^{LI}_{a'b',ab} (q',q; E) &=& \sum_{a''b''}
w^{LI}_{a'b'}(q') \tau^{LI}_{a'b',a''b''}(E)\;
\nonumber\\
&&\times h^{LI}_{a''b'',ab}\; w^{LI}_{ab}(q) ,
\label{eq:pw-2body-cont}
\end{eqnarray}
with
\begin{eqnarray}
\left[(\tau^{LI}(E))^{-1}\right]_{a'b',ab} &=& 
\delta_{a'b',ab} -  \sigma^{LI}_{a'b',ab}(E) ,
\end{eqnarray}
\begin{eqnarray}
\sigma^{LI}_{a'b',ab}(E) &=& 
{\cal B}_{ab}\, h^{LI}_{a'b',ab}
\int  dq\; q^2  
\left[w^{LI}_{ab}(q)\right]^2
\nonumber\\
&&\times
\left(
{1\over E-E_a(q)-E_b(q)+i {\Gamma_a\over 2}+i {\Gamma_b\over 2}}
\right.
\nonumber\\
&&\left.-{1\over E+E_a(q)+E_b(q)+i\epsilon}
\right) ,
\label{eq:pw-cont-self}
\end{eqnarray}
where the Bose factor ${\cal B}_{ab}$ is
${\cal B}_{ab}=1/2$ for identical particles $a$ and $b$, and 
${\cal B}_{ab}=1$ otherwise.
For cases with $\Gamma_a=\Gamma_b=0$, the above formula reduces to
\begin{eqnarray}
\sigma^{LI}_{a'b',ab}(E) &=& 
{\cal B}_{ab} \, h^{LI}_{a'b',ab} \int  dq\; q^2  
{2 M_{ab}(q) \left[w^{LI}_{ab}(q)\right]^2
\over E^2- M^2_{ab}(q)+i\epsilon}
 .
\nonumber \\
\label{eq:pw-cont-self2}
\end{eqnarray}

Second, we include bare $R$-excitation mechanisms:
\begin{eqnarray}
V^{LI}_{a'b',ab} (q',q; E) &=&
 \sum_{R} f^{LI}_{a'b',R}(q') {1\over E^2-m^2_R} f^{LI}_{R,ab}(q) 
\nonumber\\
&&
+ v^{LI}_{a'b',ab} (q',q)  .
\label{eq:pw-2body-v}
\end{eqnarray}
The bare $R\to ab$ vertex function 
${f}^{LI}_{ab,R}(q)$ [$f^{LI}_{R,ab}(q)= f^{{LI}}_{ab,R}(q)$]
is defined in Eq.~(\ref{eq:pipi-vertex})
and thus is plugged into Eq.~(\ref{eq:pipi-vertex0}).
In this section, we use the superscript $LI$ rather than $LS$ in 
Eq.~(\ref{eq:pipi-vertex}).
The resulting scattering amplitude 
from the interaction of Eq.~(\ref{eq:pw-2body-v})
is 
\begin{eqnarray}
T^{LI}_{a'b',ab} (q',q; E) &=&
 \sum_{R',R}
\bar{f}^{LI}_{a'b',R'}(q';E) \tau^{LI}_{R',R}(0,E) \bar{f}^{LI}_{R,ab}(q;E) 
\nonumber\\
&&+t^{LI}_{a'b',ab} (q',q; E) 
\ .
\label{eq:pw-2body-t}
\end{eqnarray}
The second term has already been given in Eq.~(\ref{eq:pw-2body-cont}).
The dressed $R\to ab$ vertex $\bar{f}_{ab,R}$ is
\begin{eqnarray}
\bar{f}^{LI}_{ab,R}(q;E) &=& 
f^{LI}_{ab,R}(q) 
+ \sum_{a'b'}{\cal B}_{a'b'}\int dq' q'^2\; 
2 M_{a'b'}(q')
\nonumber\\
&& 
\times{
t^{LI}_{ab,a'b'} (q,q'; E)\; f^{LI}_{a'b',R}(q')
\over E^2-M^2_{a'b'}(q') + i\epsilon}  \ ,
\nonumber\\
\bar{f}^{LI}_{R,ab}(q;E) &=& 
f^{LI}_{R,ab}(q) + \sum_{a'b'}{\cal B}_{a'b'}
\int dq' q'^2\; 
2 M_{a'b'}(q')
\nonumber\\
&&\times
{f^{LI}_{R,a'b'}(q')\; t^{LI}_{a'b',ab} (q',q; E)
\over E^2-M^2_{a'b'}(q') + i\epsilon}  
\ ,
\label{eq:dressed-vertex}
\end{eqnarray}
where we assumed 
$\Gamma_{a'}=\Gamma_{b'}=0$.
The dressed $R$ Green function
in Eq.~(\ref{eq:pw-2body-t}), $\tau^{LI}_{R',R}(p,E)$, 
is obtained from Eqs.~(\ref{eq:green-Rc})--(\ref{eq:RR-self2}) 
by setting $p=0$ and
replacing only one of 
$f^{LS}_{ab,R'}$ or $f^{LS}_{R,ab}$ with the dressed one of Eq.~(\ref{eq:dressed-vertex}).

The $S$-matrix is related to 
the partial wave amplitude in Eq.~(\ref{eq:pw-2body-t}) by
\begin{eqnarray}
s^{LI}_{ab,ab} (E)
 &&= \eta_{LI}\, e^{2i\delta_{LI}} \nonumber\\
&&= 
1 - 2\pi i\rho_{ab}\, {\cal B}_{ab} T^{LI}_{ab,ab} (q_o,q_o; E) \ ,
\label{eq:s-matrix}
\end{eqnarray}
where the phase shift and inelasticity are denoted by
 $\delta_{LI}$ and $\eta_{LI}$, respectively, and
$q_o$ is the on-shell momentum ($E=E_a(q_o)+E_b(q_o)$);
$\rho_{ab}= q_o E_a(q_o)E_b(q_o)/E$ is the phase-space factor.
The above formalism is used to
calculate 
the $f_0$, $f_2$, $D^*_0(2300)$, $Z_{c}$ and $Z_{cs}$ amplitudes for which 
Table~\ref{tab:R} specifies details.

\begin{table}
\caption{\label{tab:pipi} 
Parameter values for the $\pi\pi$ partial wave scattering model.
The $i$-th bare $R$ states ($R_i$) has a mass of $m_{R_i}$, 
and it decays into $h_1$ and $h_2$ particles 
with a coupling ($g_{h_1h_2,R_i}$)
and a cutoff ($c_{h_1h_2,R_i}$).
Couplings and cutoffs for contact interactions
are denoted by $h_{h_1h_2,h_1h_2}$ and 
$b_{h_1h_2}$, respectively.
For simplicity, we suppress
the superscripts, $LI$, of the parameters.
The mass and cutoff values are given in the unit of MeV,
and the couplings are dimensionless.
}
\begin{ruledtabular}
\begin{tabular}{lrr}
$R~\{L,I\}$&$f_0$ \{0, 0\}&$f_2$ \{2, 0\} \\\hline
$m_{R_1}$              &  1055&   1633\\
$g_{\pi\pi,R_1}$       &  22.0& $-$0.689\\
$c_{\pi\pi,R_1}$       &  1195&   1448\\
$g_{K\bar{K},R_1}$     & $-$6.29&  0.487\\ 
$c_{K\bar{K},R_1}$     &  1173&   1628\\  
$m_{R_2}$              &  1769&   --\\
$g_{\pi\pi,R_2}$       & $-$27.7&   --\\
$c_{\pi\pi,R_2}$       &  1195&   --\\
$g_{K\bar{K},R_2}$     &  17.2&   --\\
$c_{K\bar{K},R_2}$     &  1173&   --\\
$h_{\pi\pi,\pi\pi}$    &  13.7&  --\\	 
$h_{\pi\pi,K\bar{K}}$  & $-$2.76&  --\\	 
$h_{K\bar{K},K\bar{K}}$& $-$3.68&  --\\	 
$b_{\pi\pi}$           &  1195&  --\\
$b_{K\bar{K}}$         &  1173&  --\\
\end{tabular}
\end{ruledtabular}
\end{table}

\begin{table}
\caption{\label{tab:Dpi} 
Parameter values for the $D\pi$ partial wave scattering model.
See Table~\ref{tab:pipi} for the description.
}    
\begin{ruledtabular}
\begin{tabular}{lr}
$R~\{L,I\}$&$D^*_0$ \{0, 1/2\} \\\hline
$m_{R_1}$            &    2294\\
$g_{D\pi,R_1}$       &    24.9\\ 	 
$c_{D\pi,R_1}$       &    1000\\ 
$h_{D\pi,D\pi}$      &    6.26\\	 
$b_{D\pi}$           &    1000\\
\end{tabular}
\end{ruledtabular}
\end{table}

\begin{table}
\caption{\label{tab:Zc} 
Parameter values for the $Z_c$ and $Z_{cs}$ amplitude models.
The cutoff is $b_x=1000$~MeV for all channels $x$.
See Table~\ref{tab:pipi} for the description, and the text for the notations.
}    
\begin{ruledtabular}
\begin{tabular}{lr}
$R~IJ^{P(C)}$& $Z_c 11^{+-}$, $Z_{cs} {1\over 2}1^{+}$ \\\hline
$h_{[D^*\bar{D}],[D^*\bar{D}]}=h_{D^*\bar{D}^*,D^*\bar{D}^*}=h_{[D^*_s\bar{D}],[D^*_s\bar{D}]}$
     & $-$4.00  \\	 
$h_{[D^*\bar{D}],D^*\bar{D}^*}$   & $-$3.59  \\	 
$h_{[D^*\bar{D}],J/\psi\pi}$      & $-$4.23  \\	 
$h_{[D^*\bar{D}],\psi'\pi}$       &  2.77  \\	 
$h_{[D^*\bar{D}],h_c\pi}$         & 0.805  \\	 
$h_{[D^*\bar{D}],\eta_c\rho}$     & $-$5.95  \\	 
$h_{D^*\bar{D}^*,J/\psi\pi}$    &$-$0.714  \\	 
$h_{D^*\bar{D}^*,\psi'\pi}$     &  4.02  \\	 
$h_{D^*\bar{D}^*,h_c\pi}$       &  2.25  \\	 
$h_{D^*\bar{D}^*,\eta_c\rho}$   &  1.53  \\	 
$h_{[D^*_s\bar{D}],J/\psi K}$     & $-$4 (fixed)  \\	 
\end{tabular}
\end{ruledtabular}
\end{table}

\begin{table}
\renewcommand{\arraystretch}{1.6}
\tabcolsep=4.mm
\caption{\label{tab:pole-pipi} 
Pole positions ($M_{\rm pole}$) in our $\pi\pi$ and $D\pi$ scattering amplitudes. 
The Riemann sheets (RS) of the pole positions are specified by $(s_{\pi\pi},s_{K\bar{K}})$
for $\pi\pi$ and $(s_{D\pi})$ for $D\pi$;
$s_{x}=p(u)$ indicates that a pole is on the physical
 (unphysical) sheet of the channel $x$.
}    
\begin{ruledtabular}
\begin{tabular}{lccc}
$\{L,I\}$& $M_{\rm pole}$ (MeV)  & RS & name \\\hline
\multirow{3}{*}{\{0, 0\}} & $461 - 252i$ & $(up)$  & $f_0(500)$ \\
 & $994 - 11i$ & $(up)$  & $f_0(980)$ \\
 & $1426 - 204i$ & $(uu)$  & $f_0(1370)$ \\\hline
\{2, 0\} & $1245 - 100i$ & $(uu)$  & $f_2(1270)$ \\\hline
\{0, 1/2\} & $2104-100i$  & $(u)$ & $D^*_0(2300)$ \\
\end{tabular}
\end{ruledtabular}
\end{table}

We make several remarks on the $Z_{c(s)}$ amplitudes.
We consider $L=0$ in Eq.~(\ref{eq:cont-ptl}), except for the
$h_c\pi$ channel for which $L=1$. 
By considering the heavy quark spin and SU(3) symmetries~\cite{Zc3900-mldu}, we set the coupling
strengths in Eq.~(\ref{eq:cont-ptl}) as 
$h_{[D^*\bar{D}],[D^*\bar{D}]}=h_{D^*\bar{D}^*,D^*\bar{D}^*}=h_{[D^*_s\bar{D}],[D^*_s\bar{D}]}$,
where 
$[D^*\bar{D}]\equiv {D^*\bar{D}-\bar{D}^*D\over \sqrt{2}}$ and 
$[D^*_s\bar{D}]\equiv {D_s^*\bar{D}-\bar{D}^*D_s\over \sqrt{2}}$.
Also, we set 
$h_{[D^*_s\bar{D}],J/\psi K}\sim h_{[D^*\bar{D}],J/\psi\pi}$
from the SU(3).
We assume no interactions between the hidden charm channels such as 
$J/\psi \pi$, $\psi'\pi$, $h_c\pi$, and $\eta_c\rho$.
For $\eta_c\rho$ loops, we consider the $\rho$ width ($\Gamma_\rho$).
Thus, we use 
Eq.~(\ref{eq:pw-cont-self}) with $\Gamma_a=\Gamma_\rho=150$~MeV and $\Gamma_b=0$
rather than Eq.~(\ref{eq:pw-cont-self2}).

\begin{table*}
\renewcommand{\arraystretch}{1.2}
\caption{\label{tab:h} 
Parameter values for 
$C_{\beta\alpha}$ in 
Eq.~(\ref{eq:cont-ptl-oc}).
Hyphens indicate unused parameters.
}    
\begin{ruledtabular}
\begin{tabular}{c|rrrrrrrrrrr}
$\beta\ \backslash \alpha$ & 
$D\bar{D}$ &
$D^*\bar{D}$ &
$D^*\bar{D}^*$ &
${D}_1\bar{D}$ & 
${D}_1\bar{D}^*$ & 
${D}_2^*\bar{D}^*$&
$D_s\bar{D}_s$ &
$D_s^*\bar{D}_s$ &
$D_s^*\bar{D}_s^*$ &
$D_{s1}\bar{D}_s$ & 
$\Lambda_c\bar{\Lambda}_c$
\\ \hline
$D\bar{D}$                 &$-$1.02  &  0.475 &  0.385 &  --    &  6.43  &$-$1.51  &  --    &  0.316 &  --    &  --    &  --    \\
$D^*\bar{D}$ 		   &        &$-$1.02  &  0.135 &$-$1.85  &  1.28  &  --    &  --    &  --    &  --    &  --    &$-$1.27  \\
$D^*\bar{D}^*$ 		   &        &        &$-$0.460 &  --    &  --    &  5.05  &  --    &  0.468 &  --    &  --    &  1.27  \\
${D}_1\bar{D}$  	   &        &        &        &$-$20.9   &  --    &  --    &  --    &  --    &  --    &  --    & 11.3   \\
${D}_1\bar{D}^*$  	   &        &        &        &        &$-$10.2   &$-$2.02  &$-$1.50  &  --    &  --    &  --    & 18.1   \\
${D}_2^*\bar{D}^*$	   &        &        &        &        &        &  --    &$-$0.887 &  --    &  --    &  --    &$-$3.39  \\
$D_s\bar{D}_s$ 		   &        &        &        &        &        &        &$-$1.01  &$-$0.0724&  0.508 &$-$4.08  &  --    \\
$D_s^*\bar{D}_s$ 	   &        &        &        &        &        &        &        &  --    &$-$1.15  &  --    &  --    \\
$D_s^*\bar{D}_s^*$	   &        &        &        &        &        &        &        &        &$-$0.181 &$-$2.60  &$-$1.00  \\
$D_{s1}\bar{D}_s$  	   &        &        &        &        &        &        &        &        &        &$-$30.5   &  --    \\
$\Lambda_c\bar{\Lambda}_c$ &        &        &        &        &        &        &        &        &        &        &$-$20.1   \\
  \end{tabular}
\end{ruledtabular}
\end{table*}

The coupling parameters in the two-meson scattering models are
determined as follows.
Our $f_0$ and $f_2$ amplitudes are fitted to the empirical $\pi\pi$ $s$
and $d$-wave phase shifts and inelasticities~\cite{pipi-data2}.
The $D\pi$ $s$-wave ($D^*_0$) amplitude is fitted to
an amplitude based on the lattice QCD (LQCD) spectrum~\cite{D0-lqcd}.
Our fits are shown in Fig.~\ref{fig:two-body}.
The $Z_c$ amplitude is determined in the global fit. 
Numerical values of the fitting parameters
are given in Tables~\ref{tab:pipi}--\ref{tab:Zc}.

After the fits,
resonance poles are extracted from the amplitudes 
and presented in Table~\ref{tab:pole-pipi}.
The $Z_c$ poles are presented in Table~\ref{tab:pole2}.
$Z_{cs}$ pole is not shown 
because our dataset hardly constrains it.
The $f_0$ and $f_2$ pole locations are consistent with the PDG averages~\cite{pdg}.
The LQCD-based amplitude has a 
$D^*_0$ pole at $2105^{+6}_{-8}-i 102^{+10}_{-12}$~MeV~\cite{D0-lqcd},
and our $D^*_0$ pole in
Table~\ref{tab:pole-pipi} is consistent.~\footnote{
The analyses of the LQCD spectra in Refs.~\cite{D0-lqcd,D0-lqcd2}
also identified a higher pole slightly below the $D_s\bar{K}$ threshold.
}
$D^*_0$ poles from previous LQCD analyses (at physical mass) 
are not yet well determined; for example, see Fig.~2 of Ref.~\cite{D0-Yan}.
Thus, our choice of the $D\pi$ $s$-wave amplitude in Fig.~\ref{fig:two-body}(c) should
be regarded as an assumption in our coupled-channel model.
Because the present global analysis does not include data of $D\pi$ invariant-mass
distribution in $e^+e^-\to \pi D \bar{D}^*$, the fit quality would not
sensitively depend on the $D^*_0$ pole location.

\subsection{
Contact interactions between open-charm channels}
\label{sec:supp:DD}

For the interactions $v^{\rm s}$ in Eq.~(\ref{eq:ptl}),
we consider contact interactions between
11 open-charm channels ($I J^{PC}=0 1^{--}$):
$D^{(*)}\bar{D}^{(*)}$, 
$D_s^{(*)}\bar{D}_s^{(*)}$, 
$D_1(2420)\bar{D}^{(*)}$,
$D_2^*(2460)\bar{D}^{*}$, 
$D_{s1}(2536)\bar{D}_s$, and
$\Lambda_c\bar{\Lambda}_c$.
Labeling the channels with $\alpha$ ($L$-wave, total spin $S$)
and $\beta$ ($L', S'$),
our interaction potential for 
an $\alpha\to\beta$ process is given by
\begin{eqnarray}
v^{\rm s}_{\beta,\alpha} (p',p) &=& f^{L'}_\beta(p') C_{\beta\alpha}\, f^L_\alpha(p) ,
\label{eq:cont-ptl-oc}
\end{eqnarray}
where $C_{\beta\alpha}$ is a coupling constant and 
$C_{\beta\alpha}=C_{\alpha\beta}$.
The dipole form factor $f^L_\alpha$ is given by
\begin{eqnarray}
f^L_\alpha(p) = {1\over \sqrt{4E_{1\alpha}E_{2\alpha}}}\!
\left(\Lambda^2 \over \Lambda^2 + p^2\right)^{\!\!2+L/2}\!\!\!
\left(p\over m_\pi\right)^{\!\!L},
\label{eq:cont-ff}
\end{eqnarray}
where $E_{i\alpha}$ is the energy of an $i$-th particle in the channel
$\alpha$; $\Lambda=1$~GeV is used.
We consider
$\{L,S\}=\{0,1\}$ for 
$D_1(2420)\bar{D}^{(*)}$, $D_2^*(2460)\bar{D}^{*}$, 
$D_{s1}(2536)\bar{D}_s$, and $\Lambda_c\bar{\Lambda}_c$, 
and 
$\{L,S\}=\{1,0\}, \{1,1\}, \{1,0\}$ for 
$D_{(s)}\bar{D}_{(s)}$, 
$D_{(s)}^{*}\bar{D}_{(s)}$, and 
$D_{(s)}^{*}\bar{D}_{(s)}^{*}$, respectively;
$\{L,S\}=\{1,2\}$ is not considered for 
$D_{(s)}^{*}\bar{D}_{(s)}^{*}$.
$C_{\alpha\beta}$ values from the global fit are listed in 
Table~\ref{tab:h}.

We have introduced two types of contact interactions in
Eqs.~(\ref{eq:cont-ptl}) and (\ref{eq:cont-ptl-oc}). 
The contact interactions of Eq.~(\ref{eq:cont-ptl}) 
[Eq.~(\ref{eq:cont-ptl-oc})] work on 
$ab$ [$Rc$] pair in Fig.~\ref{fig:diag}(a).

\begin{table*}
\renewcommand{\arraystretch}{1.2}
\caption{\label{tab:z-d} 
Values of $d_{R'c',Rc}$ in Eq.~(\ref{eq:z-pw}) for the $Rc\to R'c'$ $Z$-diagrams.
Exchanged particles ($\bar{c}$) or their charge-conjugates
are either indicated in the parentheses
or $\bar{c}=\pi$ for other entries with nonzero $d_{R'c',Rc}$.
}    
\begin{ruledtabular}
\begin{tabular}{c|rrrrrrrrrrrrrrr}
$Rc\ \backslash R'c'$ & 
$D\bar{D}$ &
$D^*\bar{D}$ &
$D^*\bar{D}^*$ &
${D}_0^*\bar{D}^*$ &
${D}_1\bar{D}$ & 
${D}'_1\bar{D}$ &
${D}_2^*\bar{D}$&
${D}_1\bar{D}^*$ & 
${D}'_1\bar{D}^*$ &
${D}_2^*\bar{D}^*$&
$f_0J/\psi$&
$f_2J/\psi$&
$f_0\psi'$&
$f_0 h_c$&
$Z_c\pi$\\ \hline
$D\bar{D}$ &$\cdots$&$\cdots$& $-2$ & $-\sqrt{2}$ &$\cdots$&$\cdots$&$\cdots$&$\cdots$&$\cdots$& $-\sqrt{2}$ &$\cdots$&$\cdots$&$\cdots$&$\cdots$& $\sqrt{2}(D^*)$\\
$D^*\bar{D}$  & & $-1$ & $-\sqrt{2}$ & $-1$ &$\cdots$&$\cdots$& $-1$ & $-1$ & $-1$ & $-1$&$\cdots$&$\cdots$&$\cdots$&$\cdots$& $1(D^*)$\\
$D^*\bar{D}^*$&&& $-2$ &$\cdots$& $-\sqrt{2}$ & $-\sqrt{2}$ &$-\sqrt{2}$&$-\sqrt{2}$&$-\sqrt{2}$&$-\sqrt{2}$&$\cdots$&$\cdots$&$\cdots$&$\cdots$& $-\sqrt{2}(D),-2(D^*)$\\
$D_0^*\bar{D}^*$  &&&&$\cdots$& $-1$ & $-1$ & $-1$ &$\cdots$&$\cdots$&$\cdots$&$\cdots$&$\cdots$&$\cdots$&$\cdots$& $-1(D)$\\
$D_1\bar{D}$   &&&&&$\cdots$&$\cdots$&$\cdots$&$\cdots$ &$\cdots$ &$-1$ &$\cdots$&$\cdots$&$\cdots$&$\cdots$& $1(D^*)$\\
$D'_1\bar{D}$   &&&&&&$\cdots$&$\cdots$&$\cdots$ &$\cdots$ &$-1$ &$\cdots$&$\cdots$&$\cdots$&$\cdots$& $1(D^*)$\\
$D_2^*\bar{D}$ &&&&&&&$-1$&$\cdots$ &$\cdots$ &$-1$ &$\cdots$&$\cdots$&$\cdots$&$\cdots$& $1(D^*)$\\
$D_1\bar{D}^*$ &&&&&&&&$-1$ &$-1$ &$-1$ &$\cdots$&$\cdots$&$\cdots$&$\cdots$& $-\sqrt{2}(D^*)$\\
$D'_1\bar{D}^*$&&&&&&&&&$-1$ &$-1$ &$\cdots$&$\cdots$&$\cdots$&$\cdots$& $-\sqrt{2}(D^*)$\\
$D_2^*\bar{D}^*$&&&&&&&&&&$-1$ &$\cdots$&$\cdots$&$\cdots$&$\cdots$& $-1(D),-\sqrt{2}(D^*)$\\
$f_0J/\psi$ &&&&&&&&&&&$\cdots$&$\cdots$&$\cdots$&$\cdots$& $1$\\
$f_2J/\psi$ &&&&&&&&&&&&$\cdots$&$\cdots$&$\cdots$& $1$\\
$f_0\psi'$ &&&&&&&&&&&&&$\cdots$&$\cdots$& $1$\\
$f_0 h_c$ &&&&&&&&&&&&&&$\cdots$& $1$\\
$Z_c\pi$ &&&&&&&&&&&&&&& $1(J/\psi),1(\psi')$\\
\end{tabular}
\end{ruledtabular}
\caption{\label{tab:z-d2} 
Continued from Table~\ref{tab:z-d}.
}    
\begin{ruledtabular}
\begin{tabular}{c|rrrr}
$Rc\ \backslash R'c'$ & 
${D}_{s1}\bar{D}_s$ & 
$f_0J/\psi$&
$f_2J/\psi$&
$Z_{cs}\bar{K}$\\ \hline
$D_{s1}\bar{D}_s$   &$\cdots$&$\cdots$&$\cdots$& $-1/\sqrt{2}(D^*)$\\
$f_0J/\psi$    &&$\cdots$&$\cdots$& $\sqrt{2}(K)$\\
$f_2J/\psi$    &&&$\cdots$& $\sqrt{2}(K)$\\
$Z_{cs}\bar{K}$&&&& $1(J/\psi)$\\
\end{tabular}
\end{ruledtabular}
\end{table*}

\section{Particle-exchange mechanisms ($Z$-diagrams)}
\label{sec:zgraph}

\begin{figure}[b]
\begin{center}
\includegraphics[width=.15\textwidth]{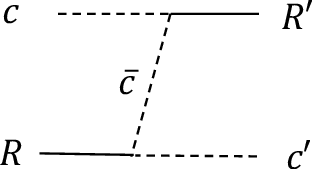}
\end{center}
 \caption{
Particle($\bar{c}$)-exchange $Rc\to R'c'$ mechanism ($Z$-diagram).
 }
\label{fig:z}
\end{figure}
As introduced in Eq.~(\ref{eq:ptl}),
we consider particle-exchange mechanisms ($Z$-diagrams) depicted in
Fig.~\ref{fig:z}.
These mechanisms are essential to satisfy the three-body coupled-channel
unitarity.
A particle($\bar{c}$)-exchange $Rc\to R'c'$ mechanism is given by
\begin{eqnarray}
Z^{\bar{c}}_{R'c',Rc}(\bm{p}_{c'},\bm{p}_{c}; E) =
{
[\Gamma_{c\bar{c},R'}(\bm{p}_c^*)]^*\,\Gamma_{c'\bar{c},R}(\bm{p}_{c'}^*)
\over E-E_c-E_{c'}-E_{\bar{c}} + i\epsilon} , 
\label{eq:z}
\end{eqnarray}
with 
$\bm{p}_{\bar{c}}=-\bm{p}_{c}-\bm{p}_{c'}$.
The $R\to c'\bar{c}$ vertices 
$\Gamma_{c'\bar{c},R}(\bm{p}_{a}^*)$ have been
defined in Eq.~(\ref{eq:pipi-vertex0}).
For a case where $c'\bar{c}$ interact via a contact interaction of
Eq.~(\ref{eq:cont-ptl}), 
Eq.~(\ref{eq:z}) is modified to include $h^{LI}_{c'\bar{c}',ab}$ and $w^{LI}_{c'\bar{c}}$ of Eq.~(\ref{eq:cont-ptl}), 
as detailed in Ref.~\cite{d-decay}.
Equation~(\ref{eq:z}) is projected onto a partial wave 
with the total angular momentum $J$, total isospin $I$, and $C=-1$:
\begin{widetext}
\begin{eqnarray}
Z^{\bar{c}\, IJ^{PC}}_{(R'c')_{l's'},(Rc)_{ls}} (p_{c'},p_c; E)
&=& d_{R'c',Rc}\,
{\sqrt{4\pi(2l+1)}\over (2I+1) (2J+1)}
\sum_{\rm (iso)spins}
(t_R t^z_R t_c t^z_c | I I^z)
(t_{R'} t^z_{R'} t_{c'} t^z_{c'} |  I I^z)
(s_R s^z_R s_c s^z_c | s s^z)
\nonumber\\
&&\times
(s_{R'} s^z_{R'} s_{c'} s^z_{c'} | s' s^{'z})
(l 0 s s^z | J M)
(l' m' s' s^{'z} | J M)
\int d\Omega_{\hat{p}} \,
Y^*_{l'm'}(\hat{p})
Z^{\bar{c}}_{R'c',Rc}(\bm{p}_{c'},p_{c}\bm{e}_z; E) , \nonumber\\
\label{eq:z-pw}
\end{eqnarray}
\end{widetext}
where $\bm{p}_c$ is taken along the $z$-axis and 
$\hat{p}=-\hat{p}_{c'}$;
 $\sum_{\rm (iso)spins}$ indicates the summation of all $z$-components
in the CG coefficients.
The use of $C=-1$ base states is accompanied by $d_{R'c',Rc}$ in Eq.~(\ref{eq:z-pw}), 
and their values along with the exchanged particles $\bar{c}$ 
are listed in Tables~\ref{tab:z-d} and \ref{tab:z-d2}.
Since the partial wave form of Eq.~(\ref{eq:z-pw})
with $IJ^{PC}=01^{--}$ is used in the present analysis,
the label $IJ^{PC}$ is suppressed in Eq.~(\ref{eq:ptl}).

\section{Pole uncertainty estimations}

\subsection{Vector charmonium poles}
\label{sec:vec-error}

In a standard procedure of estimating uncertainties of (resonance) pole values,
all fitting parameters are varied around the minimum $\chi^2$ to
generate an error matrix of the parameters.
The matrix is then used to propagate 
the parameter errors to the pole-value errors.
In our global coupled-channel analysis, however, this procedure is
practically impossible.
This is because 
the cross section calculation 
is rather time-consuming
for the four-dimensional phase-space integral in Eq.~(\ref{eq:xs-formula2}),
and we have too many 200 fitting parameters (Tables~\ref{tab:Zc},
\ref{tab:h}, \ref{tab:psi_param}--\ref{tab:Y4660_param}) in our model
to reach a convergence
in the $\chi^2$-minimization.
This problem is common in global coupled-channel analyses for nucleon
resonances~\cite{ystar,jb22}.
We thus use a practical uncertainty estimation method hinted by
Ref.~\cite{ystar}.

As discussed in Sec.~\ref{sec:fit-a},
we first adjust 200 fitting parameters
and obtain a default fit to 
the data shown in Figs.~\ref{fig:xs-opencc}--\ref{fig:hcpipi-etacrhopi}.
Then, we introduce complex parameters $\delta m_{\psi_i}$ 
into Eq.~(\ref{eq:mstar-g1}) as
\begin{eqnarray}
m_{\psi_i} \to 
m_{\psi_i} + 
\delta m_{\psi_i} .
\label{eq:res_bare}
\end{eqnarray}
To obtain an error matrix,
we select parameters that would relatively largely influence 
the pole values, and refit them to the data.
We vary 85 parameters in total:
$\delta m_{\psi_i}$;
$\psi(4660)$ BW mass and width in Eq.~(\ref{eq:y4660}); 
real scaling factors multiplied to 
the $\psi(4660)$ and $\psi(4710)$ amplitudes in Eq.~(\ref{eq:y4660})
and to the NR amplitude 
$\bar\Gamma^\mu_{R'c,\gamma^*}$ in Eq.~(\ref{eq:amp_full});
bare $\gamma^*\to \psi_i$ couplings
$g_{\psi_i}$ in Eq.~(\ref{eq:psi-prod2});
coupling constants $C^i_{(Rc)_{ls}}$ in Eq.~(\ref{eq:bare_mstar})
and $C^{\gamma^*}_{(Rc)_{ls}}$ in Eq.~(\ref{eq:nr_gamma})
for bare $\psi_i, \gamma^*\to$ open-charm channels;
coupling constants $C_{\beta\alpha}$ in Eq.~(\ref{eq:cont-ptl-oc})
that are diagonal ($\beta=\alpha$) or
whose absolute values are larger than
 10, 0.5, and  $\sqrt{10 \times 0.5}$ for
$s$-wave, $p$-wave, and $s$-wave $\leftrightarrow$ $p$-wave
interactions, respectively.
Effects of the other fixed parameters to pole uncertainty are simulated by 
${\rm Im}[\delta m_{\psi_i}]$ in Eq.~(\ref{eq:res_bare}).
The obtained error matrix is used to estimate 
the vector charmonium pole uncertainties in Table~\ref{tab:pole}
through the standard error propagation.
We note that 
the pole values from the default fit 
and those (central values) obtained from the refit
are slightly different; two reasons:
(i) The imaginary parts of $\delta m_{\psi_i}$ are degrees of
freedom not existing in the default model.
(ii) Some data points are weighted in the default fit 
so that the fit is overall reasonable.
The above error analysis is done with unweighted data.
The purpose of this error analysis is to estimate how much the pole
values can fluctuate due to the experimental errors. 
Thus, in Table~\ref{tab:pole},
the pole central values are from the default fit 
and their uncertainties are from the above uncertainty analysis.

\subsection{$Z_c$ poles}
\label{sec:zc-error}

We estimate uncertainties of the $Z_c$ poles as below.
We select parameters that
are presumably most relevant to the $Z_c$ poles,
and vary them around the default values to 
refit the data and obtain an error matrix;
the other parameters are fixed at their default-fit values. 
Then the error matrix is used to estimate 
$Z_c$-pole uncertainties through the standard error propagation.
The 63 fitting parameters in the refit are:
The $Z_c$ amplitude parameters in Table~\ref{tab:Zc};
$\delta m_{\psi_i}$ in Eq.~(\ref{eq:res_bare});
$\psi(4660)$ BW mass and width in Eq.~(\ref{eq:y4660}); 
real scaling factors multiplied to 
the $\psi(4660)$ and $\psi(4710)$ amplitudes in Eq.~(\ref{eq:y4660})
and to the NR amplitude; 
bare $\gamma^*\to \psi_i$ couplings
$g_{\psi_i}$ in Eq.~(\ref{eq:psi-prod2});
coupling constants $C^i_{(Rc)_{ls}}$ in Eq.~(\ref{eq:bare_mstar})
and $C^{\gamma^*}_{(Rc)_{ls}}$ in Eq.~(\ref{eq:nr_gamma})
for bare 
$\psi_i,\gamma^*\to 
(D_2\bar{D})_{22}, 
(h_c f_0^1)_{11}$, and 
$(Rc)_{01}$ 
with
$Rc=D^{(\prime)}_1\bar{D}^{(*)}$, 
$D_2\bar{D}^{*}$,
$J/\psi f_0^1$, and
$\psi' f_0^1$;
coupling constants $C_{\beta\alpha}$ in Eq.~(\ref{eq:cont-ptl-oc})
for which $\alpha=\beta=D_1\bar{D}^{(*)}$.
The $Z_c$ pole uncertainties are listed in Table~\ref{tab:pole2}.
The pole central values in Table~\ref{tab:pole2} are from the default fit for the reasons
discussed in Sec.~\ref{sec:vec-error}.

\section{Parameter values from the global fit and heavy-quark spin symmetry}

\subsection{Parameters from the global fit}
\label{app:para}

Parameter values determined from the global fit to the $e^+e^-\to c\bar{c}$
data are listed in Tables~\ref{tab:Zc}, \ref{tab:h}, \ref{tab:psi_param},
\ref{tab:cutoff}, and \ref{tab:Y4660_param}.

Parameters $C^{i}_{((ab)_{LI} c)_{ls}}$ in
Table~\ref{tab:psi_param} are defined as follows.
For a two-meson scattering model including contact interactions, 
we consider a direct bare $\psi \to abc$ decay where 
the $ab$ meson-pair has
an orbital angular momentum $L$
and a total isospin $I$.
This bare vertex function is given by
[cf. Eq.~(\ref{eq:bare_mstar})]
\begin{eqnarray}
F_{((ab)_{LI}c)_{ls},\psi_i}(q)\! &=& 
C^{i}_{((ab)_{LI} c)_{ls}}
\left({q\over m_\pi}\right)^l
\nonumber\\
&&\times 
\frac{ [1+q^2/(\Lambda^{i}_{((ab)_{LI} c)_{ls}})^2]^{-2-{l\over 2}} 
}
{\sqrt{4 E_c(q) m_{\psi_i}}} 
,
\label{eq:bare_mstar-contact}
\end{eqnarray}
where 
$C^{i}_{((ab)_{LI} c)_{ls}}$ and 
$\Lambda^{i}_{((ab)_{LI} c)_{ls}}$
are coupling and cutoff parameters, respectively. 
The dressed vertices [Eq.~(\ref{eq:dressed-ff})], 
dressed $\psi$ productions [Eq.~(\ref{eq:psi-prod})], 
and self energies [Eq.~(\ref{eq:mstar-sigma})] 
include the bare vertices
$F_{(Rc)_{ls},\psi_i}$ in Eq.~(\ref{eq:bare_mstar}) and also
$F_{((ab)_{LI} c)_{ls},\psi_i}$ in Eq.~(\ref{eq:bare_mstar-contact})
in a similar manner.

A remark on the cutoffs in Table~\ref{tab:cutoff} is in order.
As discussed in the main text, 
we fix most cutoffs 
[$\Lambda^i_{(Rc)_{ls}}$ in Eq.~(\ref{eq:bare_mstar})
for bare $\psi_i\to Rc$ vertices ($i=1,\cdots,5$)
and $\Lambda^{\gamma^*}_{(Rc)_{ls}}$ in Eq.~(\ref{eq:nr_gamma})
for nonresonant $\gamma^*\to Rc$ vertices]
to 1~GeV, and some of them to 0.7~GeV. 
However, cutoffs in the nonresonant 
$\gamma^*\to D_{(s)}^{(*)}\bar{D}_{(s)}^{(*)}, \Lambda_c\bar{\Lambda}_c$ vertices need to be 
adjusted to fit the 
$e^+e^- \to D_{(s)}^{(*)}\bar{D}_{(s)}^{(*)}, \Lambda_c\bar{\Lambda}_c$ data.

\subsection{Heavy-quark spin symmetry (HQSS)}
\label{app:para2}

It is interesting to examine the consistency between the parameters
determined from the global fit and those following the HQSS.
The HQSS relation among
$p$-wave $D\bar{D}-D^*\bar{D}-D^*\bar{D^*}$ coupled-channel interactions
is given in Ref.~\cite{Du-hqss}, 
and that for $s$-wave $D_1\bar{D}-D_1\bar{D}^*-D_2^*\bar{D^*}$
interactions in 
Ref.~\cite{review_guo}.\footnote{See Table~VI and Eq.~(85) in arXiv:1705.00141v3.}
We find HQSS-based short-range interactions close to those in Table~\ref{tab:h} 
by adjusting independent interaction parameters such as
$C_i$ in Eqs.~(9) and (10) of Ref.~\cite{Du-hqss}
and $F^{d(c)}_{Ij_\ell}$ in Eq.~(82) of Ref.~\cite{review_guo}.
The result is shown in Tables~\ref{tab:hqss1}\footnote{
We can relate $C_i$ in
Ref.~\cite{Du-hqss} (GeV${}^{-2}$) and those in Table~\ref{tab:hqss1} (dimensionless)
by dividing the latter by $m_\pi^2$.}
and \ref{tab:hqss2}.
In Table~\ref{tab:hqss1}, we show two cases where
the HQSS-violating effect is either not included ($C_1=C_3$) 
or included ($C_1\ne C_3$). 
Our parameter set is fairly consistent with ``HQSS($C_1\ne C_3$)'' 
in Table~\ref{tab:hqss1}
and ``HQSS'' in Table~\ref{tab:hqss2}.

Regarding the bare $\psi_i\to (Rc)_{ls}$ branchings,
the HQSS also provides relations as given by
${\cal B}^{\rm HQSS}_{\psi(S)\to\alpha}$ and ${\cal B}^{\rm HQSS}_{\psi(D)\to\alpha}$
in Table~\ref{tab:hqss3}; $\alpha\equiv (Rc)_{ls}$.
The relation between $D\bar{D}-D^*\bar{D}-D^*\bar{D^*}$ channels
and that between $D_1\bar{D}-D_1\bar{D}^*-D_2^*\bar{D^*}$ channels
are independent.
We examine to what extent the relations hold in our model
by listing in Table~\ref{tab:hqss4}
\begin{eqnarray}
R^i_{S,\alpha} \equiv {|C^i_\alpha|^2 \over 
{\cal B}^{\rm HQSS}_{\psi(S)\to\alpha}} ,
\quad
R^i_{D,\alpha} \equiv {|C^i_\alpha|^2 \over 
{\cal B}^{\rm HQSS}_{\psi(D)\to\alpha}} ,
\label{eq:br}
\end{eqnarray}
where $C^i_\alpha$ is 
the bare $\psi_i\to \alpha$ coupling constant defined in
Eq.~(\ref{eq:bare_mstar}) and given in Table~\ref{tab:psi_param}.
The HQSS indicates that,
for a given $i$,
 either of $R^i_{S,\alpha}$ or $R^i_{D,\alpha}$
to be the same among $\alpha=D\bar{D}$, $D^*\bar{D}$, and $D^*\bar{D^*}$, 
and among $\alpha=D_1\bar{D}$, $D_1\bar{D}^*$, $D_2^*\bar{D^*}$, 
respectively.
However, we do not find such a tendency in Table~\ref{tab:hqss4},
showing large HQSS-violations in our model. 
We can find arguments 
that the HQSS can be
often broken badly above the open-charm thresholds in Ref.~\cite{hqss-violation}.
Yet, an interesting future work would be to perform a global fit
that implements the HQSS constraints to some extent.

\begin{table*}
\renewcommand{\arraystretch}{1.3}
\caption{\label{tab:psi_param} 
Parameter values for $i$-th bare $\psi$ states ($i=1,\cdots,5$).
Bare $\psi_i$ masses
[$m_{\psi_i}$ in Eq.~(\ref{eq:mstar-g1})],
bare $\psi_i\to Rc$ coupling constants 
[$C^i_{(Rc)_{ls}}$ in Eq.~(\ref{eq:bare_mstar})], 
and bare $\psi_i$ photo-production couplings
[$g_{\psi_i}$ in Eq.~(\ref{eq:psi-prod2})]
are listed.
In the last column ($i=\gamma^*$), nonresonant 
$\gamma^*\to Rc$ coupling constants 
[$C^{\gamma^*}_{(Rc)_{ls}}$ in Eq.~(\ref{eq:nr_gamma})]
are listed.
Hyphens indicate parameters either unused (thus 0) or non-existent.
}    
\begin{ruledtabular}
\begin{tabular}{lrrrrrr}
 & $i=1$& $i=2$& $i=3$& $i=4$& $i=5$& $i=\gamma^*$ \\ \hline
$m_{\psi_i}$(MeV)                          &      3785&      4199&      4354&      4518&      4522&       -- \\
$C^i_{(D_1\bar{D}  )_{01}}$		   &   --     &$-$19.1    &   --     &  18.7    &  29.2    &  2.29    \\
$C^i_{(D_1\bar{D}  )_{21}}$		   &   --     &   --     &   --     &   --     &   --     &   --     \\
$C^i_{(D'_1\bar{D} )_{01}}$		   &   --     &   --     &   --     &  31.3    &  6.39    &  3.88    \\
$C^i_{(D'_1\bar{D} )_{21}}$		   &   --     &   --     &   --     &   --     &   --     &   --     \\
$C^i_{(D^*_2\bar{D})_{22}}$		   &   --     &   --     &   --     &  2.54    &   --     &$-$0.307    \\
$C^i_{(D_1\bar{D}^*  )_{01}}$		   &   --     &   --     &   --     &$-$26.1    &$-$56.7    &$-$1.70    \\
$C^i_{(D_1\bar{D}^*  )_{21}}$		   &   --     &   --     &   --     &   --     &   --     &   --     \\
$C^i_{(D'_1\bar{D}^* )_{01}}$		   &   --     &  14.5    &  35.9    &   --     &   --     &$-$0.760    \\
$C^i_{(D'_1\bar{D}^* )_{21}}$		   &   --     &   --     &   --     &   --     &   --     &   --     \\
$C^i_{(D^*_2\bar{D}^*)_{01}}$		   &   --     &   --     &   --     &   --     &$-$39.3    &  3.19    \\
$C^i_{(D^*_2\bar{D}^*)_{21}}$		   &   --     &   --     &   --     &   --     &   --     &   --     \\
$C^i_{(D\bar{D}    )_{10}}$    		   &  1.87    &$-$0.378    &$-$1.63    &$-$1.15    &$-$5.37    &$-$0.197    \\
$C^i_{(D^*\bar{D}  )_{11}}$    		   &   --     &$-$1.07    & 0.446    &$-$1.00    &$-$1.94    & 0.980    \\
$C^i_{(D^*\bar{D}^*)_{10}}$    		   &   --     &  6.87    &$-$5.43    &  1.66    &$-$2.34    & 0.418    \\
$C^i_{(D_{s1}\bar{D}_s)_{01}}$    		   &   --     &   --     &   --     &   --     &   --     &$-$0.0520   \\
$C^i_{(D_s\bar{D}_s)_{10}}$    		   &   --     &$-$0.246    &  1.18    & 0.0533   &  1.58    & 0.234    \\
$C^i_{(D_s^*\bar{D}_s)_{11}}$  		   &   --     &  1.31    &  5.00    &  3.41    &$-$3.15    & 0.557    \\
$C^i_{(D_s^*\bar{D}_s^*)_{10}}$		   &   --     &$-$3.85    &$-$3.42    & 0.0924   &$-$2.87    &  1.04    \\
$C^i_{(f_0^1 J/\psi )_{01}}$		   &   --     &$-$1.76    &  4.68    &   --     &$-$3.46    & 0.582    \\
$C^i_{(f_0^1 J/\psi )_{21}}$		   &   --     & 0.209    &$-$0.556    &   --     & 0.486    &$-$0.0157   \\
$C^i_{(f_0^2 J/\psi )_{01}}$		   &   --     &   --     &$-$8.63    &   --     &   --     &   --     \\
$C^i_{(f_0^2 J/\psi )_{21}}$		   &   --     & 0.214    &$-$0.413    &   --     &   --     &$-$0.0437   \\
$C^i_{((\pi\pi)_{00}J/\psi)_{01}}$		   &   --     &$-$0.532    &  2.78    &   --     &$-$0.606    &   --     \\
$C^i_{((\pi\pi)_{00}J/\psi)_{21}}$		   &$-$0.0298   & 0.0356   &$-$0.160    &   --     & 0.111    &   --     \\
$C^i_{((K\bar{K})_{00}J/\psi)_{01}}$ 	   &   --     &   --     &$-$0.592    &$-$0.174    &$-$0.232    & 0.0347   \\
$C^i_{((K\bar{K})_{00}J/\psi)_{21}}$	   &   --     &   --     &$-$0.0296   &   --     &   --     &$-$0.0729   \\
$C^i_{(f_2 J/\psi)_{01}}$			   &   --     &   --     &   --     &$-$0.311    &$-$1.91    &   --     \\
$C^i_{(f_2 J/\psi)_{21}}$			   &   --     &   --     &   --     &   --     &   --     &   --     \\
$C^i_{(f_0^1 \psi' )_{01}}$		   &   --     &$-$6.09    &  7.27    &  11.4    &$-$8.78    &$-$0.858    \\
$C^i_{(f_0^1 \psi' )_{21}}$		   &   --     &   --     &  1.24    &$-$1.31    &$-$0.329    &   --     \\
$C^i_{(f_0^2 \psi' )_{01}}$		   &   --     &  6.37    &$-$10.5    &$-$12.2    &  16.2    &$-$0.541    \\
$C^i_{(f_0^2 \psi' )_{21}}$		   &   --     &$-$2.75    &$-$1.01    &$-$1.96    &$-$3.05    &$-$0.286    \\
$C^i_{((\pi\pi)_{00}\psi')_{01}}$		   &   --     &$-$1.15    &  1.68    &  2.26    &$-$1.89    &$-$0.236    \\
$C^i_{((\pi\pi)_{00}\psi')_{21}}$		   &   --     & 0.300    & 0.216    & 0.0751   & 0.399    & 0.0414   \\
$C^i_{(f_0^1 h_c )_{11}}$			   &   --     &   --     &$-$2.14    &  1.31    &  1.74    &   --     \\
$C^i_{(f_0^2 h_c )_{11}}$			   &   --     &   --     &   --     &   --     &   --     &   --     \\
$C^i_{((\pi\pi)_{00}h_c)_{11}}$		   &   --     &$-$0.190    &   --     & 0.584    &   --     &   --     \\
$C^i_{(J/\psi \eta)_{11}}$			   & 0.101    & 0.456    & 0.127    &$-$0.347    &$-$0.414    &   --     \\
$C^i_{(J/\psi \eta')_{11}}$		   &   --     &$-$0.0659   &$-$0.229    &$-$0.0989   & 0.130    &   --     \\
$C^i_{(\omega\chi_{c0})_{01}}$     	   &   --     &$-$1.20    & 0.556    &$-$2.53    &$-$3.28    &   --     \\
$C^i_{(\Lambda_c\bar{\Lambda}_c)_{01}}$       &   --     &   --     &   --     &   --     &   --     &  1.22    \\
$g_{\psi_i}$	  			   &  25.1    &  22.1    &$-$109.    &  115.    &$-$47.3    &       -- \\
\end{tabular}
\end{ruledtabular}
\end{table*}

\begin{table}
\renewcommand{\arraystretch}{1.4}
\caption{\label{tab:cutoff}
Numerical values of cutoffs (unit: MeV) 
for bare $\psi_i\to Rc$ vertices
[$\Lambda^i_{(Rc)_{ls}}$ in Eq.~(\ref{eq:bare_mstar})]
and nonresonant bare $\gamma^*\to Rc$ vertices
[$\Lambda^{\gamma^*}_{(Rc)_{ls}}$ in Eq.~(\ref{eq:nr_gamma})].
We use cutoffs not dependent on the label `$i$' in $\psi_i$.
Cutoffs not shown are fixed to 1000~MeV.
}    
\begin{ruledtabular}
\begin{tabular}{lr}
$\Lambda^{\gamma^*}_{(D\bar{D}    )_{10}}$             &   1373\\
$\Lambda^{\gamma^*}_{(D^*\bar{D}  )_{11}}$	            &   1395\\
$\Lambda^{\gamma^*}_{(D^*\bar{D}^*)_{10}}$	            &   1999\\
$\Lambda^{\gamma^*}_{(D_s\bar{D}_s)_{10}}$	            &   1052\\
$\Lambda^{\gamma^*}_{(D_s^*\bar{D}_s)_{11}}$           &    700\\
$\Lambda^{\gamma^*}_{(D_s^*\bar{D}_s^*)_{10}}$         &    903\\
$\Lambda^{\gamma^*}_{(\Lambda_c\bar{\Lambda}_c)_{01}}$ &   1838\\\hline
$\Lambda^i_{(f_0^1 \psi' )_{01}}$                  &    700 (fixed)\\
$\Lambda^i_{(f_0^1 \psi' )_{21}}$		   	&    700 (fixed)\\
$\Lambda^i_{(f_0^2 \psi' )_{01}}$		   	&    700 (fixed)\\
$\Lambda^{\gamma^*}_{(f_0^2 \psi' )_{21}}$		&    700 (fixed)\\
$\Lambda^{\gamma^*}_{((\pi\pi)_{00}\psi')_{21}}$	&    700 (fixed)\\
\end{tabular}
\end{ruledtabular}
\end{table}

\begin{table}
\renewcommand{\arraystretch}{1.4}
\caption{\label{tab:Y4660_param} 
Numerical values for parameter in $\psi(4660)$ and $\psi(4710)$ 
BW amplitudes of Eq.~(\ref{eq:y4660}); $\psi_6=\psi(4660)$
and $\psi_7=\psi(4710)$.
 }    
\begin{ruledtabular}
\begin{tabular}{lrr}
 & $i=6$& $i=7$ \\ \hline
$m_{\psi_i}$(MeV)                          &   4655             &   4710 (fixed)\\	  
$\Gamma_{\psi_i}$(MeV)                     &    134	     &    152\\	  
$\phi_{\psi_i}$			       	   &       2.02         &     $-$2.87   \\
$C^i_{(D_1\bar{D}^*  )_{01}}/g_{\psi_i}$		&      0.523         &     $-$0.559  \\
$C^i_{(D_{s1}\bar{D}_s)_{01}}/g_{\psi_i}$		&      0.0178	     &      --       \\
$C^i_{(D_s\bar{D}_s)_{10}}/g_{\psi_i}$   		&      0.00          &     $-$0.0157\\ 
$C^i_{(D_s^*\bar{D}_s)_{11}}/g_{\psi_i}$ 		&      0.00663	     &     $-$0.0221\\ 
$C^i_{(D_s^*\bar{D}_s^*)_{10}}/g_{\psi_i}$		&      0.0365	     &      0.0588\\	  
$C^i_{(\Lambda_c\bar{\Lambda}_c)_{01}}/g_{\psi_i}$ &     $-$0.279       &      --       \\
$C^i_{(f_0^1 \psi' )_{01}}/g_{\psi_i}$             &     $-$0.253       &      --       \\
$C^i_{((\pi\pi)_{00}\psi')_{01}}/g_{\psi_i}$       &     $-$0.0473	     &      --       \\
\end{tabular}
\end{ruledtabular}
\end{table}

\begin{table}
\renewcommand{\arraystretch}{1.4}
\caption{\label{tab:hqss1} 
Short-range interaction parameters, 
$C_{\alpha\beta}$ in Eq.~(\ref{eq:cont-ptl-oc}),
from our global fit and those based on HQSS.
The $p$-wave channels are labeled by $\alpha (\beta)=1$:$D\bar{D}$;
 2:$D^*\bar{D}$; 3:$D^*\bar{D}^*(S=0)$; 
4:$D^*\bar{D}^*(S=2)$. See the text for details.
 }    
\begin{ruledtabular}
\begin{tabular}{lrrr}
$C_{\alpha\beta}$    & Global fit  &  HQSS ($C_{1}=C_{3}$) &  HQSS ($C_{1}\ne C_{3}$)  \\\hline
$C_{11}$        &  $-$1.023  &    $-$0.748 &     $-$0.987\\
$C_{12}$        &   0.475  &     0.160 &      0.298\\
$C_{13}$        &   0.384  &     0.171 &      0.426\\
$C_{14}$        &   --     &    $-$0.431 &     0.558 \\
$C_{22}$        &  $-$1.022  &    $-$1.205 &     $-$1.022\\
$C_{23}$        &   0.135  &    $-$0.092 &     $-$0.172\\
$C_{24}$        &   --     &    0.749  &   $-$0.645  \\
$C_{33}$        &  $-$0.459  &    $-$0.550 &     $-$0.495\\
$C_{34}$        &    --    &     0.249 &    $-$0.322\\
$C_{44}$        &    --    &    $-$1.476 &    $-$1.175\\\hline
$C_{1}$     &    --       &    $-$0.451       &  $-$0.249\\
$C_{2}$     &    --       &    $-$2.279       &  $-$0.387\\
$C_{3}$     &    --       &    $-$0.451       &  $-$2.252\\
$C_{4}$     &    --       &    $-$0.799       &  $-$0.791\\
\end{tabular}
\end{ruledtabular}
\end{table}

\begin{table}
\renewcommand{\arraystretch}{1.4}
\caption{\label{tab:hqss2} 
Continued from Table~\ref{tab:hqss1}.
The $s$-wave channels are labeled by 
$\alpha (\beta)=5$:$D_1\bar{D}$; 6:$D_1\bar{D}^*$; 7:$D_2^*\bar{D}^*$. 
 }    
\begin{ruledtabular}
\begin{tabular}{lrr}
$C_{\alpha\beta}$    & Global fit  &  HQSS \\\hline
$C_{55}$       &     $-$20.9  &    $-$18.0      \\
$C_{56}$       &        0   &     8.96      \\
$C_{57}$       &        0   &     3.78      \\
$C_{66}$       &     $-$10.2  &    $-$11.5      \\
$C_{67}$       &     $-$2.02  &    $-$3.08     \\
$C_{77}$       &        0   &    $-$6.70     \\\hline
$F^{d}_{01}$ &    --   &      $-$5.32\\
$F^{d}_{02}$ &    --   &      $-$12.8\\
$F^{c}_{01}$ &    --   &      0.365\\
$F^{c}_{02}$ &    --   &       12.8\\
\end{tabular}
\end{ruledtabular}
\end{table}

\begin{table}
\renewcommand{\arraystretch}{1.4}
\caption{\label{tab:hqss3} 
HQSS-based
relative branchings
for $S$ and $D$-wave charmonium decays
to open-charm channels $\alpha$,
denoted by
${\cal B}^{\rm HQSS}_{\psi(S)\to\alpha}$ and ${\cal B}^{\rm HQSS}_{\psi(D)\to\alpha}$,
respectively.
 }    
\begin{ruledtabular}
\begin{tabular}{ccccc}
$\alpha(l=1)$ & $D\bar{D}$& $D^*\bar{D}$& $D^*\bar{D}^*(S=0)$ & $D^*\bar{D}^*(S=2)$\\\hline
${\cal B}^{\rm HQSS}_{\psi(S)\to\alpha}$ &$1\over 12$ & $1\over 3$&
	     ${1\over 36}$& ${5\over 9}$\\
${\cal B}^{\rm HQSS}_{\psi(D)\to\alpha}$ &$5\over 12$ & ${5\over 12}$ &
	     ${5\over 36}$&${1\over 36}$\\
\end{tabular}
\begin{tabular}{cccc}
$\alpha(l=0)$ & $D_1\bar{D}$ & $D_1\bar{D}^*$ & $D_2^*\bar{D}^*$  \\\hline
${\cal B}^{\rm HQSS}_{\psi(S)\to\alpha}$ &0&0&0\\
${\cal B}^{\rm HQSS}_{\psi(D)\to\alpha}$ & ${5\over 8}$&	 ${5\over 16}$&  ${1\over 16}$\\
\end{tabular}
\end{ruledtabular}
\end{table}

\clearpage

\begin{table}
\renewcommand{\arraystretch}{1.4}
\caption{\label{tab:hqss4} 
Ratios $R^i_{S,\alpha}$ and $R^i_{D,\alpha}$ 
defined in Eq.~(\ref{eq:br}).
 }    
\begin{ruledtabular}
\begin{tabular}{crrr}
$\alpha(l=1)$ & $D\bar{D}$ & $D^*\bar{D}$ & $D^*\bar{D}^*(S=0)$ \\\hline
$R^1_{S,\alpha}$ &           42   &        --      &     -- \\
$R^1_{D,\alpha}$ &            8   &        --      &     -- \\
$R^2_{S,\alpha}$ &            2   &         3      &   1701 \\
$R^2_{D,\alpha}$ &           0.3  &         3      &    340 \\
$R^3_{S,\alpha}$ &           32   &        0.6     &    1060\\
$R^3_{D,\alpha}$ &            6   &        0.5     &     212\\
$R^4_{S,\alpha}$ &           16   &         3      &     99 \\
$R^4_{D,\alpha}$ &            3   &         2      &     20 \\
$R^5_{S,\alpha}$ &          346   &        11      &    198 \\
$R^5_{D,\alpha}$ &           69   &         9      &     40 \\
\end{tabular}
\begin{tabular}{crrr}
$\alpha(l=0)$ & $D_1\bar{D}$ & $D_1\bar{D}^*$ & $D_2^*\bar{D}^*$  \\\hline
$R^1_{D,\alpha}$ &            --   &         --   &     --  \\
$R^2_{D,\alpha}$ &          581    &        --    &     --  \\
$R^3_{D,\alpha}$ &            --   &         --   &     --  \\
$R^4_{D,\alpha}$ &          558    &     2185     &     --  \\
$R^5_{D,\alpha}$ &         1364    &    10280     &   24662 \\
\end{tabular}
\end{ruledtabular}
\end{table}



\end{document}